\documentclass[a4paper,fleqn]{cas-sc}

\usepackage[authoryear]{natbib}
\usepackage[dvips]{epsfig}
\usepackage{graphicx}
\usepackage[scriptsize]{subfigure}
\usepackage{amsfonts}
\usepackage{amssymb}
\usepackage{amsmath}
\usepackage{rotating}
\usepackage{longtable}  
\usepackage{helvet}
\usepackage{epsf}
\usepackage{scalefnt}
\usepackage{placeins}
\usepackage{pst-all}
\usepackage{algorithm}% http://ctan.org/pkg/algorithms
\usepackage{algpseudocode}% http://ctan.org/pkg/algorithmicx
\usepackage{rotating} % Rotating table

\usepackage{mathtools}

\usepackage{booktabs}
\usepackage{multirow}
\usepackage{tabularx}

\newcommand{\cm}[1]{}

\usepackage{amsthm}

\usepackage[gen]{eurosym}
\usepackage{hyperref}
\usepackage{nomencl}
\usepackage{etoolbox}

\renewcommand\nomgroup[1]{%
  \item[\bfseries
  \ifstrequal{#1}{A}{Sets and indices}{%
  \ifstrequal{#1}{B}{Parameters}{%
  \ifstrequal{#1}{C}{Variables}{}}}%
]}

\def\argmin{\text{argmin}}

\usepackage{xcolor}

\renewcommand{\baselinestretch}{1.0}
\def\tsc#1{\csdef{#1}{\textsc{\lowercase{#1}}\xspace}}
\tsc{WGM}
\tsc{QE}
\ExplSyntaxOn
\cs_set:Npn \__first_footerline:
{
  \group_begin:
  \small
  \sffamily \__short_authors: 
  { \rmfamily \itshape  }
  \group_end:
}
\ExplSyntaxOff

\begin{document}
\let\WriteBookmarks\relax
\def\floatpagepagefraction{1}
\def\textpagefraction{.001}

\shorttitle{A hybrid model for day ahead electricity price forecasting}    

\shortauthors{Watermeyer et al.}  

\title [mode = title]{A hybrid model for day-ahead electricity price forecasting: Combining fundamental and stochastic modelling }  


%\tnotetext[1]{We are grateful to xxx for their valuable comments.} 


\author[1]{ Mira Watermeyer}[]
\cormark[1]

%\fnmark[3]

\ead{mira.watermeyer@kit.edu}

\ead[url]{https://www.as.ior.kit.edu/}

\author[2]{Thomas Möbius}[]

\cormark[2]

%\fnmark[3]

\ead{thomas.moebius@b-tu.de}

\ead[url]{https://www.b-tu.de/fg-energiewirtschaft/}

\author[1]{ Oliver Grothe}[]

\author[2]{ Felix Müsgens}[]

% \credit{<Credit authorship details>}

\affiliation[1]{organization={Karlsruhe Institute of Technology (KIT)},
            addressline={Institute for Operations Research}, 
            city={Karlsruhe},
%          citysep={}, % Uncomment if no comma needed between city and postcode
            postcode={76131}, 
            country={Germany}}

\affiliation[2]{organization={Brandenburg University of Technology (B-TU)},
            addressline={Chair of Energy Economics}, 
            city={Cottbus},
%          citysep={}, % Uncomment if no comma needed between city and postcode
            postcode={03046}, 
            country={Germany}}            

% \author[<aff no>]{<author name>}[<options>]

% Footnote of the second author

% Email id of the second author
% \ead{}


% % Credit authorship

% % Address/affiliation
% \affiliation[<aff no>]{organization={},
% %          citysep={}, % Uncomment if no comma needed between city and postcode

% % Corresponding author text
% \cortext[1]{Corresponding author}

% % Footnote text


\begin{abstract}
The accurate prediction of short-term electricity prices is vital for effective trading strategies, power plant scheduling, profit maximisation and efficient system operation. However, uncertainties in supply and demand make such predictions challenging. We propose a hybrid model that combines a techno-economic energy system model with stochastic models to address this challenge. The techno-economic model in our hybrid approach provides a deep understanding of the market. It captures the underlying factors and their impacts on electricity prices, which is impossible with statistical models alone. The statistical models incorporate non-techno-economic aspects, such as the expectations and speculative behaviour of market participants, through the interpretation of prices. The hybrid model generates both conventional point predictions and probabilistic forecasts, providing a comprehensive understanding of the market landscape. Probabilistic forecasts are particularly valuable because they account for market uncertainty, facilitating informed decision-making and risk management. Our model delivers state-of-the-art results, helping market participants to make informed decisions and operate their systems more efficiently.
\end{abstract}


%\item We provide a multi-step hybrid model that combines techno-economic and stochastic energy models. 
%\item The model generates accurate point and probabilistic forecasts for day-ahead electricity prices.  
%\item We prove that techno-economic energy system models can contribute to short-term price forecasting, especially when paired with stochastic models, to improve accuracy.
%\item The short-term forecasts' accuracy is comparable to purely statistical state-of-the-art models. 
%\item Probabilistic forecasts enable power plant operators to quantify the likelihood of prices becoming negative at any given hour.
%%\item data availability

% Keywords

\maketitle

% Main text
\section{Introduction}

Accurate forecasting in the energy sector is crucial for multiple stakeholders, including industry practitioners, researchers and policymakers. The effectiveness of financial and operational decisions and regulatory interventions depends on accurate predictions of future developments in relevant areas. 
As a result, the forecasting of electricity prices has become a key area of focus \citep{Weronreview2014}. With companies facing increasingly intense competition due to deregulation and liberalisation in the electricity sector, day-ahead price forecasts and insight into the next day's market situation are essential to the development of bidding strategies and production plans that maximise a company’s profit margins and ensure a reliable grid operation. 
Quantifying uncertainty has become increasingly important in recent years due to the growth of renewable energies and the need to integrate them alongside an increase in infrastructural challenges and fluctuating commodity prices, raising uncertainty in the energy market \citep{Nowotarski2018, Hong2016, Hong2020}. Probabilistic forecasts help in the planning and operation of energy systems, allowing for the assessment of uncertainty and the development of future strategies against the background of various probable future events \citep{Amjady2006EnergyPF}. 

Our paper presents a novel, open-source hybrid model that forecasts day-ahead electricity prices punctually and probabilistically by combining two main methodological streams: techno-economic energy system modelling and stochastic modelling. Techno-economic models are fundamental energy system models that determine (partial) market equilibria through the bottom-up optimisation of an energy system. They can explain actual developments and reflect structural breaks by identifying techno-economic interdependencies in energy markets. However, when estimating prices in the short term (e.g., day-ahead, intraday), these models exhibit larger and more systematic errors than other model classes. Stochastic models, on the other hand, learn from history and are developed and trained with historical data, enabling them to capture fluctuations and uncertainties in the market, especially in the short term. They offer high flexibility and the ability to specify forecast ranges and distributions. Still, they can only capture structural breaks and changes in external influences ex-post due to their dependence on historical data. 

Our proposed hybrid model combines the strengths of techno-economic energy system models and stochastic models to develop a more robust and accurate approach to forecasting electricity prices on the day-ahead market. The model retains the structural statements of techno-economic energy system models -- and, thus, insights into the driving market mechanisms -- while incorporating stochastic short-term structures and distribution functions to account for uncertainty. The model uses state-of-the-art methods to generate point and probabilistic price forecasts. These probabilistic price forecasts, with probabilities for each potential price scenario, are increasingly valuable to the industry (e.g., when assessing the probabilities of negative prices or when assessing the overall risk level of the price forecast). 

\begin{figure}[htbp]
	\centering
	\includegraphics[width=0.9\linewidth]{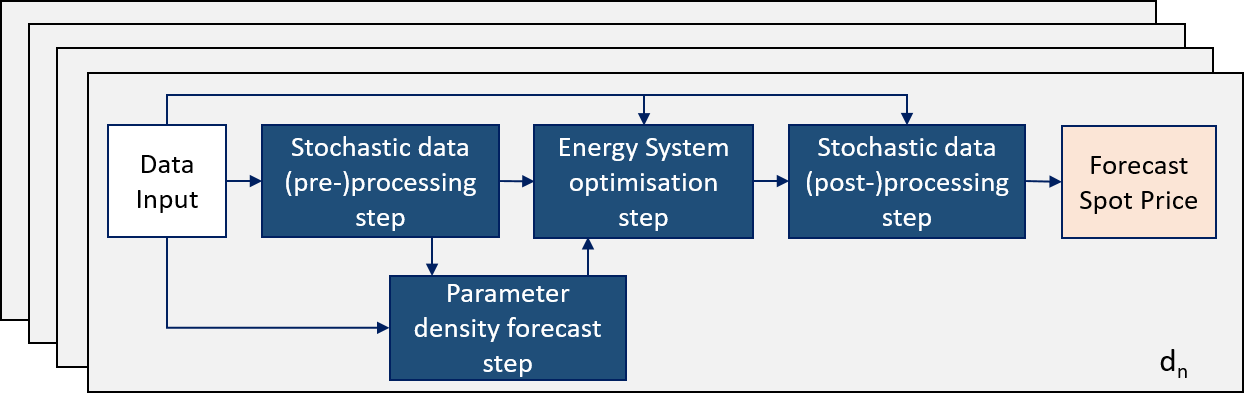}
	\caption{Scheme of the hybrid model.}
	\label{H_fig_scheme_hybrid}
\end{figure}

The model is schematically illustrated in Figure \ref{H_fig_scheme_hybrid}. It employs a rolling-window approach. In each iteration, it forecasts day-ahead prices exclusively through the use of data known prior to the day-ahead market’s closure, accurately reflecting the knowledge of stakeholders making decisions in these markets. The model is repeatedly applied each day ($d_n$) to generate forecasts for the following 24 hours of the day-ahead market. Each daily forecast includes four steps -- stochastic data pre-processing, parameter density forecast, energy system optimisation and stochastic data post-processing -- to produce point and probabilistic forecasts.

The first step, stochastic data pre-processing, aims to improve the accuracy of input data in advance of the energy system optimisation step and generates the basis for the parameter density forecast step. In the second step, the parameter density forecast generates prediction intervals for selected input parameters of the third step, energy system optimisation, enabling us to account for uncertainty in the operational decisions of market participants. This step considers the improved input data from the first and second steps. It forms a stochastic optimisation model that minimises total system costs, identifies the equilibrium between supply and demand and determines the hourly marginal system costs, which can be interpreted as price estimators.

These price estimators are the initial values in the fourth and final step, stochastic data post-processing. The errors of the price estimators are mapped using a multidimensional model in which seasonal effects and structures are captured using a combination of univariate and multivariate approaches, resulting in an enhanced price forecast. By modelling and improving the price forecast error, the model calculates forecast intervals and probability densities for the forecast prices, providing a quantification of uncertainty. Finally, our model can combine the strengths of both method classes to achieve excellent state-of-the-art price forecasts, including both point and probabilistic forecasts, to capture the stochastic uncertainty of the market. 

This paper contributes to the literature in three main ways. First, it presents a novel hybrid model that provides a general framework to combine techno-economic and stochastic energy models. Since the model’s source code and algebra are available in full online, other researchers can apply our methodology and extend it to different time periods and electricity markets around the world. Second, it proves that techno-economic energy system models can contribute to short-term price forecasting, especially when paired with stochastic models for the sake of error improvement. Our hybrid model delivers highly accurate day-ahead price forecasts on top of the insights that techno-economic models provide. We demonstrate its value with an empirical analysis based on European data with a focus on Germany, the largest European electricity market. Third, our hybrid model provides probabilistic forecasts in addition to point forecasts, enabling power plant operators to, for example, quantify the likelihood of prices becoming negative at any given hour. 

The remainder of this chapter is organised as follows. Section \ref{H_sec.lit} reviews the existing literature. Afterwards, we provide all of the information necessary to use our model and replicate this study in Section \ref{H_sec.data}. Section \ref{H_sec.method} describes our methodology. Following, Section \ref{H_sec.results.Hybrid} presents and evaluates the results of our hybrid model, while Section \ref{H_sec.analysis model steps} closely analyses the individual model steps. Finally, Section \ref{H_sec.concl} offers some concluding remarks.

\section{Literature}
\label{H_sec.lit}
Research on electricity prices has garnered the interest of many scholars due to the complexities and extraordinary challenges of achieving high accuracy in forecasting. They have developed and refined numerous methodological approaches to achieve high accuracy and adapt to changes in the electricity market. The number of relevant publications has increased rapidly over the last two decades. 

\citet{Weronreview2014} offers a detailed review of several approaches to forecasting electricity prices, including the following five model classes: multi-agent models, fundamental models, reduced-form models, statistical models, and artificial intelligence models. Previously, \cite{AGGARWAL200913} provide an overview of the methods used in electricity price forecasting. However, their focus was on stochastic time-series, causal and artificial intelligence-based models. \cite{Weron2019} and \cite{Hong2020} present a general review of and outlook on energy forecasting. The most recent overview of forecasting theory and practice comes from \cite{PETROPOULOS2022}, who provide an overview of a wide range of theoretical models, methods, principles and approaches to preparing, organising and evaluating forecasts. In addition, they provide several real-world examples of how these theoretical concepts are applied.

Many publications conduct time-series analysis based on time-series models, which are particularly suitable for short-term electricity price forecasting. Time series models constitute a special subtype of regression model in which target variables $y$ are represented, among other things, by past values of the time series as regressors $x$. They include autoregressive moving-average (ARMA), generalised autoregressive conditional heteroscedastic (GARCH) and Markov regime-switching (MS) models. \cite{Steinert2019}, for example, develop an auto-regressive model with 24 individual models -- one for each hour of the day -- that also incorporates electricity futures prices to produce hourly electricity price forecasts. \cite{Nowotarski2016} refer to the decomposition of values into different components, which is common in time-series analysis, and show that the quality of time-series models benefits greatly from decomposing a set of electricity prices into a long-term seasonal component and a stochastic component, modelling them independently and combining their forecasts. 
In an extensive study, \cite{Ziel2018} compare two options for the type of time-series modelling used with high-frequency data sets. They compare models with univariate model frameworks, with one of these models being constructed for the entire time series, featuring models with multivariate model frameworks, in which each hour of a day is modelled separately and independently. This is initiated by the organisation of electricity markets as day-ahead auctions, as in the U.S. or Europe. Their study shows no clear dominance by one framework, suggesting that the combination of both modelling approaches could improve predictive accuracy.

While \cite{Steinert2019}, \cite{Nowotarski2016} and \cite{Ziel2018} focus on day-ahead electricity prices in general, \cite{Christensen2012} use a nonlinear variant of the autoregressive conditional risk model to predict price peaks, treating them collectively as a discrete-time point process, \cite{Manner2014} an approach based on the autoregressive conditional hazard model, and \cite{MANNER2016255} the mapping of inter-regional linkages between different electricity markets in a dynamic multivariate binary choice model to predict electricity price spikes. \cite{Garcia2005} develop a GARCH model to predict day-ahead electricity prices, while \cite{HICKEY2012307} evaluate the accuracy of ARMAX-GARCH models in forecasting short-term prices in the U.S., determining that model choice depends largely on location, horizon and regulation, with asymmetric power auto-regressive conditional heteroskedastic (APARCH) models being more appropriate in deregulated markets and GARCH models being better for regulated markets. \cite{Bordignon2013} develop a linear regression model to account for relationships between prices and various price drivers, using a time-varying parameter (TVR) and an MS model to capture peaks and discontinuities. Other examples of applying MS models include \cite{KOSATER2006943} for the German market and \cite{Bierbrauer2004} for the Nordic market. Notably, in a recent paper, \cite{Mari2022} uses deep learning-based regime-switching models to predict electricity prices. 

Parameter-rich ARX models represent a special type of time-series model. The Lasso estimated autoregressive (LEAR) model introduced by \cite{Uniejewski2016} is further developed by \cite{lago2021forecasting}. 

To provide a set of best practices for evaluating future model developments in electricity price forecasting and comparing state-of-the-art statistical and deep-learning methods, \cite{lago2021forecasting} define a deep neuronal network (DNN) and a LEAR model based on the latest findings. Together with various evaluation metrics, these models are accessible in a Python toolbox to evaluate new algorithms. Accordingly, we compare our hybrid model with the statistical benchmark model. 

Deep-learning models (e.g., artificial neural network (ANN), DNN, long short-term memory (LSTM) network, recurrent neural network (RNN), feed-forward neural network) are used in an increasing share of electricity price forecasts. In addition to the cited benchmark model, \cite{Panapakidis2016}, as an example, study ANNs using different inputs and ANN typologies. These authors characterise such models as having comprehensive functionality and a high degree of flexibility. In their analysis of the impact of different markets on one another, \cite{Lago2018} develop a DNN that considers interconnected markets’ characteristics to boost forecasting accuracy. Notably, they show that predicting the price of two markets simultaneously enhances forecast accuracy. \cite{Amjady2006fuzzy} develops a fuzzy neuronal network that forecasts hourly electricity prices for the Spanish day-ahead market. 
Notably, the combination of deep learning and time-series models can be found in the nonlinear autoregressive neural network of \cite{MARCJASZ20191520}. 

To compare time-series and neural network models using external regressors, \cite{LEHNA2022105742} use four different forecasting approaches to the German day-ahead electricity market: a seasonal integrated autoregressive moving average ((S)ARIMA(X)) model, an LSTM neural network, a convolutional neural network LSTM (CNN-LSTM) and an extended two-stage multivariate vector autoregressive (VAR) model. While the LSTM model achieves the highest average accuracy, the two-stage VAR model has advantages at shorter prediction horizons. A combination of both methods outperforms each of the individual models in terms of accuracy.

The methods presented so far are fundamentally based on historical day-ahead electricity price time series. Another approach entails using models that simulate the actions of individual market participants (agents). \cite{Qussous2022}, for example, developed an agent-based model to derive day-ahead prices and simulate the bidding strategies of market participants. To evaluate their model, they reproduce day-ahead electricity prices in the 2016–2019 German bidding zone. Compared to other techno-economic approaches to short-term electricity price forecasting, this agent-based model achieves the highest accuracy. Consequently, we compare the hybrid model presented in this paper with this model. 

Due to their explanatory character in identifying an efficient market outcome and their comprehensive modelling of the entire electricity system, techno-economic energy system models have been employed for the ex-post analysis of electricity prices. \cite{Muesgens2006} and \cite{Borenstein2002} replicate day-ahead prices to assess the existence of market power in Germany and the U.S., respectively. \cite{KELES2013213} investigate on the importance of adequate wind power feed-in time series to obtain better results in electricity price simulation. \cite{HIRTH2013218} determine the market value of renewables and \cite{SENSFU20083086} quantify the merit order effect computing day-ahead prices. The merit order effect describes the displacement of fossil fuel generation by renewable energy sources due to their lower marginal costs and the subsequent decline in total electricity costs. \cite{Pape2016} analyse to what extent day-ahead and intraday electricity prices can be explained and represented by techno-economic energy system models. Notably, however, they show that this method has significant weaknesses in explaining short-term electricity prices compared to other methods. 

In contrast to ex-post analyses of electricity prices, which examine actual prices that have already occurred, this article focuses on ex-ante forecasts. Techno-economic energy system models, used for ex-ante prediction, have the key disadvantage they do not use recent historical prices to benchmark their price estimators. As a result, they may struggle to explain random short-term variations compared to econometric models that "learn from the past". However, energy system models possess an advantage in that they are based on established economic theory and replicate the workings of markets. As such, they are able to predict prices independently of past data and are less prone to structural changes in the market and other similar factors.
In line with that, we did not find techno-economic energy system model applications to forecasts for ex-ante predictions of short-term electricity prices. However, there are such uses in ex-ante predictions of long-term electricity prices, where random short-term variations are less inherent \citep[see, e.g.,][]{MUSGENS2020104107, GREEN20103211, LAMONT20081208}. In addition, technical-economic energy system models have been employed in bodies of literature that extend beyond price estimates (e.g., to determine the value of demand response \citep{MISCONEL2021117326, KIRCHEM2020114321}, to identify an optimal transmission-expansion plan \citep{VANDERWEIJDE20122089}, to support decision-making at the municipal level \citep{SCHELLER2019444}, to analyse the effect of power-to-gas \citep{LYNCH20191197}, to evaluate policy instruments to reduce CO\textsubscript{2} emissions \citep{SGARCIU2023113375}). Additionally, \cite{PLAGA2023120384} thoroughly examines how energy system models can account for climate uncertainty. A comprehensive overview of energy system modelling can be found in \cite{VENTOSA2005897}.

In recent years, hybrid methods have garnered significant attention in electricity price forecasting. Hybrid models are those that combine two or more distinct methods. They aim to use the combined strengths of the employed methods while mitigating their individual weaknesses to achieve better overall results. Many hybrid methods have been developed that combine a wide variety of methods. \cite{Aggarwal2017}, for example, present a hybrid approach that uses a wavelet transform, a time-series time-delay neural network and an error-predicting algorithm to predict day-ahead electricity prices in the ISO New England market. \cite{CHANG2019115804} combine an Adam-optimised LSTM neural network to generate electricity prices with a wavelet transform to decompose an electricity price time series into several series of electricity prices. A combination of an empirical wavelet transform, a support vector regression, a bi-directional LSTM and a Bayesian optimisation is proposed by \cite{CHENG2019653}. \cite{NAZAR2018214} apply a three-stage hybrid model to the DK2 area of Nordpool and the Spanish power market. The first stage features a wavelet and Kalman machines to decompose price data into different frequency components. The second stage uses an adaptive neuro-fuzzy inference system (ANFIS) to forecast price frequency components. In the third stage, the output of the second stage is fed into the ANFIS to boost forecasting accuracy. A wavelet transform and an ARMA are paired with a kernel-based extreme learning machine by \cite{YANG2017291}, and with a radial basis function neural network by \cite{Olamaee2016}. \cite{ZHANG2020114087} propose a hybrid model based on variational mode decomposition, self-adaptive particle swarm optimisation, SARIMA and a deep belief network for short-term electricity price forecasting.

Most of the hybrid models mentioned above use statistical and deep learning methods. However, a few applications also combine a techno-economic energy system model with another approach. For example, \cite{DEMARCOS2019240} detail a short-term hybrid electricity price forecasting model for the Iberian market that combines a techno-economic cost-generation optimisation model with an ANN. \cite{Gonzalez2012} propose two hybrid approaches based on a techno-economic electricity market model. Focusing on the day-ahead market in the UK, they combine this model type separately with two other models: a linear autoregressive model with exogenous data on price drivers (ARX model) and a nonlinear logistic regression model with a smooth transition (LSTR model), which is a regime change in times of structural change. Their results support the idea of incorporating fundamental information for better price forecasting. Particularly in highly volatile periods, the nonlinear hybrid model achieves better results. 
In \cite{Preprocessing}, our previous study, we introduced a techno-economic market model tailored to the day-ahead market and combined it with a stochastic model to enhance day-ahead load forecast accuracy in the estimation of day-ahead electricity prices. We highlighted the positive effects of better load forecasts on the day-ahead price estimators generated with an energy system model. However, this approach merely represents a first step; it does not fully realise the potential of a hybrid model, which seamlessly integrates the strengths of both the techno-economic and stochastic models. 

The literature on electricity price forecasting mostly focuses on developing point forecasting methods for the day-ahead market. However, in recent years, there has been a growing interest in probabilistic forecasting methods \citep{Hong2020}. The Global Energy Forecasting Competition (GEFCom2014) \citep[see][]{Hong2016} served as a catalyst for this trend, and many studies have been published on this topic in the time since. \cite{Nowotarski2018} provide a comprehensive overview of the different approaches used in this field. A hybrid model combining point and probabilistic forecasting in four steps was developed by \cite{MACIEJOWSKA20161051} for the GEFCom2014.

Common approaches to probabilistic electricity price forecasting include using time-series models, such as ARIMA, GARCH and exponential smoothing (ETS) \citep[e.g.,][]{WERON2008744} and using deep learning models. Bootstrapping is widely used in combination with deep learning approaches \citep[e.g.,][]{Chen2012, Wan2014, Rafiei2017, KHOSRAVI2013120}. On top of deep learning, \cite{Zhao2008} use a support vector machine (SVM) to estimate prediction intervals and density forecasts, and \cite{Zou2006} use an extended ARIMA model to do the same. An econometric model for probabilistic forecasting is proposed by \cite{Panagiotelis2008}. \cite{MANNER2019143} use vine-copula models to forecast quantiles for a vector of day-ahead electricity prices from interconnected electricity markets, while \cite{Kaechele2022} propose an approach based on copula techniques that entails generating multivariate probabilistic forecasts by modelling cross-hour dependencies. Considering these dependencies in probabilistic forecasts is uncommon, in contrast to point forecasts. However, including them in the methodology for generating probabilistic price forecasts can enhance forecast accuracy. 

Historical simulation and distribution-based prediction intervals are other popular approaches. Historical simulation estimates risk and generates prediction intervals in the simulation of multiple scenarios using historical data; it then uses the results to estimate the probability of different outcomes \citep[e.g.,][]{WERON2008744, Nowotarski2015}. Distribution-based prediction intervals are calculated based on the distribution of historical data \citep[e.g.,][]{MisiorekTrueckWeron2006, Zhao2008, DUDEK20161057, MACIEJOWSKA2016957, Panagiotelis2008}. A theoretical introduction to the generation of prediction intervals based on distribution and historical simulation is provided by \cite{WeronLoadbook2006}. 

Quantile regression averaging (QRA) is a method that has risen in prominence recently in probabilistic electricity price forecasting. It combines predictions from multiple quantile regression models, each of which is trained to predict a different quantile of the response variable. This method was first formally introduced by \cite{Nowotarski2015} and has since continued to be applied and developed further due to its accuracy \citep[e.g.,][]{MACIEJOWSKA2016957, Nowotarski2014, MARCJASZ2020466, UNIEJEWSKI2019171, Uniejewski2021}. 

Despite the rising prominence of probabilistic forecasts in various models, there is still a general lack of approaches that combine probabilistic forecasting with techno-economic energy system models. This paper aims to fill this gap in the literature. By adapting and developing an energy system model specifically for the short-term electricity market and combining this model with common stochastic models through multiple steps, we can leverage the strengths of both models and open up the field of short-term electricity price forecasting for energy system models. Having already highlighted the positive effects of combining a stochastic model (for better load forecasts) with an energy system model (for the day-ahead market, developed by \cite{Preprocessing}), these building blocks are included in the hybrid model. We demonstrate that a multi-layer hybrid model makes point and probabilistic price forecasting with techno-economic and stochastic models possible.

\section{Data}
\label{H_sec.data}

In our study, we develop a hybrid model that integrates stochastic modelling approaches and energy system optimisation to forecast wholesale electricity prices. Notably, this energy system optimisation requires multiple inputs. Table~\ref{H_table_parameter_data} provides an overview of the necessary input data. In this section, we provide more details on how the data is obtained and applied.

\begin{table}[htbp]
	%\bigskip
	\begin{center}
        \caption{Overview of required data}
	
%	\footnotesize
		\begin{tabular}{|l|l|}
		    \hline
			\multicolumn{1}{|c|}{Parameter} &
			\multicolumn{1}{c|}{Source} \\
			\hline

CO\textsubscript{2} prices	& \cite{Sandbag2020}		\\
\hline
Control power procurement & \cite{Regelleistung}	\\
\hline 
%Curtailment costs for RES	& own assumption		\\
%\hline
Efficiency of generation capacities	&	\cite{Schroeder2013}, 	\\
 & \cite{OPSDb}	\\
\hline
Efficiency losses at partial load	&	\cite{Schroeder2013} \\
\hline
Electricity demand (load)	& \cite{EntsoeTPe}		\\
\hline
Energy-power factor (for storages) &	own assumption: 9	\\
\hline
Fuel prices	&	\cite{Destatis2020},	\\
(Lignite, nuclear, coal, gas, oil)	&	\cite{EntsoS}, \cite{EntsoS}, \\
&  \cite{EEX2021}	\\
\hline
Generation and storage capacity	&	\cite{BNetzA2021}, \cite{UBA2020}, \cite{EBC2021},  \\
& \cite{EntsoeTPa}, \\
&  \cite{OPSDa}   \\
\hline
Generation by CHP units	&	\cite{EC2021}	\\
\hline
Historic electricity generation	&	\cite{EntsoeTPd}	\\
\hline
Load shedding costs	&	own assumption: 3,000 €/MWh	\\
\hline
Minimum output levels  &  \cite{Schroeder2013}	\\
\hline
NTCs	&	\cite{EntsoeTPf}, \\
&  \cite{JAO2021}	\\
\hline
Variable O\&M costs  &  \cite{Schroeder2013}	\\
\hline
Power plant outages	&	\cite{EntsoeTPb}	\\
\hline
RES feed-in	& \cite{EntsoeTPc}		\\
\hline
RES curtailment costs 	&  own assumption: 20 €/MWh		\\
\hline
Start-up costs	&	\cite{Schroeder2013}	\\
\hline
Seasonal availability of hydro power &   \cite{EntsoeTPd}	\\
\hline
Temperature (daily mean)	&	\cite{OPSDb}	\\
\hline
Water value 	&  \cite{EntsoeTPd}, 	\\
& \cite{EntsoeTPg}	\\
\hline

    \end{tabular}
	\label{H_table_parameter_data}
	\end{center}
\end{table}

\begin{figure}[htbp]
	\centering
	\includegraphics[width=0.7\linewidth]{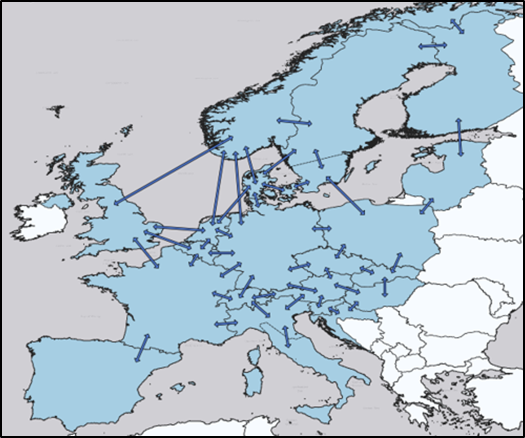}
	\caption{Geographical scope of the energy system model.}
	\label{H_fig_mapESM}
\end{figure}

Although our modelling approach can be applied to many markets, our empirical exercise focuses exclusively on the German spot market. However, the high level of integration among European electricity markets and the resulting interdependencies require a comprehensive representation of these markets, particularly during the energy system optimisation step. Figure \ref{H_fig_mapESM} shows the geographical scope of the collected data and the interconnection among European markets. We consider the bidding zones of most of the EU’s 27 member states\footnote{Bulgaria, Cyprus, Greece, Iceland, Ireland, Malta and Romania are omitted.} as well as Norway, Switzerland and the United Kingdom.\footnote{Note that we aggregate the bidding zones of Spain and Portugal to a single ‘Iberian’ market and the bidding zones of Lithuania, Estonia and Latvia to a single ‘Baltic’ market. Additionally, note that we consider the distinct bidding zones within countries. However, we aggregate the following zones: in Norway, zones NO1–NO5; in Sweden, zones SE1–SE3; and in Italy, all zones but IT-North.} Unless stated otherwise, the collected data is from 2016 to 2020.

Electricity demand is represented by hourly values for the system’s electrical load, of which both a day-ahead forecast and the actual values are published by the respective transmission system operators (TSOs) and provided by \cite{EntsoeTPe}. The collected load data for the Germany-Luxembourg bidding zone represents 2015–2020. In energy system models, electricity demand is usually considered volatile and inflexible in the short term. However, there is typically an option to shed load amid supply scarcity. In our application, we assume the cost of load shedding to be 3,000~\euro{}/MWh, as this was the maximum bidding price prior to September 2022\footnote{The maximum bidding price was increased to 4,000~\euro{}/MWh on 20 September 2022.}.

The availability of intermittent renewable energy, namely onshore wind, offshore wind and photovoltaics (PV), depends on meteorological conditions and varies from hour to hour. The feed-in data of these energy sources are provided as hourly day-ahead forecasts by \cite{EntsoeTPc}. 
Despite weather dependency, renewable energy electricity generation can still be intentionally curtailed. Acknowledging the various support schemes for renewable generation in Germany and Europe, which prevent renewable sources from being shut down immediately when negative prices occur, this study assumes a curtailment cost of 20 \euro{}\text{$/MWh_{el}$}.

For conventional thermal generation, we distinguish between ten technologies and divide further by age if their technical parameters (especially those that impact efficiency) have changed significantly over time. We use 30 capacity clusters to group power plants based on technology and date of commissioning. We derive technology- and age-based efficiencies from \cite{OPSDa} data and assign them to the corresponding capacity clusters. 

Moreover, we assign the clusters minimum output levels and efficiency losses in part-load operations based on \cite{Schroeder2013}. Hence, supply that follows fluctuations in demand and renewables is incentivised to shut down due to physical barriers (minimum output levels) and economic incentives (efficiency losses). The capacity, fuel type, generation technology and date of commissioning for units in the German market are derived from \cite{BNetzA2021}, \cite{UBA2020} and \cite{OPSDa}. For the remaining markets considered in our study, we use data from \cite{EntsoeTPa}, \cite{OPSDa} and \cite{EBC2021}. 

Power plant efficiency and the costs of fuel and CO\textsubscript{2} emissions form the variable generation costs of conventional thermal technologies. For fuel costs, we apply daily gas prices that are provided by \cite{EEX2021}, monthly coal prices taken from \cite{Destatis2020} and monthly oil prices from \cite{Destatis2020}. Fuel costs for nuclear and lignite are derived from \cite{EntsoS} and are assumed not to vary over the time horizon of our study. Prices for CO\textsubscript{2} certificates are taken as weekly data from \cite{Sandbag2020}. 

Due to the time and additional fuel used by power plants to heat up during a start-up process, fuel and CO\textsubscript{2} prices also impact the cost of starting up a power plant. Further data regarding start-ups (e.g., secondary fuel usage, depreciation) are derived from \cite{Schroeder2013}. 

Electricity generation relies on both the installed capacity and technical availability of power plants. As a result, we consider all scheduled and unscheduled outages that were known before the closure of the day-ahead market. Information on hourly outages is obtained from \cite{EntsoeTPb}.

In most electricity markets, combined heat and power (CHP) plants are used, where electricity generation and heat supply are interconnected and reliant on each other. To account for this relationship, we apply a must-run condition to all CHP units to ensure operation at a minimum output level, as defined by the heat demand. These output levels are established in two steps. First, we calculate an hourly heat-demand factor consisting of temperature-dependent (spatial heating) and temperature-independent (warm water and process heat) components. The temperature-driven heat demand is calculated using heating degree days derived by mean temperature data obtained from \cite{OPSDb}, while the temperature-independent heat demand is obtained from hourly and daily consumption patterns provided by \cite{Hellwig2013}. Second, we allocate annual electricity generation volumes from CHP plants to each hour of the year based on the hourly heat-demand factor. The data on annual technology-specific electricity generation from CHP units is sourced from \cite{EC2021}.

Control power is essential for system operators to ensure frequency stability at all times. The day-ahead market is affected by the market for control power  provision, meaning that the capacities reserved for control power cannot be placed on the day-ahead market. The amount of control power to be procured is an average of the tender results  taken from \cite{Regelleistung}.

In addition to conventional thermal technologies and intermittent renewables, we consider waste, biomass, energy storage, hydro-reservoirs and run-of-river hydroelectricity. Since the operation of both waste and biomass has largely been historically constant (compare with \cite{EntsoeTPd}), we implement both technologies as base-load. 

The energy storage units are divided into high capacity-to-energy ratios and low capacity-to-energy ratios. Storage units with high capacity-to-energy ratios actively charge and discharge. We exclusively consider a subset of pumped storage plants (PSPs) in this group. The overall turbine capacity of these PSPs is detailed by \cite{EntsoeTPa}, and the efficiency of a storage cycle is around 75~\% \citep{Schroeder2013}. For these PSPs, we assume a capacity-to-energy ratio of 1/9 \citep[see][]{Dena2010}. This means that the plant can generate electricity at full load for nine hours before its storage is emptied. Storage units with low capacity-to-energy ratios comprise long-term PSPs as well as hydro-reservoirs. They are assigned a variable generation cost in the model (i.e., an opportunity cost for water consumption). Using historical electricity prices from \cite{EntsoeTPg} and the observed generation and pumping activities in the respective hour from \cite{EntsoeTPd}, we construct a step-wise merit order for long-term PSPs and hydro-reservoirs. Run-of-river and high-capacity-to-energy PSPs are assigned a monthly availability factor derived from the historical generation data from \cite{EntsoeTPd}. We assume that 70~\% of pump storage capacity belongs to high-capacity-to-energy PSPs, while the remaining 30~\% of pump storage capacity belongs to low-capacity-to-energy PSPs. 

European electricity markets are highly interconnected through cross-border transmission capacities. Thus, the German electricity market is also integrated into the European electricity system, with a total interconnector capacity of 27~GW, representing more than 30~\% of the German peak load.\footnote{Note that the availability of the interconnectors depends on various factors (e.g., congestion within a market zone).} Annual aggregated exports, accounting for approximately 13\,\% of German consumption in 2019, and imports, making up around 7\,\% of consumption in the same year, are both significant. In our energy system model, net transfer capacities (NTCs) constrain transmission between market zones. We implement hourly day-ahead forecasts for NTCs that are made available by \cite{EntsoeTPf} and \cite{JAO2021}.

While parameterised data sets interest numerous stakeholders, replicating these is a tremendous effort \citep[see, e.g.,][]{Schroeder2013, Kunz2017Electricity}; thus, our input data can be found in the supplementary materials.

\section{Methodology}
\label{H_sec.method}
The implementation of the entire hybrid model is open-source; all data we could make public are available on GitHub at the following link: \hyperlink{https://github.com/ProKoMoProject/A-hybrid-model-for-day-ahead-electricity-price-forecasting-combining-fundamental-and-stochastic-mod}{https://github.com/ProKoMoProject/A-hybrid-model-for-day-ahead-electricity-price-forecasting-combining-fundamental-and-stochastic-mod}.

In this section, we present the components of the hybrid model, starting with the methodologies for the data pre-processing and parameter density forecast steps in Sections \ref{H_sec.datapreprocessing} and \ref{H_sec.ParameterDensityForecast}. Next, we introduce the dispatch market model used to generate the first price estimators in the energy system optimisation step in Section \ref{H_sec.ESM}. Finally, we detail the model for the stochastic post-processing step in Section \ref{H_sec.EIM}.

\subsection{Stochastic Data Pre-Processing Step}
\label{H_sec.datapreprocessing}

Modelling the electricity market with a dispatch model requires an understanding of the several various fundamental variables, including demand (represented by load), which is a crucial element. 
The TSOs provide the actual load and a day-ahead load forecast, which has the potential to be improved, as demonstrated by, for example, \cite{Maciejowska2021} and \cite{Preprocessing}. To enhance this forecast, we use a stochastic model for data pre-processing. Additionally, we model a two-day-ahead load forecast to capture power plant start-ups and shutdowns based on the load of the second following day, using the TSOs' day-ahead load forecast as a starting point.

\subsubsection*{Day-Ahead Load Forecast}
We use the approach initially presented in \cite{Preprocessing} to improve load forecasting. Thus, we occasionally refer to it for detailed specifications and analyses. We propose a purely endogenous time-series approach: a model for the TSOs' load forecast error $\varepsilon_t$ that depends only on past values of the forecast error itself and, in turn, on the TSOs' load forecast $\hat{l_t}$. The forecast error is the difference between the actual load data $l_t$ and the TSOs' load forecast $\hat{l_t}$. Designing a model for the error and forecasting it enables us to improve the TSOs' load prediction. The resulting load prediction $\hat{l_t}^*$ at time $t$ is given by the following equation:  
\begin{equation}
	\begin{aligned}
		\hat{l_t}^* = \hat{l_t} + \hat{\varepsilon_t},
	\end{aligned}
\end{equation}
where $\hat{\varepsilon_t}$ is our forecasted TSOs' load prediction error. Thus, $\hat{l}^*$ is an improved load forecast in which we adjust the original forecast for the predictable structure of its error. The sub-index $t$ denotes consecutive hours.

To model and forecast the forecast error $\varepsilon_t$, we use a decomposition model and decompose the error time series into the sum of a seasonal component and a stochastic component (see \cite{Luetkepohl2005, Hyndman21, BoxJenkins2016} for comprehensive introductions to time series models):
\begin{equation}
	\label{H_eqn_ecm_formula}
	\begin{aligned}
		\varepsilon_t = SC_t + RC_t, \\
	\end{aligned}
\end{equation}
where $SC_t$ is a seasonal and $RC_t$ is the remaining, stochastic component at time $t$.

Capturing a weekly season, the seasonal component $SC_t$ for time $t$ is defined by Eq. (\ref{H_eqn_Sc_t}) with $HS^{h,wd}$ being the average of TSOs' forecast errors for hour $h = 1, ..., 24$ and day $wd = 1$ (Monday), \ldots, $7$ (Sunday) and $HoW_t^{h,wd}$, describing dummy variables to address the hour of the day and the day of the week:
\begin{equation}
	\label{H_eqn_Sc_t}
	\begin{aligned}
		SC_t = \sum_{h=1}^{24} \sum_{wd=1}^{7} HoW_t^{h,wd} \cdot HS^{h,wd}, \\
	\end{aligned}
\end{equation}
with 
\begin{equation}
	\label{H_eqn_HS}
	\begin{aligned}
		HS^{h,wd} \coloneqq \frac{\sum_{s=t-h-l_w-23}^{t-h-24} \varepsilon_s \cdot HoW_s^{h,wd}}{\sum_{s=t-h-l_w-23}^{t-h-24} HoW_s^{h,wd}},\\
	\end{aligned}
\end{equation}

\begin{center}
	%\label{H_eqn_HoW}
	\small \[ 
	HoW_t^{h,wd} = 
	\begin{cases} 
		\text{1,} & \text{if $t$ is the h-th hour of the wd-th day of the week,}\\ 
		\text{0,} & \text{otherwise.}
	\end{cases}
	\] \end{center}

For the residual component $RC_t = \varepsilon_t - SC_t$ of the time series, we propose an econometric SARMA $(1,1)$x$(1,1)_{24}$ model given by the following equation:
\begin{equation}
	\label{H_eqn_Sarma}
	\begin{aligned}
		RC_t =& \phi_0 + \phi_1 \cdot RC_{t-1} + \phi_{24} \cdot RC_{t-24} - \phi_1 \phi_{24} \cdot RC_{t-25} \\
		& + \omega_1 \cdot \psi_{t-1} + \omega_{24} \cdot \psi_{t-24} + \omega_1 \omega_{24} \cdot \psi_{t-25}\\
		& + \psi_t,
	\end{aligned}
\end{equation}
where the innovations are assumed to be homoscedastic and normally distributed. This model contains an additional 24-hour seasonal component, making it stochastic, flexible and dependent on the values of past hours and days. In contrast, $SC_t$ describes a static seasonality. 

The model is estimated over a calibration window of one year. Within the hybrid model, the window is constantly rolled over by full days, with the forecast of the next day’s 24 hours being made recursively. If no actual data are available due to time points in the future, we use forecast values for the model variables.

\subsubsection*{Two-Day-Ahead Load Forecast}

Power plants make operational start-up and shut-down decisions based on the current day’s demand, the demand from the day before and the expected demand on the next day. To account for this in the dispatch model, a forecast of load consumption is needed two days in advance. We use the modelling and forecast of the current load $l_t$ (meaning the TSOs’ day-ahead load forecast $\hat{l_t}$) as a starting point from which to forecast the day-ahead forecast for the second following day, resulting in a two-day-ahead forecast $\hat{l}^{2DA}_t = \hat{\hat{l}}_t$.

For the model, we propose an econometric SARMA $(1,1)$x$(1,1)_{24}$ model with an additional exogenous variable, the TSOs' load forecast at lag 168:
\begin{equation}
	\label{H_eqn_Sarma2DA}
	\begin{aligned}
		\hat{\hat{l}}_t =& \phi_0 + \phi_1 \cdot \hat{l}_{t-1} + \phi_{168} \cdot \hat{l}_{t-168} + \phi_{24} \cdot \hat{l}_{t-24} - \phi_1 \phi_{24} \cdot \hat{l}_{t-25} \\
		& + \omega_1 \cdot \psi_{t-1} + \omega_{24} \cdot \psi_{t-24} + \omega_1 \omega_{24} \cdot \psi_{t-25}. %\\
		%& + \psi_t. 
	\end{aligned}
\end{equation}
The model’s innovations $\psi_t$ are assumed to be homoscedastic and normally distributed, meaning that $\psi_t \sim N(0,\sigma^2_\varepsilon)$. The model features 24-hour seasonality. Since we also observe weekly seasonality, we include the TSOs' load forecast at the same hour one week earlier, $\hat{l}_{t-168}$, as a regressor.   

We calibrate and estimate the model based on window length $l_w$, which contains one year of historical data. 
The estimated model is used to recursively (i.e., on an hourly basis) predict the values of each hour of the next day. Since we rely on an autoregressive time-series model, day-ahead load forecasts from 168 hours to one hour before the predicted hour are in the model as explanatory variables for the two-day-ahead prediction. However, this means that some values are unavailable when the forecast is made. They are replaced with recursively forecasted values based on the most recently available observations.

Given the increased uncertainty associated with forecasting two days ahead, our hybrid model incorporates a parameter density forecast that considers various scenarios for the level and development of load, utilising the hourly two-day-ahead load forecast as an input variable.

\subsection{Parameter Density Forecast Step}
\label{H_sec.ParameterDensityForecast}

To account for uncertainty in two-day-ahead load predictions, scenarios are calculated at the 5\,\% and 95\,\% quantiles using QRA. It describes a method for determining quantiles of predictive cumulative distribution functions, which can then be used to construct prediction intervals. A prediction interval is calculated using the $(\alpha/2)$-th and $(1 - \alpha/2)$-th quantile of the predictive cumulative distribution function, $\alpha \in (0,1)$, as the lower and upper bound of the interval. QRA is based on quantile regression and aims to model quantiles of real-valued variables that depend on explanatory variables \citep[see, e.g.,][]{Koenker1978}. It employs point predictions to explain the $q$-th quantile of the conditional distribution of the observation, setting a fixed $q \in (0,1)$. Here, quantile regression uses a vector of regressors $X_{t} = [1, \hat{l}_t^{2DA}]$, including a value of one for the intercept and the two-day-ahead point prediction for the load at given time $t$, to calculate the two-day-ahead load prediction in the $q$-th quantile ($Q_{l_{t}}(q\vert \cdot)$) (conditional on additional information):
\begin{align}
	Q_{l_{t}}(q\vert X_{t}) = X_{t}\beta_{q}
	\label{H_eqn_QR}
\end{align}

Thereby,  $\beta_{q}$ is a vector of parameters for the $q$-th quantile. QRA estimates $\beta_{q}$ and, thus, the $q$-th quantile by minimising the pinball loss function of the respective $q$-th quantile, given by
\begin{align}
	%\hat{\beta}_{q} = &\underset{\beta}{\argmin} \Bigl[ \sum_{\lbrace t: l_{t}\geq X_{t}\beta\rbrace} q\vert l_{t} - X_{t}\beta\vert + \sum_{\lbrace t: l_{t} < X_{t}\beta\rbrace} (1-q)\vert l_{t} - X_{t}\beta\vert \Bigr]  \\
	%\end{gather*}
	%\begin{align}
	\hat{\beta}_{q} = \underset{\beta}{\argmin} \Bigl[ \sum_{t: l_t \geq X_t\beta} (q - \textbf{1}_{l_{t}< X_{t}\beta} ) (l_{t} - X_{t}\beta) \Bigr], 
	\label{H_eqn_QRA}
\end{align}
where $\textbf{1}$ is the indicator function \citep[see, e.g.,][]{Nowotarski2015, Nowotarski2018}.
%Here, $Q_{l_{t}}(q\vert \cdot)$ is the two-day-ahead load prediction in the $q$-th quantile (conditional on additional information), and $\beta_{q}$ is a vector of parameters for the $q$-th quantile, where $\textbf{1}$ is the indicator function. 

We use the two-day-ahead load prediction $\hat{l}_{t}^{2DA}$ and the corresponding load from a one-year historical period to estimate the unknown parameter $\beta_{q}$. After calculating the 5\,\% and 95\,\% quantiles, the scenarios cover 90\,\% of possible load forecast values with $\alpha = 0.9$. With the parameter density forecast and the point forecast of expected value, this approach provides three possible scenarios for the two-day-ahead load: an expected scenario, described with the two-day-ahead load forecast, a low scenario described with the hourly estimated 5\,\% quantile, and a high scenario described with the hourly estimated 95\,\% quantile. Motivated by optimal integration rules in \cite{Grothe2013} (and setting $\kappa = 2$, as recommended in that work), we weight the expected value with $2/3$ and each quantile with $1/6$ to include the scenarios in the \emph{em.power dispatch} model described below.

\subsection{Energy System Optimisation Step}
\label{H_sec.ESM}

This section presents the energy system model \emph{em.power dispatch}, which generates wholesale day-ahead price estimators in the hybrid model’s energy system optimisation step. The model considers a detailed representation of the key techno-economic aspects of an integrated European electricity sector, including transmission restrictions between markets, electricity production by CHPs, energy storage and control power provision. For all considered market zones, our model determines the optimal dispatch decisions for various generation and storage technologies, the most effective use of cross-border transmission capacities and the short-run marginal system costs\footnote{Technically, the dual variable of demand constraint derives the short-run marginal system cost, also called the ‘shadow price’.}, which determine the price estimator for the day-ahead market in hourly resolution. 

Since our research focuses on day-ahead price forecasts, the energy system model is developed to reflect the level and quality of information available to market participants on the day before delivery. The model is formulated as a linear optimisation problem minimising total system costs. Ensuring the linear formulation of a highly complex system, we form capacity clusters, parameterising them as described in Section \ref{H_sec.data}. Within each capacity cluster, capacity units can be started up, and electricity can be produced in marginal increments \citep[see][]{Muesgens:2006}. This approach has two key advantages. First, it reduces computational requirements. Second, the problem is differentiable at each point, and the dual variable of the demand constraint can be interpreted as a wholesale market price estimator. Additionally, the accuracy of modelling large energy systems remains reasonably high for our purpose \citep{MuesgensandNeuhoff:2006}.

\begin{figure}[htbp]
	\centering
	\includegraphics[width=0.6\linewidth]{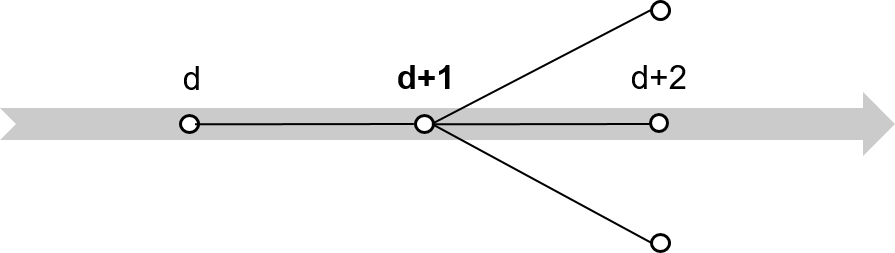}
	\caption{Illustration of the rolling window.}
	\label{H_figure_rolling_window}
\end{figure}

We implement the resultant imperfect forecasts with two model features. First, we implement a rolling window approach that repeatedly solves three days ($d \in D= \{d,d+1,d+2\}$), as shown in Figure \ref{H_figure_rolling_window}. In this setting, the 24 hours of the target day are represented by the second day of the horizon ($d+1$). This follows the EPEX spot market organisation, in which 24 hourly day-ahead prices are determined at 12 p.m. on the day before delivery ($d$). In addition to the target day d+1, we include the day before ($d$) and the day after ($d+2$). By considering three days in our rolling window, we reduce the problems of starting and ending values, particularly those stemming from power plant start-ups and pump storage plants. Second, we account for the increased uncertainty of the two-day-ahead estimate of key parameters. While parameters for the day $d$ are fully known, and forecasts for $d+1$ are made available by the European network transmission system operators for electricity (Entso-e), the realisation of key input parameters exhibits higher uncertainty in $d+2$. Therefore, we implement probabilistic forecast intervals only in $d+2$, as shown in Figure \ref{H_figure_rolling_window}. All other days are provided through one scenario.
The resulting stochastic rolling window is then repeatedly solved using only data available to market participants when they need to make their decisions. In each model run, we extract the information for the 24 hours of our ‘target day’ -- the day ahead.

The optimisation problem is repeatedly solved each day. For the reader’s convenience, we provide a nomenclature in the appendix. Note that all endogenous variables are written in upper case, and all exogenous parameters are written in lower case.  

The objective function in Eq. (\ref{H_eqn:TotalCost}) minimises total system costs ($TC$) consisting of the expected value of all operational costs ($OC$) across all three days of a rolling window $d^*$. Our empirical exercise considers three scenarios $s$ to be equally likely to appear. 

\begin{equation}
	\label{H_eqn:TotalCost}
	\begin{aligned}
		\displaystyle \min TC =  \mathbb{E}_{s}[OC(s)] \\
	\end{aligned}
\end{equation}

The operational costs in Eq. (\ref{H_eqn:OC}) contain all of the short-term costs that generation units face. We include costs at full-load operation ($vc_{i,n,d,h}^{FL}$), additional costs for units that operate at partial load ($vc_{i,n,d,h}^{ML}-vc_{i,n,d,h}^{FL}$) and start-up costs ($sc_{i,n,d,h}$). Note that we apply a linear unit commitment formulation and that all units must produce at least a certain minimum output level (see Eqs. (\ref{H_eqn:GenMax}) -- (\ref{H_eqn:startup2})). Additionally, we account for load-shedding costs ($voll$) and penalty payments for curtailing renewables ($curtc$), as discussed in Section \ref{H_sec.data}.

\begin{equation}
	\label{H_eqn:OC}
	\begin{aligned}
		\displaystyle OC_s =&  \sum_{i,n,d,h} G_{i,n,d,h,s} \cdot vc_{i,n,d,h}^{FL} + \sum_{i,n,d,h} SU_{i,n,d,h,s}\cdot sc_{i,n,d,h} \\
		&+ \sum_{i,n,d,h} (P_{i,n,d,h,s}^{on} - G_{i,n,d,h,s})\cdot (vc_{i,n,d,h}^{ML}-vc_{i,n,d,h}^{FL})\cdot g_i^{min} / (1-g_i^{min})  \\
		&+ \sum_{stl,n,d,h}  G_{stl,n,d,h,s} \cdot wv_{stl,n,d,h} - \sum_{stl,n,d,h}  CL_{stl,n,d,h,s}\cdot wv_{stl,n,d,h}   \\
		& + \sum_{hr,n,d,h}  G_{hr,n,d,h,s}\cdot wv_{hr,n,d,h} + \sum_{n,d,h} SHED_{n,d,h,s}\cdot voll  \\
		& + \sum_{res,n,d,h}  CURT_{res,n,d,h,s}\cdot curtc \\
	\end{aligned}
\end{equation}

As we apply our model with a rolling window, each model run considers three days $d$ with 24 hours each day, meaning 72 hours per daily model run. Modelling an additional day before and after the target day seems appropriate for storage units operated on a daily cycle. However, storage units (both PSPs and seasonal storage units without pumps) have a storage cycle longer than three days. Therefore, PSPs are divided into low-capacity-to-energy storage, which operates storage cycles longer than three days, and high-capacity-to-energy storage operating one or more storage cycles within a three-day horizon. The dispatch of the latter is determined endogenously. Low-capacity-to-energy PSPs are assigned a water value ($wv_{stl,n,d,h}$) that is implemented as a variable cost factor for electricity generation ($G_{stl,n,d,h,s}$) and electricity consumption ($CL_{stl,n,d,h,s}$).

Moving beyond pumped storage plants, hydro reservoirs have a natural water feed-in and do not have pumps installed. However, the water budget for electricity generation is limited by seasonal inflow volumes. We also apply a water value ($wv_{hr,n,d,h}$) to account for the opportunity costs of electricity generation from hydro reservoirs ($G_{hr,n,d,h,s}$).

Market clearing is ensured by Eq. (\ref{H_eqn:EnergyBalance}). For all 72 hours of the given rolling window, demand ($l_{n,d,h,s}$) must equal the sum of generation ($G_{i,n,d,h,s}$), load shedding ($SHED_{n,d,h,s}$) and electricity imports ($FLOW_{nn,n,d,h,s}$) minus the electricity consumption of mid-term energy storage ($CM_{stm,n,d,h,s}$) and long-term energy storage ($CL_{stl,n,d,h,s}$) as well as electricity exports ($FLOW_{n,nn,d,h,s}$). The dual variable of the demand constraint Eq. (\ref{H_eqn:EnergyBalance}) is used as an hourly day-ahead wholesale electricity price estimator.
\begin{equation}
	\label{H_eqn:EnergyBalance}
	\begin{aligned}
		\displaystyle l_{n,d,h,s} = &\sum_i G_{i,n,d,h,s} - \sum_{stm\subset I} CM_{stm,n,d,h,s} -\sum_{stl\subset I} CL_{stl,n,d,h,s} \\
		&+ \sum_{nn} (FLOW_{nn,n,d,h,s} - FLOW_{n,nn,d,h,s}) + SHED_{n,d,h,s} \\
		& \forall  n,nn\in N, d\in D, h\in H, s\in S   \\ 
	\end{aligned}
\end{equation}

Note that we apply the improved point forecast for the load created in Section \ref{H_sec.datapreprocessing} to Germany in d and d+1, as we focus on price predictions for the German market. For all other markets, we implement the original \mbox{Entso-e} forecasts. For d+2 in Germany, we apply the probabilistic load forecast presented in Section \ref{H_sec.ParameterDensityForecast}. For the remaining markets, we use the actual realisation of the previous week as a more naive estimator.

Electricity generation by a capacity cluster is limited by upper and lower bounds. The upper bound is formalised in Eq. (\ref{H_eqn:GenMax}). It ensures that electricity generation does not exceed the running capacity ($P_{i,n,d,h,s}^{on}$) in the cluster. The potential to generate electricity by running capacity is further limited by the reserve for positive control power provision ($PCR_{i,n,bp,s}$ and $SCR_{i,n,bs,s}^{pos}$). The lower bound is presented in Eq. (\ref{H_eqn:GenMin}). It states that running capacities must operate at a minimum power level, including the capacity reserved for negative control power provision ($PCR_{i,n,bp}$ and $SCR_{i,n,bs,s}^{neg}$). Note that primary control power provision ($PCR_{i,n,bp,s}$) in Germany is symmetrical (i.e., a unit must provide both positive and negative primary control power). Different positive and negative control power products were introduced for secondary control power. We do not include minute reserve requirements in the model for two reasons. First, fast-reacting units (e.g., hydro, open-cycle gas turbines) can be started up to provide positive minute reserve without being dispatched. Second, both positive and negative reserves can be provided by multiple market players other than the power plants included in the model (e.g., demand flexibility, P2X units, emergency power generators). The hours that belong to bidding blocks are mapped for primary control power by $bp$ and for secondary control power by $bs$.

\begin{equation}
	\label{H_eqn:GenMax}
	\begin{aligned}
		G_{i,n,d,h,s}  \leq  P_{i,n,d,h,s}^{on} - PCR_{i,n,bp|h\in bp,s} - SCR_{i,n,bs|h\in bs, s}^{pos} \\
		\forall bp\in BP, bs\in BS, i\in I, n\in N, d\in D, h\in H, s\in S \\ 
	\end{aligned}
\end{equation}

\begin{equation}
	\label{H_eqn:GenMin}
	\begin{aligned}
		P_{i,n,d,h,s}^{on}\cdot g_i^{min} + PCR_{i,n,bp|h\in bp,s} + SCR_{i,n,bs|t\in bs,s}^{neg}  \leq   G_{i,n,d,h,s}  \\
		\forall bp\in BP, bs\in BS, i\in I, n\in N, d\in D, h\in H, s\in S \\ 
	\end{aligned}
\end{equation}

The installed capacity limits the running capacity of a power system ($cap_{i,n,d,h}$) in combination with either the availability factor ($af_{i,n,d,h}$) or power plant outages ($out_{i,n,d,h}$), as shown in Eq. (\ref{H_eqn:PonMax}). For thermal generation capacities, we use hourly power plant outages. Renewables are provided with an hourly availability factor, while hydroelectric units are provided with a monthly availability factor.

\begin{equation}
	\label{H_eqn:PonMax}
	\begin{aligned}
		\displaystyle P_{i,n,d,h,s}^{on} \leq  cap_{i,n,d,h} \cdot af_{i,n} - out_{i,n,d,h} \hspace{1cm}  \forall i\in I, n\in N, h\in H, d\in D, s\in S \\ 
	\end{aligned}
\end{equation}

Eq. (\ref{H_eqn:startup}) tracks start-up activities ($SU_{i,n,d,h,s}$) that increase running capacity from one hour to another. Due to the non-negativity condition, start-ups are either positive or zero. Eq. (\ref{H_eqn:startup2}) tracks start-up activities from the last hour of a day to the first hour of the following day.

\begin{equation}
	\label{H_eqn:startup}
	\begin{aligned}
		\displaystyle P_{i,n,d,h,s}^{on} - P_{i,n,d,h-1,s}^{on}  \leq  SU_{i,n,d,h,s}  \hspace{1cm}  \forall i\in I, n\in N, h\in H, d\in D, s\in S \\ 
	\end{aligned}
\end{equation}
\begin{equation}
	\label{H_eqn:startup2}
	\begin{aligned}
		\displaystyle P_{i,n,d,hfirst,s}^{on} - P_{i,n,d-1,hlast,s}^{on}  \leq  SU_{i,n,d,hfirst,s}  \hspace{1cm}  \forall i\in I, n\in N, d\in D, s\in S \\ 
	\end{aligned}
\end{equation}
The difference between available feed-in from intermittent renewables and their actual generation defines the curtailment of renewables, as shown in Eq. (\ref{H_eqn:RES}):

\begin{equation}
	\label{H_eqn:RES}
	\begin{aligned}
		\displaystyle &cap_{res,n,d,h}\cdot pf_{res,n,d,h} = G_{res,n,d,h,s} + CURT_{res,n,d,h,s} \\   &\forall n\in N, res\in I, d\in D, h\in H, s\in S. \\ 
	\end{aligned}
\end{equation}

Some power plants are active in both the heat market and the electricity market. Thus, the model implements a must-run condition for such units on the electricity market, which varies over time (e.g., higher in the winter season due to space heating). Depending on hourly heat demand, Eq. (\ref{H_eqn:CHP}) states that the output of a combined heat and power unit is at least equal to the electricity generation linked to the heat production ($chp_{i,n,d,h}$):

\begin{equation}
	\label{H_eqn:CHP}
	\begin{aligned}
		\displaystyle chp_{i,n,d,h} \leq  G_{i,n,d,h,s} \hspace{1cm}  \forall i\in I, n\in N, d\in D, h\in H, s\in S. \\
	\end{aligned}
\end{equation}

Eq. \ref{H_eqn:Trade} constrains cross-border electricity transfer ($FLOW_{n,nn,d,h,s}$) via net transfer capacity ($ntc_{n,nn,d,h}$):

\begin{equation}
	\label{H_eqn:Trade}
	\begin{aligned}
		\displaystyle FLOW_{n,nn,d,h,s} \leq  ntc_{n,nn,d,h} \hspace{1cm}    \forall  n,nn\in N, d\in D, h\in H, s\in S. \\ 
	\end{aligned}
\end{equation}

Eq. (\ref{H_eqn:StorageLevel}) describes the state of the storage level of mid-term storage units. A storage level decreases with electricity generation ($G_{stm,n,d,h}$) and increases with charging ($ST_{stm,n,d,h,s}^{in}$). The efficiency of an entire storage cycle ($\eta_{stm}$) is assigned to the charging process. Eq. (\ref{H_eqn:StorageLevel2}) ensures the functionality of the storage mechanism between two days:

\begin{equation}
	\label{H_eqn:StorageLevel}
	\begin{aligned}
		&\displaystyle SL_{stm,n,d,h,s} =  SL_{stm,n,d,h-1,s} - G_{stm,n,d,h,s} + CM_{stm,n,d,h,s} \cdot \eta_{stm} \\
		&\forall  stm\in I, n\in N, d\in D, h\in H, s\in S, \\ 
	\end{aligned}
\end{equation}

\begin{equation}
	\label{H_eqn:StorageLevel2}
	\begin{aligned}
		\displaystyle SL_{stm,n,d,hfirst,s} = & SL_{stm,n,d-1,hlast,s} - G_{stm,n,d,hfirst,s} \\
		& + CM_{stm,n,d,hfirst,s} \cdot \eta_{stm} \\
		\forall  stm\in I, n\in N, d\in D, s\in S. \\ 
	\end{aligned}
\end{equation}

The maximum energy storage capacity ($SL_{stm,n,d,h,s}$) of storage units with high capacity-to-energy ratios is defined by their maximum installed turbine capacity divided by their capacity-to-energy ratio ($cer$), as shown in Eq. (\ref{H_eqn:maxSL}):

\begin{equation}
	\label{H_eqn:maxSL}
	\begin{aligned}
		\displaystyle SL_{stm,n,d,h,s} \leq  cap_{stm,n,d,h} \cdot cer \hspace{1cm}  \forall  stm\in I, n\in N,  d\in D, h\in H, s\in S. \\ 
	\end{aligned}
\end{equation}

Eq. (\ref{H_eqn:maxTurbine}) restricts both turbine capacity and pumping capacity, with pumping capacity assumed to be 10~\% lower than the turbine capacity:

\begin{equation}
	\label{H_eqn:maxTurbine}
	\begin{aligned}
		&\displaystyle G_{stm,n,d,h,s} + 1.1\cdot CM_{stm,n,d,h,s} \leq  cap_{stm,n,d,h}  \\
		&\forall  stm\in I, n\in N, d\in D, h\in H, s\in S.  \\ 
	\end{aligned}
\end{equation}

At the beginning and at the end of each model run, all storage units with high capacity-to-energy ratios must be filled with at least 30~\% of their energy level:

\begin{equation}
	\label{H_eqn:SLstart}
	\begin{aligned}
		\displaystyle SL_{stm,n,dfirst,hfirst,s} =  0.3\cdot cap_{stm,n,d,h} \hspace{1cm} \forall stm\in I, n\in N, s\in S,    \\ 
	\end{aligned}
\end{equation}
\begin{equation}
	\label{H_eqn:SLend}
	\begin{aligned}
		\displaystyle SL_{stm,n,dlast,hlast,s} =  0.3\cdot cap_{stm,n,dlast,hlast}  \hspace{1cm}   \forall  stm\in I, n\in N, s\in S. \\ 
	\end{aligned}
\end{equation}

Storage plants with low capacity-to-energy ratios are not subject to a storage mechanism. However, these units’ electricity generation and consumption are restricted to their installed capacity by:

\begin{equation}
	\label{H_eqn:MaxStoreLong}
	\begin{aligned}
		\displaystyle G_{stl,n,d,h,s} +  CL_{stl,n,d,h,s} \leq  cap_{stl,n,d,h} \hspace{1cm}   \forall n\in N, stl\in I, d\in D, h\in H, s\in S. \\ 
	\end{aligned}
\end{equation}

Eqs. (\ref{H_eqn:CP_primary}), (\ref{H_eqn:CP_sec_pos}) and (\ref{H_eqn:CP_sec_neg}) ensure the control power provision for primary, positive secondary and negative secondary control power.

\begin{equation}
	\label{H_eqn:CP_primary}
	\begin{aligned}
		\displaystyle \sum_i PCR_{i,n,bp,s} = pr_{n} \hspace{1cm} \forall  bp\in BP, n\in N, s\in S \\ 
	\end{aligned}
\end{equation}

\begin{equation}
	\label{H_eqn:CP_sec_pos}
	\begin{aligned}
		\displaystyle \sum_i SCR_{i,n,bs,s}^{pos} = sr_{n}^{pos} \hspace{1cm}  \forall  bs\in BS, n\in N,  s\in S \\ 
	\end{aligned}
\end{equation}

\begin{equation}
	\label{H_eqn:CP_sec_neg}
	\begin{aligned}
		\displaystyle \sum_i SCR_{i,n,bs,s}^{neg} = sr_{n}^{neg} \hspace{1cm} \forall  bs\in BS, n\in N,  s\in S \\ 
	\end{aligned}
\end{equation}

The non-negativity constraint ensures that the individual variables do not show negative values and is given by:

\begin{equation}
	\label{H_eqn:nonnegative}
	\begin{aligned}
		\displaystyle 0 \leq  & CL_{stl,n,d,h,s}, CM_{stm,n,d,h,s}, CURT_{res,n,d,h,s}, \\
		& G_{i,n,d,h,s}, FLOW_{n,nn,d,h,s}, P_{i,n,d,h,s}, PCR_{i,n,bp,s}, SCR_{i,n,bs,s}^{neg}, SCR_{i,n,bs,s}^{pos}, \\ 
		& SHED_{n,d,h,s}, SL_{stm,n,d,h,s},  SU_{i,n,d,h,s}   \\
		& \forall  n,nn\in N, bp\in BS, bs\in BS, i\in I, n\in N, h\in H, d\in D, s\in S. \\ 
	\end{aligned}
\end{equation}

\subsection{Stochastic Data Post-Processing Step}
\label{H_sec.EIM}

In this step, we use a stochastic post-processing technique to refine the estimators produced by the techno-economic energy system model. Specifically, we forecast the errors $\varepsilon_t$ of the day-ahead price estimators $\hat{P_t}$ obtained from the energy system optimisation step, either by a time series based point forecast $\hat{\varepsilon_t}$ or by inferring the forecast errors distribution functions to generate probabilistic day-ahead price predictions. We incorporate exogenous variables such as renewable energy feed-in and weather, as well as lags of the forecast error itself into the time-series forecast of $\varepsilon_t$. 

We start with the improved point prediction $\hat{P_t}^*$ at time $t$ which is given by the following equation:

\begin{equation}
	\begin{aligned}
		\hat{P_t}^* = \hat{P_t} + \hat{\varepsilon_t},
	\end{aligned}
\end{equation}

where $\hat{P_t}$ is the price prediction from the last step, and $\hat{\varepsilon_t}$ is our model’s forecasted price prediction error. Thus, $\hat{P}^*$ constitutes an improved price forecast in which we adjust the forecast from the last step for the stochastic but predictable structure in its error.

We employ two model frameworks -- univariate and multivariate -- as this approach has been proven to be useful in past research \cite{Ziel2018}. In the univariate framework, we interpret the forecast error time series as one high-frequency time series in an hourly resolution. In the multivariate framework, we split the time series into 24 individual time series, one for each hour, making them in a daily resolution. 

For the post-processing setup, subindexes ${h,d}$ will denote hours one through 24 of day $d$, with $d$ being consecutive days. So, ${P}_{1,1}$, for instance, is the actual day-ahead price of the first hour of the first day of the considered period, and ${\varepsilon}_{5,432}$ is the error of the price estimator of the fifth hour of the 432nd day. This fits best because it enables us to observe a realisation of 24 prices for the hours of the next day simultaneously for electricity prices. Please note that if $h-1$ were equal to or less than zero, we would need to shift one day backwards. Likewise, if $h+1$ were greater than 24, we would need to shift one day forward.

We chose a standard econometric time-series model for the univariate framework. It consists of endogenous (i.e., autoregressive with moving average structures) and exogenous variables, all of which are integrated into a regression model given by Eq. (\ref{H_eq_uvmodel}). 
To address several seasonal structures included in the time series of the prices’ forecast errors $\varepsilon_{h,d}$, we use the first and second observation backwards $\varepsilon_{h-1,d}, \varepsilon_{h-2,d}$ as well as the first back error $\psi_{h-1,d}$ of the estimated model. Additionally, we use the observation one day before (daily structure) $\varepsilon_{h,d-1}$ and one week before (weekly structure) $\varepsilon_{h,d-7}$ as endogenous explanatory variables. Considering daily effects, we include the minimum and maximum forecast errors for the day before $\varepsilon_{min,d-1}, \varepsilon_{max,d-1}$. To account for the strong effects of forecast errors on public holidays, we use a dummy variable $hol_{h,d}$ for public holidays as another factor. Additionally, we include an hourly wind forecast $X_{h,d}$.

\begin{equation}
	\label{H_eq_uvmodel}
	\begin{aligned}
		\varepsilon_{h,d} =& \phi_0 + \phi_1 \cdot \varepsilon_{h-1,d} + \phi_2 \cdot \varepsilon_{h-2,d} + \phi_{3} \cdot \varepsilon_{h,d-1} + \phi_4 \cdot \varepsilon_{h,d-7} + \phi_5 \cdot \psi_{h-1,d}\\
		& + \omega_1 \cdot \varepsilon_{min,d-1} + \omega_{2} \cdot \varepsilon_{max, d-1} + \omega_3 \cdot hol_{h,d} + \omega_4 \cdot X_{h,d}\\
		& + \psi_{h,d} 
	\end{aligned}
\end{equation}

As in the univariate framework, we use the well-known time-series model ARX in the multivariate framework. However, the autoregressive component refers to values of the same hour on previous days. The endogenous variables, $\varepsilon_{h, d-1}$ and $\varepsilon_{h, d-7}$, are the forecast errors at the same hour one day prior and seven days prior, respectively. The exogenous variables are the same as in the univariate framework: minimum and maximum forecast errors for the day before, a dummy variable for public holidays and an hourly wind forecast. Thus, the model for the multivariate framework is given by the following equation:

\begin{equation}
	\begin{aligned}
		\varepsilon_{h,d} = & \phi_0 + \phi_1 \cdot \varepsilon_{h, d-1} + \phi_{2} \cdot \varepsilon_{h, d-7} \\
		& + \omega_1 \cdot \varepsilon_{min,d-1} + \omega_{2} \cdot \varepsilon_{max, d-1} + \omega_3 \cdot D_{h, d} + \omega_4 \cdot X_{h,d}\\
		& + \psi_{h,d}, 
	\end{aligned}
\end{equation}
where $\phi_{i}, \omega_i$ describes the coefficients that need to be estimated. The innovations $\psi$ are assumed to be homoscedastic and normally distributed in both frameworks, meaning that $\psi_{h,d} \sim N(0,\sigma^2_\epsilon)$. 

Since we rely on an autoregressive time-series model, we need day-ahead spot prices from the last hours before prediction time as explanatory variables. In the multivariate framework, only one step into the future is forecasted at a time due to the separate modelling of each hour. In the univariate framework, 24 values are forecast for the future, with the hours of the next day predicted recursively (i.e., on an hour-by-hour basis). Unavailable variables are replaced with recursively forecasted variables based on the most recent available observations.

In line with approaches previously shown to be effective in the literature \citep[see, e.g.,][]{Marcjasz2018}, we vary the presented post-processing models by using different window lengths to estimate the model set-up and prevent random choice. We determine the calibration window for 44, 48 and 52 weeks. By using three window lengths and two model frameworks, we end up with six individual sub-models. These sub-models are used to predict the values of the hours $h$ of the next day $d$, and we denote them by $\hat{P}_{h,d}^{uv44},$ $\hat{P}_{h,d}^{uv48},$ $\hat{P}_{h,d}^{uv52},$ $\hat{P}_{h,d}^{mv44},$ $\hat{P}_{h,d}^{mv48},$ $\hat{P}_{h,d}^{mv52}.$ 

Our final improved point forecast $\hat{P_t}^*$ is obtained by taking the arithmetic average of the six prices. Despite the potential appeal of using seemingly more sophisticated methods, such as calculating optimal weights via linear regression based on past data, these methods resulted in predictors with higher root mean squared errors (RMSE) and mean absolute errors (MAE), even when we used rolling windows that look into the future. This is mainly due to the additional estimation noise that is introduced when using such methods and may lead to inefficiencies as discussed, e.g., in the context of financial literature in \cite{demiguel2009optimal}. As a result, we stick with the simpler, yet more robust method of averaging the six individual forecasts.

We now move on to generating probabilistic day-ahead price predictions. To achieve this, we use the six forecasts generated by the individual sub-models to estimate the cumulative distribution function $F_{P_{h,d}}$ of the day-ahead prices. The estimated function $\hat{F}_{P_{h,d}}$ serves as our probabilistic forecast for the price of the next day. We represent the distribution $\hat{F}_{P_{h,d}}$ in terms of its quantiles. Specifically, we employ quantile regression to model the conditional $q$-th quantile of the cumulative distribution function of the day-ahead prices, where $q$ is a value between 0 and 1. This modelling is accomplished by utilising the six individual point predictions and the following equation:

\begin{align}
	Q_{P_{h,d}}(q\vert X_{h,d}) = X_{h,d}\beta_{q}, 
	\label{H_eqQR_q}
\end{align}

where $X_{h,d} = [1, \hat{P}_{h,d}^{uv44}, \hat{P}_{h,d}^{uv48}, \hat{P}_{h,d}^{uv52}, \hat{P}_{h,d}^{mv44}, \hat{P}_{h,d}^{mv48}, \hat{P}_{h,d}^{mv52}]$ is the vector of regressors containing a value of 1 for the intercept and the six individual point predictions for the day-ahead price at time ${h,d}$. $\beta_{q}$ is again estimated by minimising the pinball loss function. To determine the predictive distribution, we forecast multiple quantiles of the distribution.

The coefficients of the regressors are estimated by a calibration window of one year with a distinction made between peak and off-peak hours. Peak hours are defined as those between 8 a.m. and 8 p.m. from Monday to Friday, while off-peak hours are all remaining hours. This distinction is made because peak hours are characterised by high demand for electricity and, therefore, often exhibit higher day-ahead prices. For each quantile $q$, we estimate two parameter vectors $\beta_{q}$ (see Eq. (\ref{H_eqn_QRA})). Estimating one parameter vector $\beta_{q}$ for all day hours proved to be less accurate. Additional information on this matter can be provided upon request. 

In summary, our hybrid model includes the following steps. To predict the next day, we first enhance the TSO’s day-ahead load forecast and predict the two-day-ahead load forecast in the stochastic data pre-processing step. We then calculate the two-day-ahead load scenarios in the parameter density forecast step and include them in the \emph{em.power dispatch} in the energy system optimisation step to generate the first price estimators for the day-ahead spot market. Finally, in the final stochastic post-processing step, we improve these price estimators with stochastic methods and conduct probabilistic price forecasts. This sequence is repeated continuously, day by day, for all points in time in our observation period. The hybrid model's rolling window approach means that we always use the most up-to-date available data.

\section{Hybrid Model Results}
\label{H_sec.results.Hybrid}

In this section, we present the electricity price forecasts of our hybrid model for Germany from January 2016 to December 2020\footnote{More precisely, prices cover the German-Austria-Luxembourg bidding zone from January 2016 to September 2018 and the German-Luxembourg bidding zone from October 2018 to December 2020.}. Since the day-ahead market is organised in an hourly resolution, our hybrid model calculates point and probabilistic forecasts for each hour of the following day. As the central point of this paper, we present the overall results of the model (i.e., the point and probabilistic price predictions) and qualitatively detail their place in the literature. 

We start by comparing the point forecasts of our hybrid model to those in the literature. With an annual average RMSE of 7.38~\euro{}/MWh and MAE of 4.60~\euro{}/MWh over the five years from 2016 to 2020, our model aligns well with previous studies. An expert model developed by \cite{Ziel2018} forecasted electricity prices with an overall MAE of 5.01~\euro{}/MWh for 2012 to 2016. Using an autoregressive model with exogenous variables, \cite{Maciejowska2021} achieved an RMSE of 8.43~\euro{}/MWh and MAE of 5.92~\euro{}/MWh for 2016 to 2019. For the same period, \cite{Qussous2022} obtained an RMSE of 11.21~\euro{}/MWh and MAE of 7.89~\euro{}/MWh, presenting an agent-based power market-simulation model with rule-based bidding strategies and, thus, a non-equilibrium-oriented techno-economic market model aimed at reproducing day-ahead electricity prices. Since they provided extensive information on the error measures, we can use these models for a detailed comparison. However, as the time periods in these studies are overlapping with but not identical to our observation period, we also perform a detailed comparison with the LEAR model developed by \cite{lago2021forecasting}. The LEAR model’s code is freely available, so we can extend it with data from the years up to 2020. Furthermore, its day-ahead price forecasts are among the most precise in the literature, and its authors are leading scholars in the field of price forecasting. Thus, we can perform a year-by-year comparison with our results. Table \ref{H_table_finalresult} presents the forecast accuracy of the hybrid model developed in this paper, the agent-based market simulation model and the LEAR model, showing the annual RMSE and MAE. It can be seen that the agent-based model has the highest error, which can be attributed to the general difficulties of techno-economic models in making short-term forecasts (the results of the \emph{em.power dispatch} model without further post-processing steps are compiled in Section \ref{H_sec.results.esm} and point in the same direction). In contrast, the LEAR model and the proposed hybrid model presented here exhibit similar error measures without larger gaps. Although the LEAR model more often takes the lead, its advantage is limited, as it is a statistical model primarily designed for generating price forecasts. Comparatively, the hybrid model's forecast encompasses the entire market state represented by the energy system model, including information on additional parameters of interest to market participants (e.g., CO\textsubscript{2} emissions, international electricity exchange, and power plant utilization).\footnote{Note that while this work focuses exclusively on prices, this model can also be informative about other factors, as discussed in the literature review.}

\begin{table}[htbp]
		%\bigskip
		\begin{center}
        \caption{RMSE and MAE of day-ahead electricity price forecast through the presented hybrid model and the LEAR model in [€/MWh]}
	
	%\footnotesize
		\begin{tabular}{|l|r|r|r|r|r|r|}
			\hline
			\multicolumn{1}{|c|}{} & \multicolumn{3}{c|}{RMSE} & \multicolumn{3}{c|}{MAE} \\ \hline
           	       	&	Hybrid	&	Agent	&	LEAR	&	Hybrid	&	Agent	&	LEAR		\\	\hline
            All years	&	7.38	&	11.21	&	7.24	&	4.60	&	7.89	&	4.38		\\	\hline
            2016	&	5.82	&	8.83	&	5.55	&	3.48	&	6.54	&	3.30		\\	\hline
            2017	&	8.79	&	13.01	&	7.91	&	5.25	&	9.44	&	4.56		\\	\hline
            2018	&	7.28	&	11.69	&	6.96	&	5.07	&	8.88	&	4.84		\\	\hline
            2019	&	7.05	&	10.91	&	7.95	&	4.43	&	6.69	&	4.53		\\	\hline
            2020	&	7.63	&		&	7.54	&	4.77	&		&	4.65		\\	\hline

		\end{tabular}
	\label{H_table_finalresult}
 \end{center}
\end{table}

\begin{figure}[htbp]
	\centering
	\includegraphics[width=0.9\linewidth]{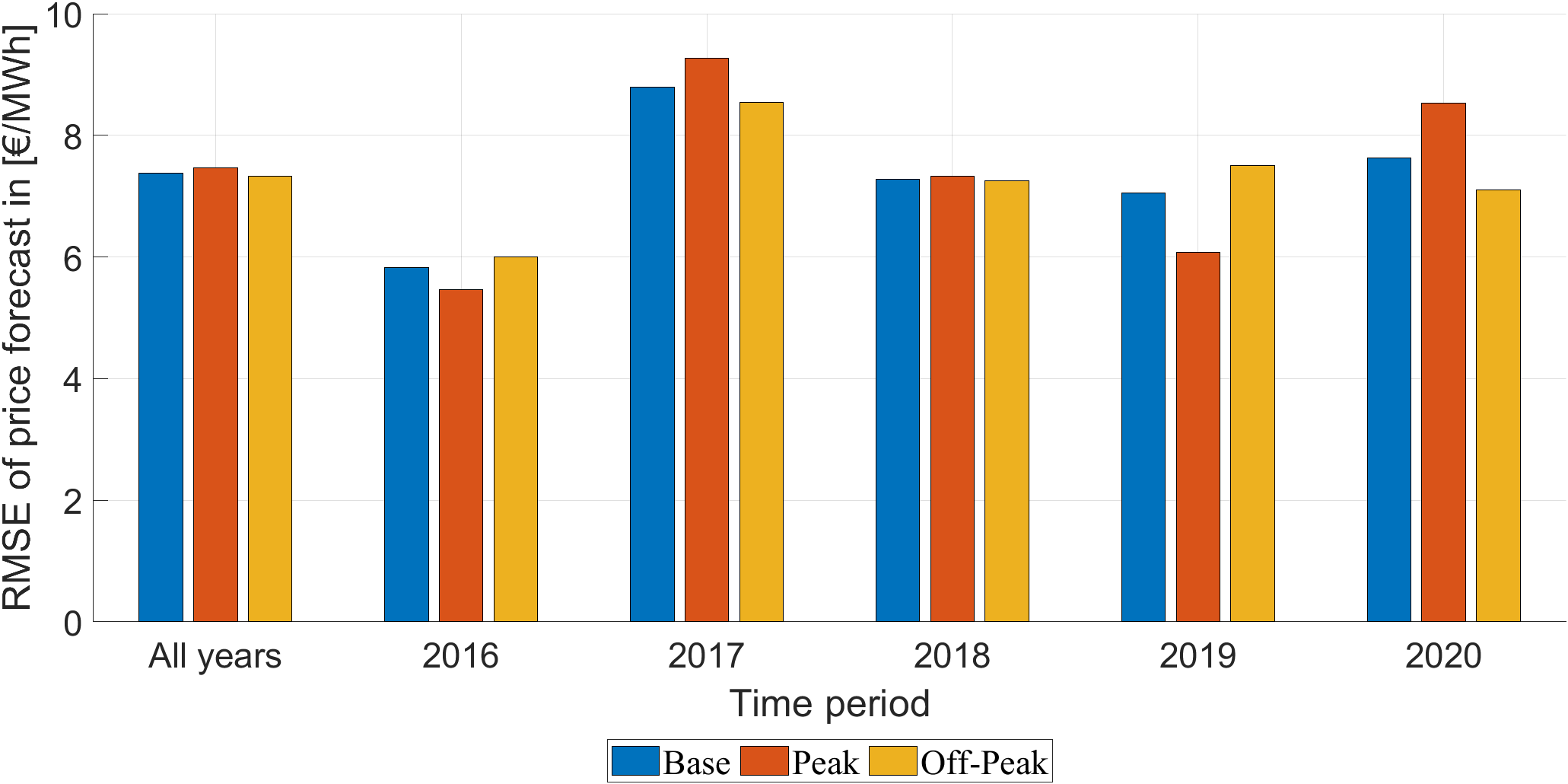}
	\caption{RMSE for base, peak and off-peak hours.}
	\label{H_fig_RMSE_BPOP}
\end{figure}

\begin{figure}[htbp]
	\centering
	\includegraphics[width=0.9\linewidth]{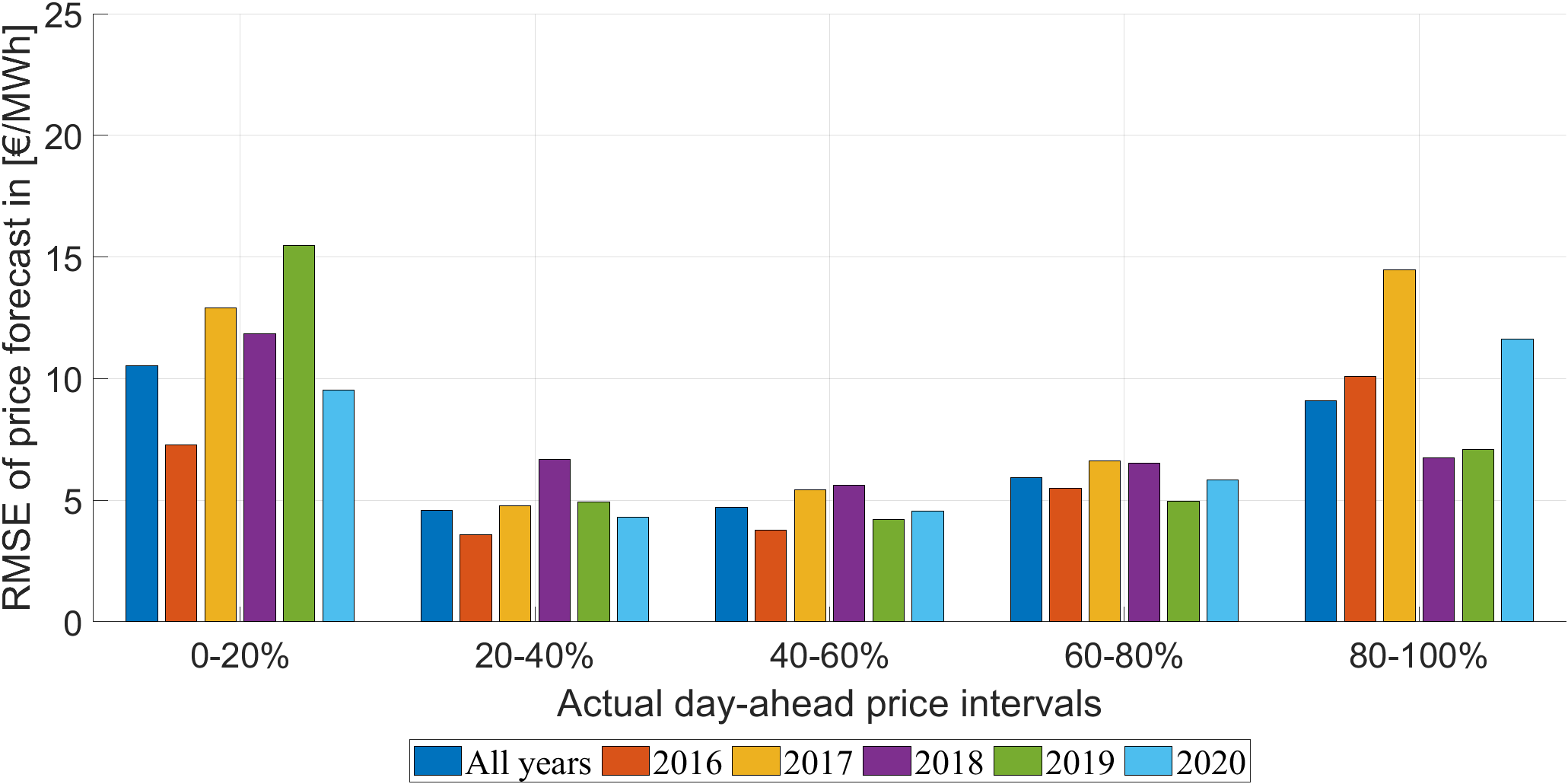}
	\caption{RMSE for hours at different day-ahead price quantiles.}
	\label{H_fig_RMSE_pricequantiles}
\end{figure}

A more detailed evaluation of the prediction errors of the hybrid model is provided in Figures \ref{H_fig_RMSE_BPOP} and \ref{H_fig_RMSE_pricequantiles}, which show the RMSE for different criteria, such as base, peak and off-peak hours, and the hours of actual day-ahead price quantiles, respectively. Again, peak hours are those between 8 a.m. and 8 p.m. from Monday to Friday; off-peak hours are the remaining hours. Base hours describe all hours of a day, regardless of the day of the week or hour of the day. 

While the errors of 7.47~\euro{}/MWh in peak hours and 7.33~\euro{}/MWh in off-peak hours seem quite similar over the whole period, Figure \ref{H_fig_RMSE_pricequantiles} provides more insight. We separate the predicted hours into five groups presenting the hours for each confidence interval of realised day-ahead prices in 20~\% steps. Therefore, we calculate and evaluate the RMSE for each group. The figure suggests a relationship between the error in price forecasts and the price level. Throughout the years, the highest RMSE has consistently been identified in the hours with the lowest 20~\% and highest 20~\% of prices. Based on economic theory, this finding can be explained by start-up costs and their impact on hourly prices. Assuming perfect foresight, \cite{Kuntz2007} have shown that start-up costs are added to fuel costs exclusively during the hour of highest demand in a cycle because additional capacity must be started-up for that hour, which is not needed in any other hour. During the hour of lowest demand, start-up costs are deducted from variable production costs because power plants save costs on re-starts when allowed to continue operations throughout that hour. In contrast, start-up costs do not influence prices during any other hour of a load cycle. The \emph{em.power dispatch} model follows this economic theory when determining wholesale electricity prices based on the shadow prices of the demand constraint. However, in reality, bidders on the day-ahead market face uncertainties with regard to which hour has the highest and lowest residual demand and what magnitude start-up costs have for that day. While uncertainty is always present, its impact is likely higher when start-up costs need to be considered in addition to fuel costs and thus increase price volatility around the highest and lowest price hours. Therefore, negative and positive price peaks are harder to capture and forecast than intermediate price levels. Note that this increased uncertainty in these market conditions affects all point forecasting models. Corresponding figures for the LEAR model, with a very similar pattern, are available in Appendix, Figures \ref{H_fig_RMSE_BPOP_LEAR} and \ref{H_fig_RMSE_pricequantiles_LEAR}. %on request from the authors.

We now turn to the analysis of our probabilistic forecasts. We represent the forecast distribution of the day-ahead prices for each forecast hour with quantile estimates in 5~\% steps. The quantile estimation for different quantile levels then enables the calculation of forecast intervals, which determine the probability of the day-ahead price being within a certain range.

\begin{figure}[htbp]
	\centering
	\includegraphics[width=0.9\linewidth]{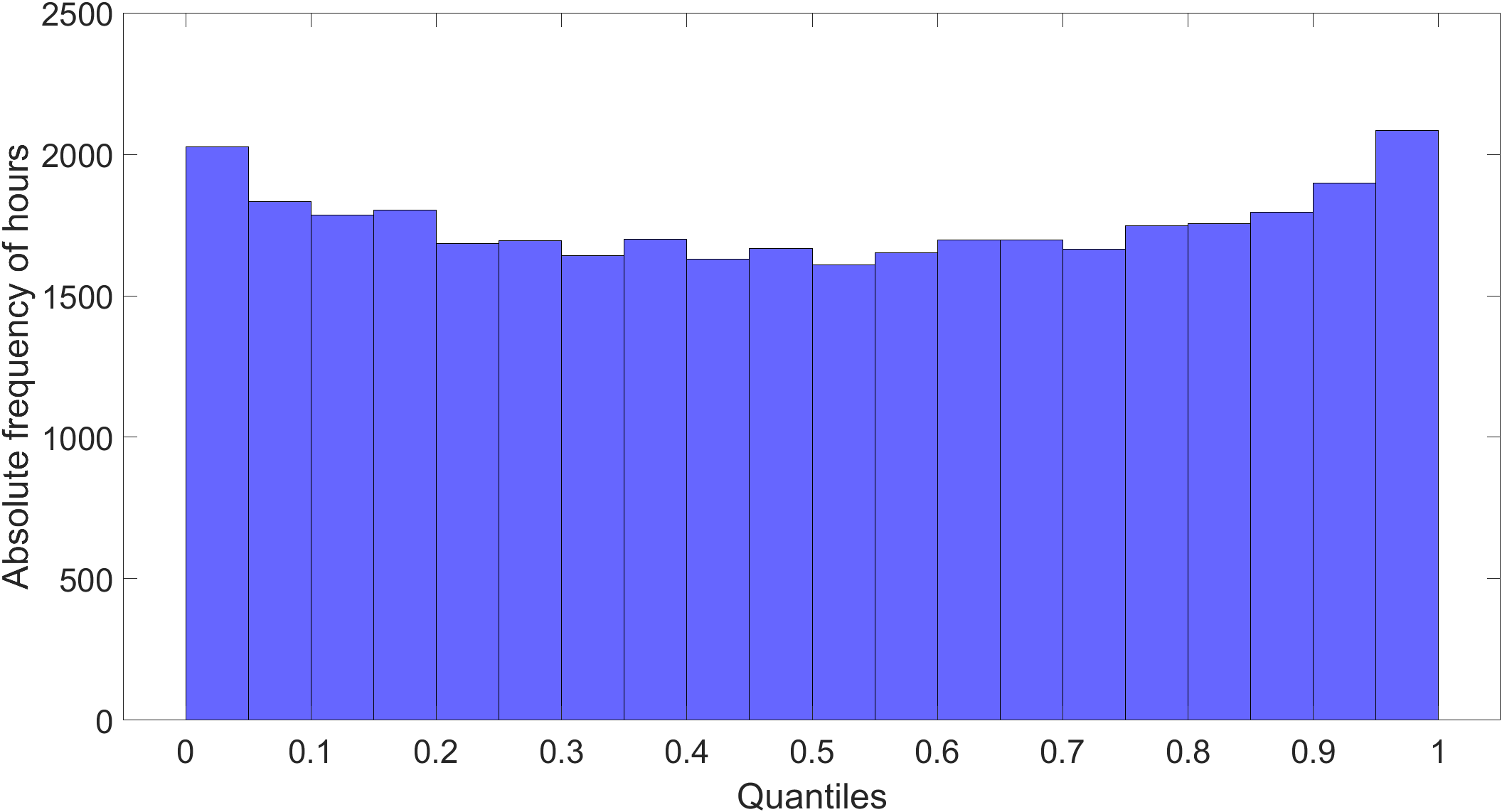}
	\caption{Number of actual prices included in the quantiles estimated in the forecast period from 2017 to 2020.}
	\label{H_fig_histogram_qra_price}
\end{figure}

The first thing to check about probabilistic forecasts is whether they are well calibrated \citep[see, e.g.,][]{gneiting2007probabilistic}, which means whether forecasted probabilities  and observed frequencies coincide. Therefore, Figure \ref{H_fig_histogram_qra_price} shows the actual frequency of all hours in which the day-ahead prices are above or below the forecast quantile limits. A forecast is considered calibrated if the predicted probabilities match the observed frequencies of the target variable over time. Thus, a fully calibrated forecast in a laboratory environment would correspond to a uniform distribution. The histogram below shows slightly increasing frequencies towards the outside but forms a good approximation of a uniform distribution generated by randomly drawn numbers. This serves as a quality check for our hybrid model and ensures that the predicted probabilities reflect the true probabilities of the day-ahead prices.

\begin{figure}[htbp]
	\centering
	\subfigure{\includegraphics[width=0.49\textwidth]{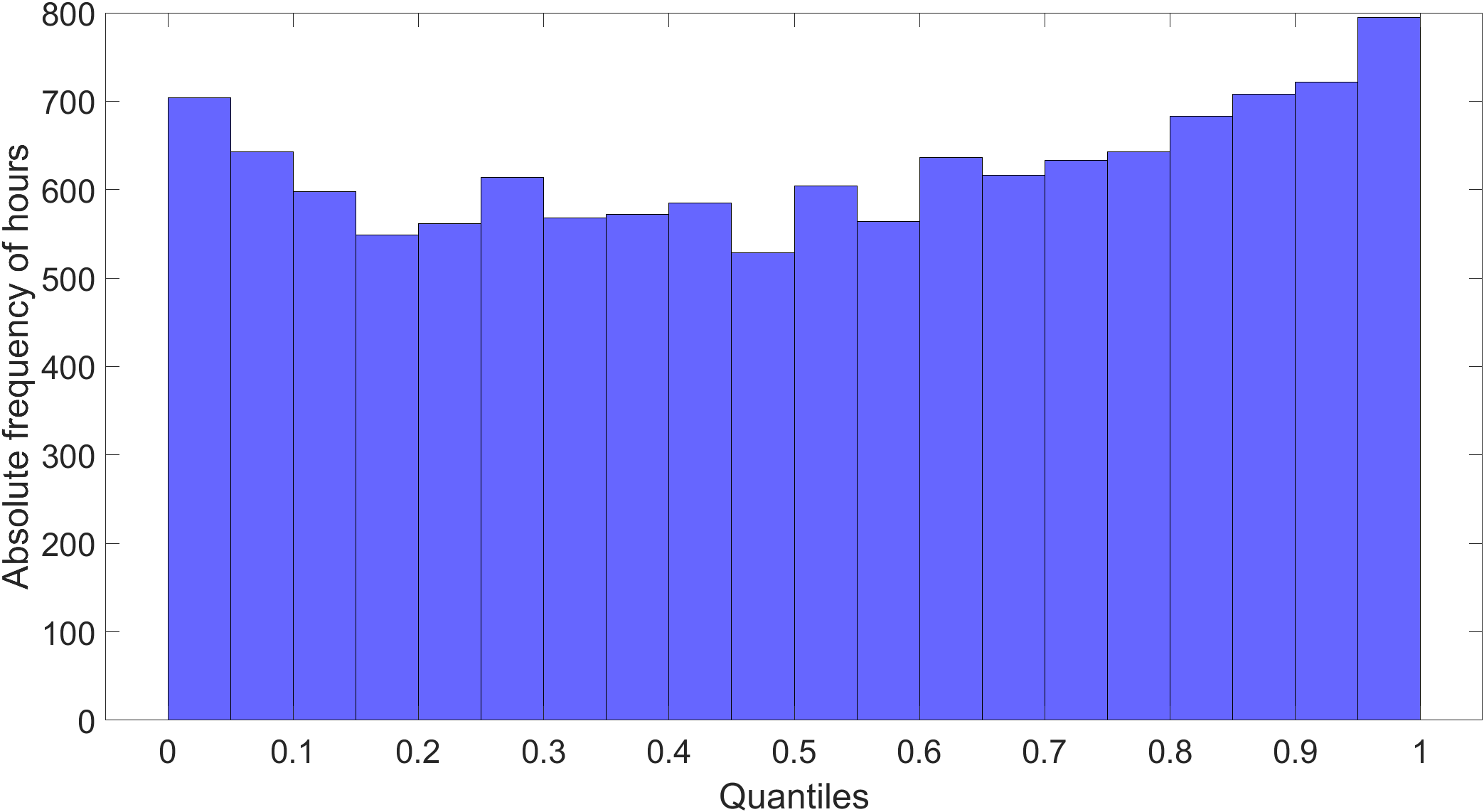}}
	\subfigure{\includegraphics[width=0.49\textwidth]{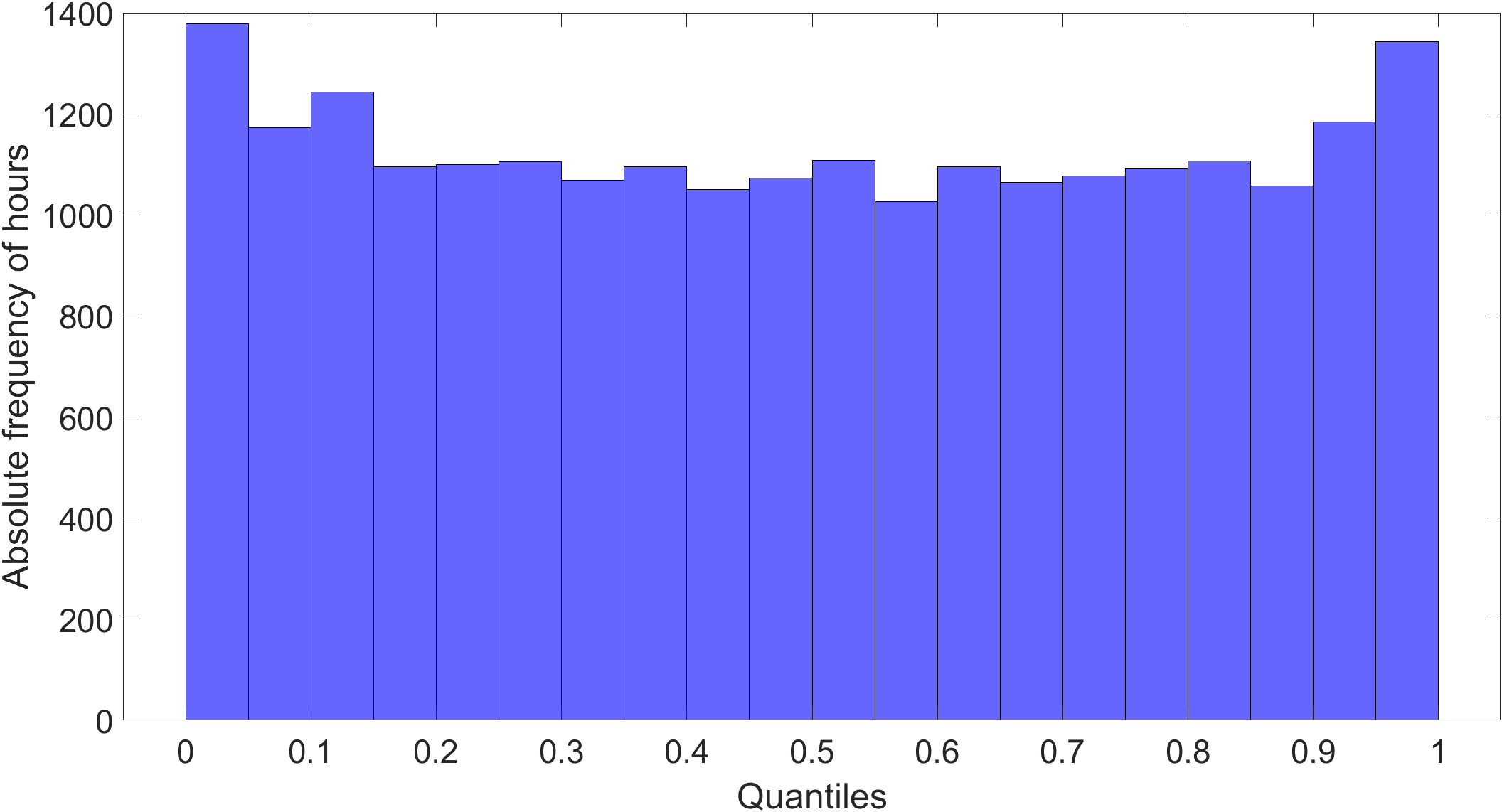}}
	\caption{Number of actual prices included in the quantiles estimated in the 2017–2020 forecast period, separated for peak (left) and off-peak (right) hours.}
	\label{H_fig_histogram_qra_price_pop}
\end{figure}

Since peak hours often have higher day-ahead prices than off-peak hours, our hybrid model predicts quantiles with two QRA estimations for each quantile, one for peak hours and one for off-peak hours. Thus, the probabilistic forecasts should be assessed for calibration separately in two disjunctive sub-sets. Figure \ref{H_fig_histogram_qra_price_pop} illustrates the calibration of the quantiles for peak hours on the left and off-peak hours on the right. During off-peak hours, there is a higher frequency in the outermost 5~\% quantiles on both sides (i.e., at high and low price levels). During peak hours, more hours exceed the quantile values, especially on the right side of the median. For example, 6.3~\% of the hours exceed the 95~\% quantile value. With a higher frequency in the low 5~\% quantile, the lowest prices can be attributed to the price-reducing influence of high electricity generation from renewable energy sources. This is consistent with the proportion of peak hours that show high demand for electricity but also a high feed-in of renewable energies due to high solar radiation or wind speeds and, thus, a high supply.

The calibration analysis of the predicted quantiles validates the calculated probabilistic forecasts. In the following, we evaluate the hourly distribution of the width of the forecast intervals to obtain more precise and quantified information about market uncertainty and the forecast accuracy that depends on it. Furthermore, we use probabilistic forecasts to calculate the probability of negative prices. Thus, we address the economic interest in including probabilistic forecasts in trading strategies and power plant deployment planning.

\begin{figure}[htbp]
	\centering
	\includegraphics[width=0.9\linewidth]{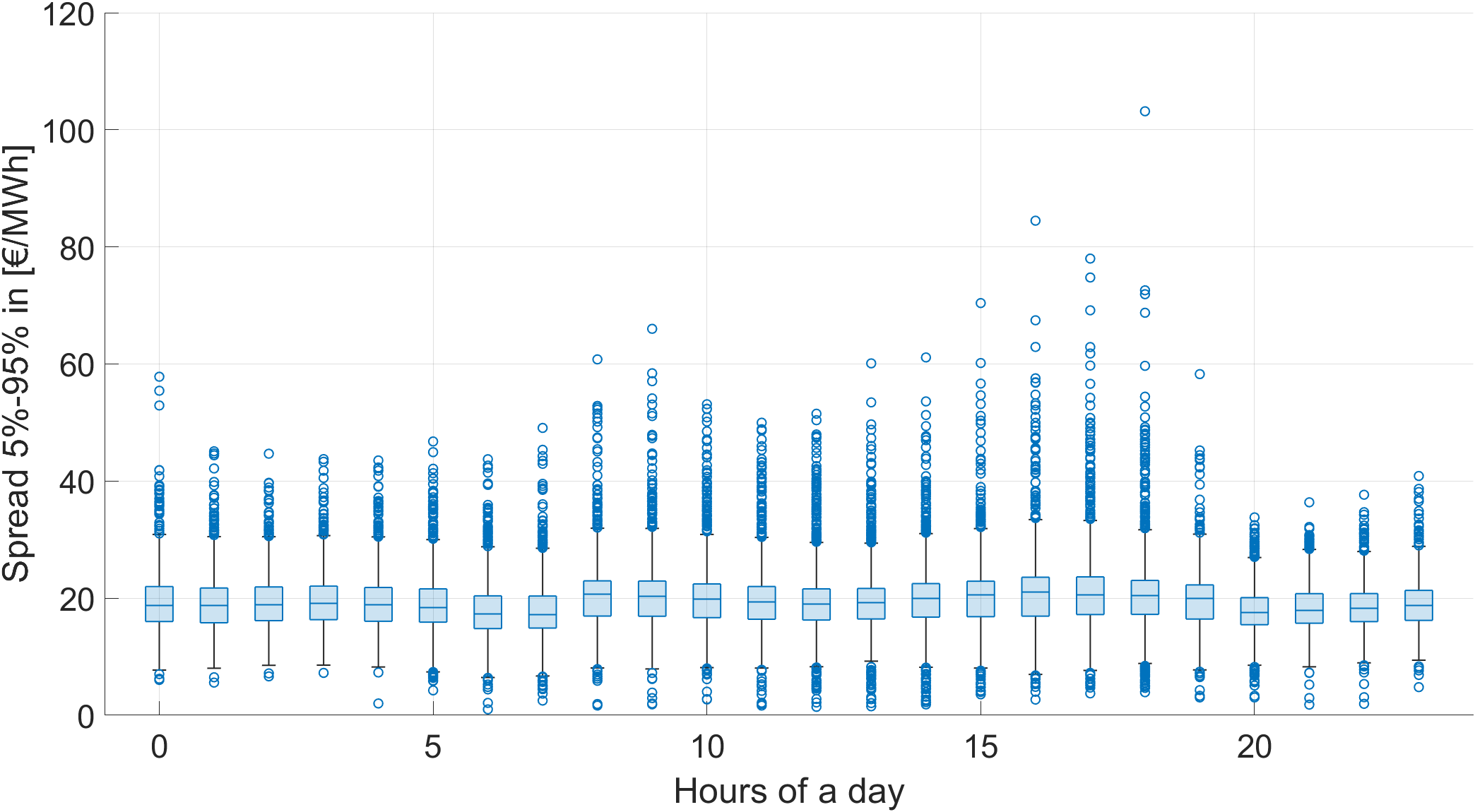}
	\caption{Box plot of the 90~\% prediction interval for each hour of the day.}
	\label{H_fig_boxplot_spread595_hourly}
\end{figure}

During peak hours, the uncertainty and fluctuations of the day-ahead prices are higher than during off-peak hours. This can be explained by a steeper merit order at high-load levels, the occurrence of start-ups during peak hours and the uncertainty of generation from renewable sources (especially PV plants, which produce more during peak hours). Our hybrid model captures this uncertainty with its probabilistic forecasts. Figure \ref{H_fig_boxplot_spread595_hourly} presents box plots that visualise the distribution of the spread between the estimated 95~\% quantile and the estimated 5~\% quantile for each hour of the day (the width of the 90~\% prediction interval. From 8~a.m. to 8~p.m. (daytime), the median spread is wider and outliers are higher than they are at night, reflecting the higher uncertainty and wider range of prices. Although the wider spread of the prediction interval covering 90~\% of the potential prices shows that these hours are more difficult to forecast than the night hours, the hybrid model can forecast them as accurately as it can the night hours. Figure \ref{H_fig_RMSE_BPOP} shows comparable point forecast accuracy for peak and off-peak hours over the entire period. Additionally, the probabilistic forecasts effectively mirror the market’s uncertainty.

\begin{figure}[htbp]
	\centering
	\includegraphics[width=0.9\linewidth]{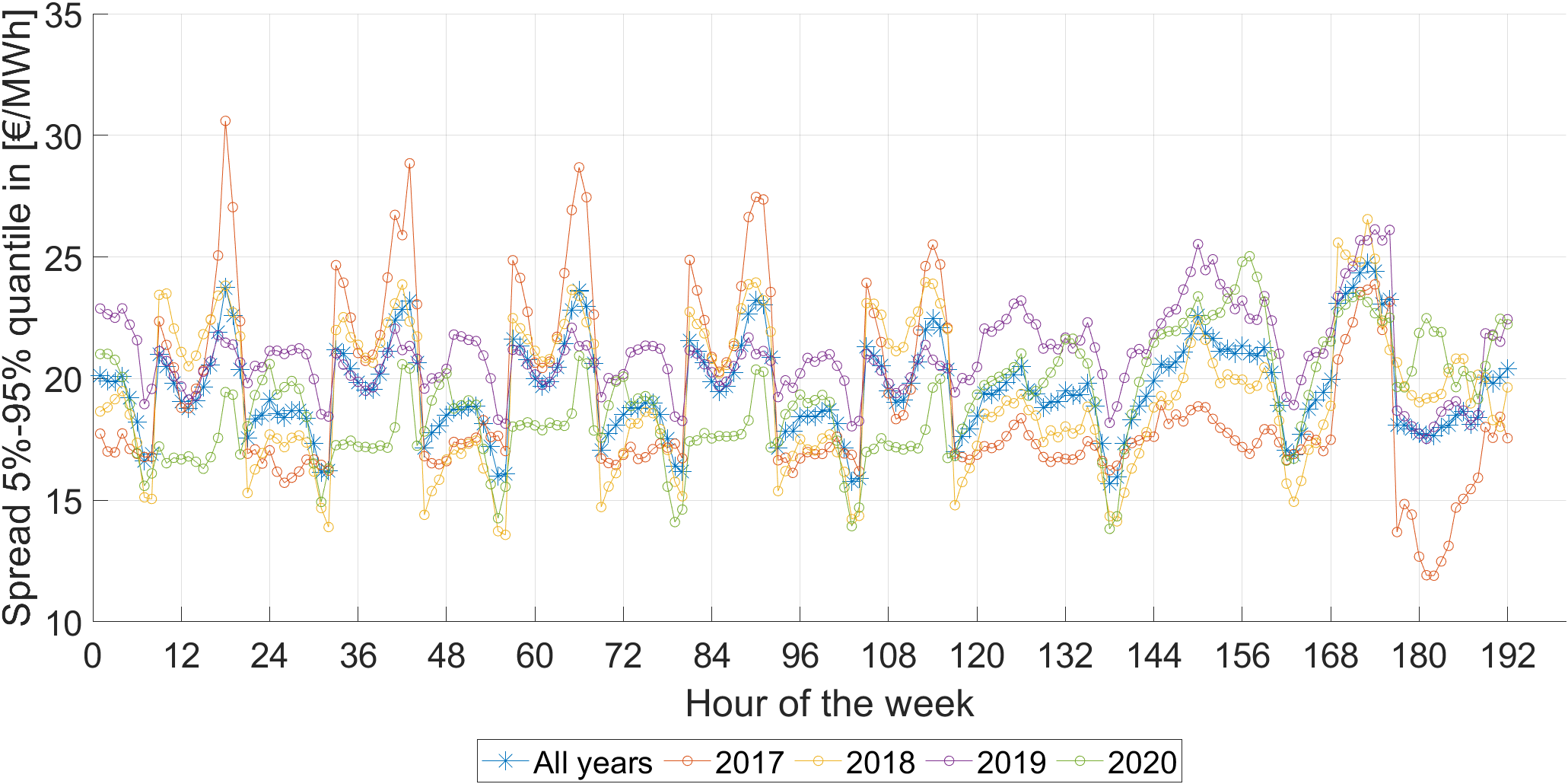}
	\caption{Average spread between the predicted 5~\% and 95~\% quantiles for the hours of the week.}
	\label{H_fig_spread595averaged_how}
\end{figure}

Figure \ref{H_fig_spread595averaged_how} shows the average width of the 90~\% prediction interval for each hour of the week, illustrating how easy or difficult it is to forecast the day-ahead electricity price for each individual hour of the week. A wider spread indicates greater uncertainty and less predictability, as the price may take on a wider range of potential values. As previously stated, daytime hours generally exhibit wider intervals than nighttime hours. However, more nuanced patterns are also discernible.

On weekdays (Monday to Friday), the spread of the prediction interval follows a distinct daily course featuring three concave curves with (local) maxima   at 3 a.m., 8 a.m. and 4 p.m. The interval is relatively narrow from midnight to 7 a.m. From 7 a.m. to 8 a.m., the interval increases abruptly, posing a major challenge for the forecasting of electricity prices, especially between 4 p.m. and 6 p.m. On weekends (Saturday and Sunday), the interval width is more evenly distributed across all hours but remains high overall and still exhibits a notable decrease between 7 p.m. and 9 p.m. Therefore, predicting day-ahead electricity prices requires careful consideration of the hour of the day and the day of the week on top of all other factors that can affect market dynamics.

\begin{figure}[htbp]
	\centering
	\includegraphics[width=0.9\linewidth]{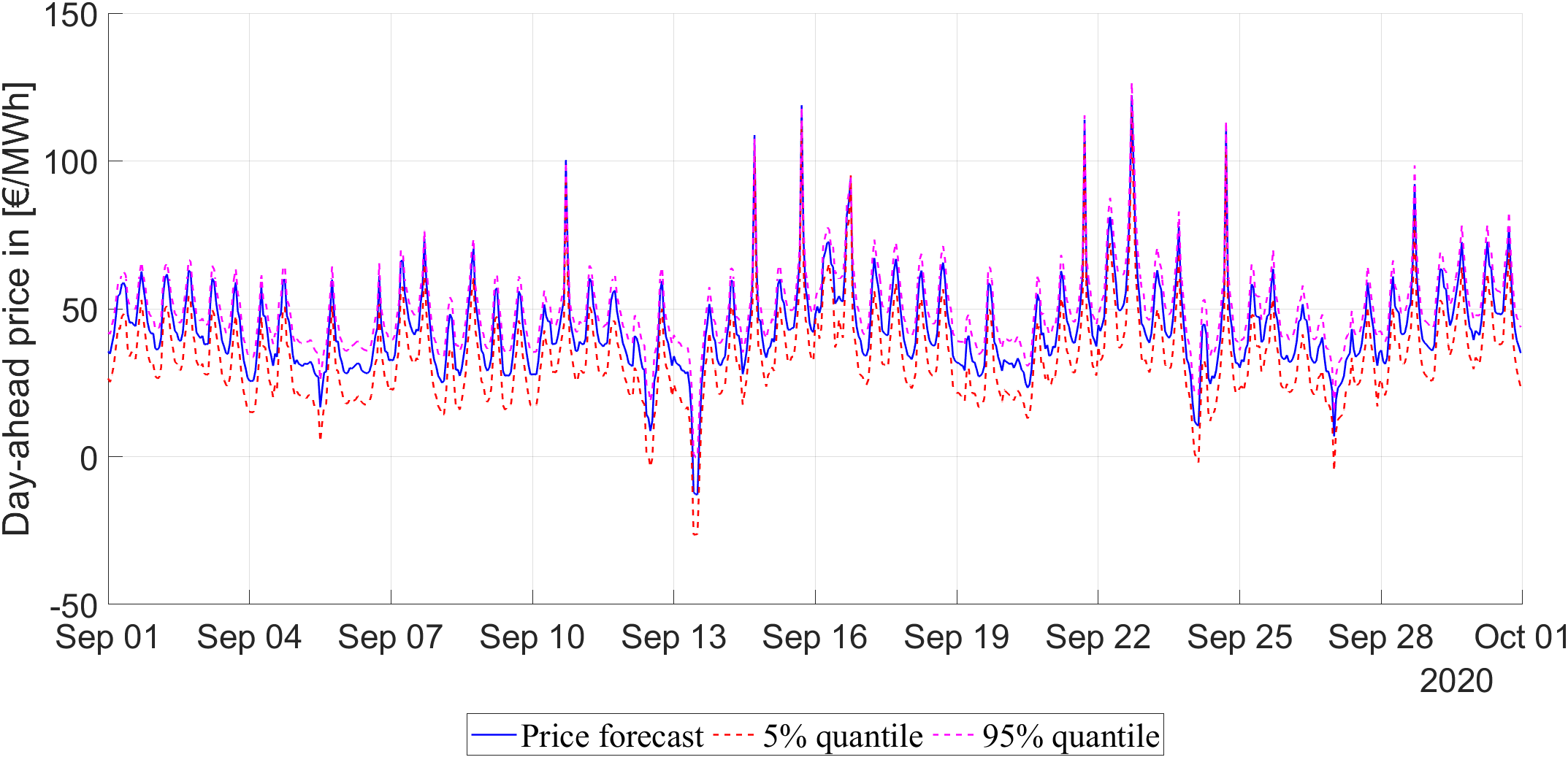}
	\caption{Day-ahead price forecast, lower-bound forecast (5~\% quantile) and upper-bound forecast (95~\% quantile).}
	\label{H_fig_density_price_course}
\end{figure}

Figure \ref{H_fig_density_price_course} shows an example of how the probabilistic forecasts account for uncertainty. It shows the course of lower and upper limits in combination with the forecast’s expected price (the price's point prediction). The presented limits correspond to the estimated 5~\% and 95~\% quantiles, respectively, and indicate the range of possible outcomes. More precisely, actual day-ahead electricity price has a 90~\% chance of lying within this range. Using the density forecast and predicted bounds, we can make informed risk estimates and probability statements, which can be useful for decision-making amid uncertainty.

\begin{figure}[htbp]
	\centering
	\includegraphics[width=0.9\linewidth]{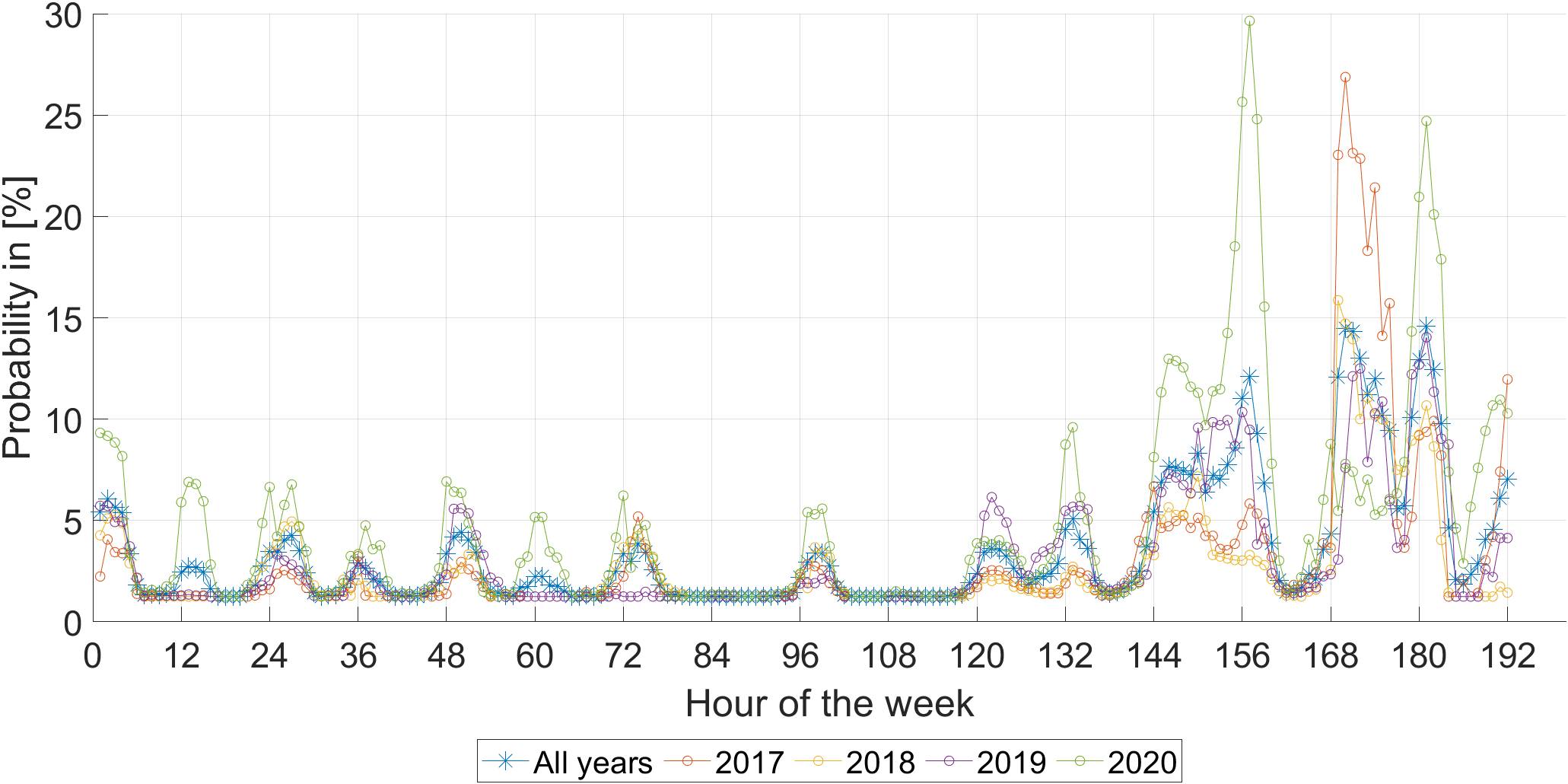}
	\caption{Probability of negative day-ahead prices.}
	\label{H_fig_density_price_negprob_how}
\end{figure}

Figure \ref{H_fig_density_price_negprob_how} illustrates the probabilities of negative prices for each hour of the week, as estimated by the density forecast. Notably, negative prices can occur when, for example, increases in generation by renewable energy sources or a large share of must-run capacities (e.g., CHP) force conventional power plant units to shut down, leading to start-up costs when these capacities are needed again. The merit order effect of renewable energies and the regulation of renewable energies, in particular payments for production besides the wholesale electricity market, can exacerbate this situation.  

Figure \ref{H_fig_density_price_negprob_how} reveals that the probability of negative prices is highest on Sundays and public holidays, reaching 10 to 15~\% in some hours. This reflects the increased occurrence of negative prices in these hours in the actual day-ahead prices. This is due to the fact that electricity demand is relatively low during these periods, increasing the likelihood of negative price events. The probability of negative prices is also relatively high (above 5~\%) in the early hours of Monday, followed by a slightly higher probability of negative prices in the early hours of the other days of the week. This pattern reflects the low-load behaviour of electricity demand in the early hours combined with the typically high wind feed-in during the night and early morning, which can lead to price drops. On Sundays and holidays, the daytime generation from PV and continuous wind feed-in can increase the likelihood of negative prices due to reduced demand.

The probability values for weekdays (Monday to Friday) are similar in their level and pattern, as these days have comparable fundamental parameters. With the probabilities determined by the density forecasts, bidding strategies, portfolios hedges and power plant usage can be optimised. The risks of negative prices or large price spreads can be incorporated into strategies through the probability statements. 

In summary, the results show the high-quality point and probabilistic price forecasts of the hybrid model, which also offers two additional advantages over previous approaches. The integration of a techno-economic energy system model allows for a deeper understanding of (energy) markets, and the use of stochastic approaches enables probabilistic price forecasts to quantify uncertainty in these markets and provide valuable information on the probability of negative prices and other market phenomena, helping market participants to make informed decisions and mitigate risks.

\section{Analysis of Individual Model Steps}
\label{H_sec.analysis model steps}

The hybrid model presented in this paper is multi-layered. In this section, we offer some insights into the individual components (steps) of the hybrid model. Section \ref{H_sec.results.datapreprocessing} shows the results of the day-ahead %and two-day-ahead 
load forecasts. %, which are transferred together with scenarios for the two-day-ahead load to the \emph{em.power dispatch}. 
Section \ref{H_sec.results.esm} provides an evaluation of the price estimators that follow the energy system optimisation step. Finally, section \ref{H_sec.results.postprocessing} points out the impact of the six individual price estimators' error forecasts.

\subsection{Load Forecasts}
\label{H_sec.results.datapreprocessing}

This section presents the results of the load forecasts, where we compare the RMSE and MAE of TSOs’ load forecasts with the improved load forecasts in Table \ref{H_table_result_load_point}, showing a significant improvement of 19~\% in RMSE and 27~\% in MAE.  
We refer to \cite{Preprocessing} for a more detailed analysis.
\begin{table}[htbp]
		%\bigskip
		\begin{center}
        \caption{RMSE and MAE for the original TSO day-ahead load forecast (TSO) and the improved day-ahead load forecast (Impr.), given in [$MWh$].}

	\small
		\begin{tabular}{|l|l||r|r|r|r|r|r|r|r|}
			\hline
            		&	year	&	all	&	2016	&	2017	&	2018	&	2019	&	2020	\\
\hline															
            	&	TSO	&	2,335.49	&	2,596.48	&	1,802.61	&	2,360.51	&	2,454.70	&	2,383.23	\\
            RMSE	&	Impr.	&	1,881.64	&	1,872.44	&	1,483.45	&	2,043.87	&	1,665.76	&	1,859.29	\\
            	&	\% Improvement	&	19.43	&	27.89	&	17.71	&	13.41	&	32.14	&	21.98	\\
            \hline													
            	&	TSO	&	1,797.05	&	2,028.44	&	1,396.45	&	1,726.67	&	1,951.00	&	1,881.84	\\
            MAE	&	Impr.	&	1,320.83	&	1,302.17	&	1,106.12	&	1,372.55	&	1,283.17	&	1,405.71	\\
            	&	\% Improvement 	&	26.50	&	35.80	&	20.79	&	20.51	&	34.23	&	25.30	\\
			\hline
			
		\end{tabular}
		\label{H_table_result_load_point}
	\end{center}
\end{table}

\subsection{Price Estimators After the Energy System Optimisation Step}
\label{H_sec.results.esm}

In optimising the energy system model, our hybrid model generates price estimators derived from the dual variable of demand constraint (see Section \ref{H_sec.ESM}). Computing the dual variable, the model optimises the energy sector’s complex technical and economic interdependencies. In addition to the costs of electricity generation, the model considers unit commitment decisions, such as the start-up and shutdown of generation units, storage operation, limitations on electricity transport to and from neighbouring countries, heat-supply requirements and the provision of control power. All of these techno-economic interdependencies determine the resulting electricity prices and are essential in our model, as they signal both high price peaks and low or even negative electricity prices observable on the market. 

Table \ref{H_table_descstat_esm} presents the descriptive statistics of the price estimation errors as well as the RMSE and MAE of these price estimators after the energy optimisation step. For all years, the RMSE is 9.50~\euro{}/MWh, and the MAE is 6.00~\euro{}/MWh. The lowest errors can be observed in 2016, and the highest errors can be observed in 2017. 
In comparison to state-of-the-art electricity price forecasting models in the literature, the error measurements are larger, and the errors still show a structural behaviour. For most years, the error’s mean and median are both negative values, meaning that the \emph{em.power dispatch} calculates prices higher than the observed prices on the market. With a value of -15.30~\euro{}/MWh for the error’s 5~\% quantile and a value of 9.60~\euro{}/MWh for the error’s 95~\% quantile, 90~\% of the error values lie in this interval.  

\begin{table}[htbp]
		%\bigskip
		\begin{center}
        \caption{Descriptive statistics of the error of energy system optimisation step in [€/MWh].}
	
	%\footnotesize
		\begin{tabular}{|l|r|r|r|r|r|r|}
			\hline
				    &	all years	&	2016	&	2017	&	2018	&	2019	&	2020	\\ \hline
            mean	&	-2.17	&	-2.23	&	-3.06	&	0.02	&	-0.73	&	-4.84	\\ \hline
            median	&	-1.75	&	-2.24	&	-2.75	&	0.71	&	-0.23	&	-3.94	\\ \hline
            minimum	&	-143.85	&	-143.85	&	-102.45	&	-81.45	&	-80.01	&	-79.05	\\ \hline
            maximum	&	105.73	&	50.62	&	105.73	&	55.19	&	42.87	&	86.76	\\ \hline
            5\%-quantile	&	-15.30	&	-11.81	&	-17.77	&	-13.24	&	-12.78	&	-19.50	\\ \hline
            95\%-quantile	&	9.60	&	7.47	&	10.68	&	11.40	&	9.75	&	6.05	\\ \hline
            Std.	&	9.25	&	7.26	&	11.62	&	8.61	&	8.05	&	9.31	\\ \hline
            RMSE	&	9.50	&	7.60	&	12.01	&	8.61	&	8.09	&	10.49	\\ \hline
            MAE	&	6.00	&	4.83	&	7.09	&	5.91	&	5.07	&	7.13	\\ \hline

		\end{tabular}
  	\label{H_table_descstat_esm}
	\end{center}

\end{table}

\begin{figure}[htbp]
	\centering
	\includegraphics[width=0.9\linewidth]{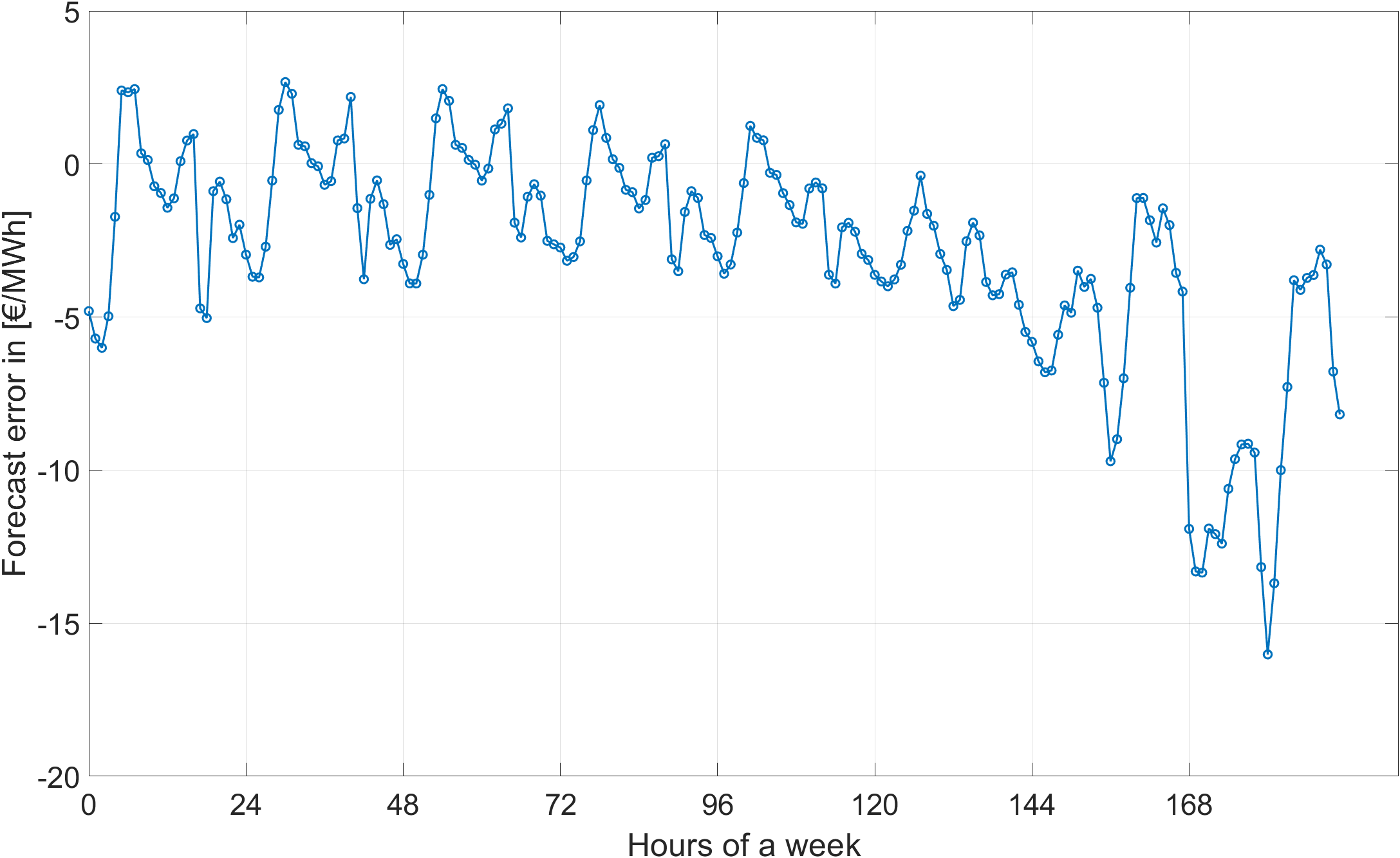}
	\caption{Mean price estimator errors for the hours of a week (including public holidays at hours 168–192) after the energy system optimisation step in [\euro{}/MWh].}
	\label{H_fig_error_esm_hourwise192}
\end{figure}

\begin{figure}[htbp]
	\centering
	\includegraphics[width=0.9\linewidth]{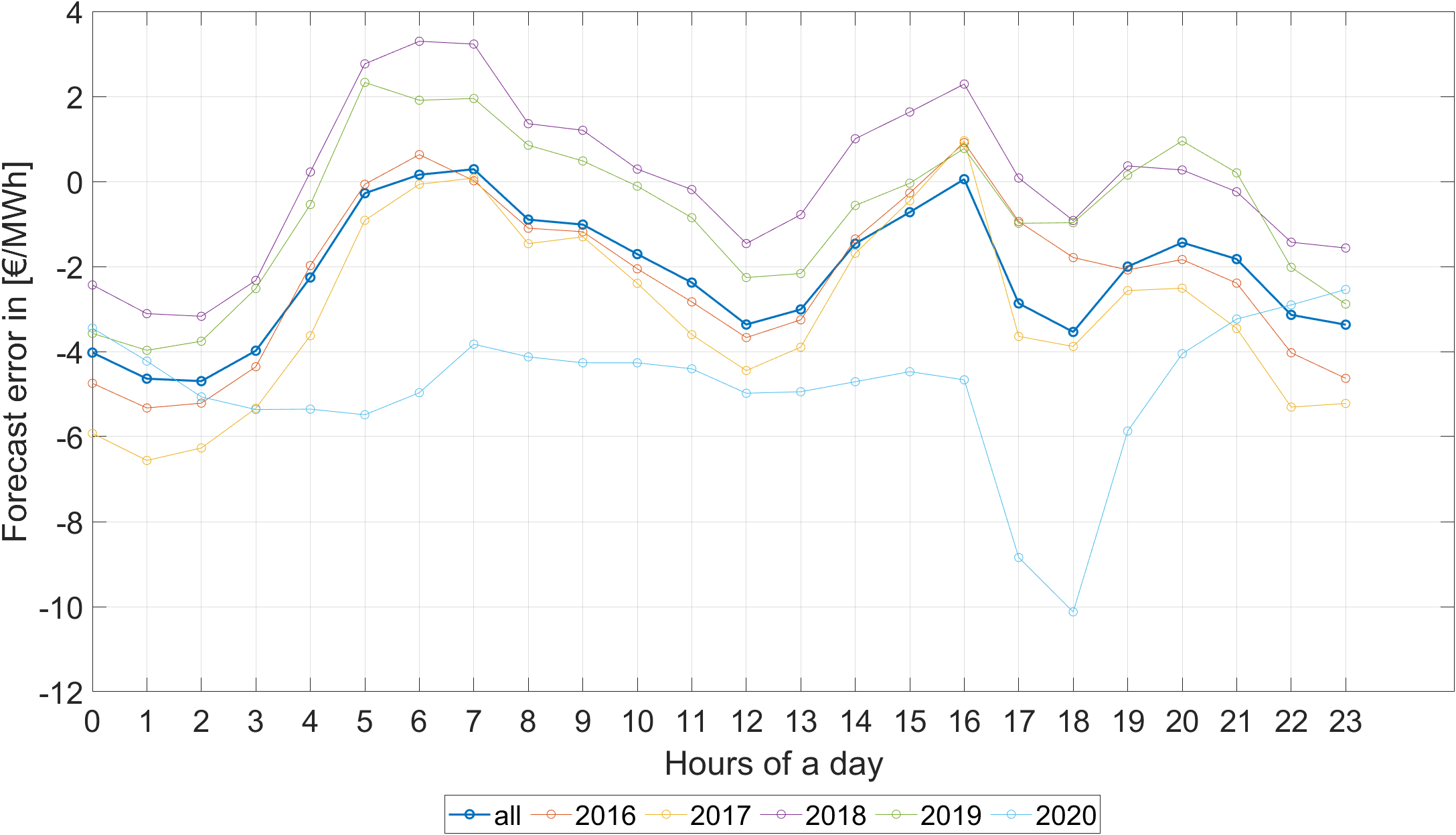}
	\caption{Mean price estimator errors for the hours of a day in each year after the energy system optimisation step in [\euro{}/MWh].}
	\label{H_fig_error_esm_hourwise}
\end{figure}

Figure \ref{H_fig_error_esm_hourwise192} plots the error structure of the entire period divided into the hours of a week. Negative values indicate an average overestimation of the prices in the corresponding hour, while positive values indicate an average underestimation of the prices in the corresponding hour. 
The figure shows that the errors tend to increase over the weekend, while there is a distinct daily pattern observed (see also Figure \ref{H_fig_error_esm_hourwise}). 

The model tends to underestimate prices during the night while overestimating them in the morning and late afternoon, which is a consistent pattern across all years from 2016 to 2019. While the error structure in 2020 shows some differences, possibly attributed to the COVID-19 pandemic, the model's RMSE per hour of the day does not significantly differ from previous years. We also observe a midday increased error rate and an evening error peak between 4 p.m. to 6 p.m., which is more pronounced in the 2020 data. 

Note, however, that we are evaluating here the short-term price prediction of a techno-economic energy system model, which we expect, due to its inability to learn from history, to map the general market situation at these time horizons, but not to form very accurate price estimators. 
Therefore, the model performs reasonably well with predictable patterns in the errors, which indicates a high potential for the data post-processing step. The effect of this step is presented in the following chapter.

\begin{figure}[htbp]
	\centering
	\includegraphics[width=0.9\linewidth]{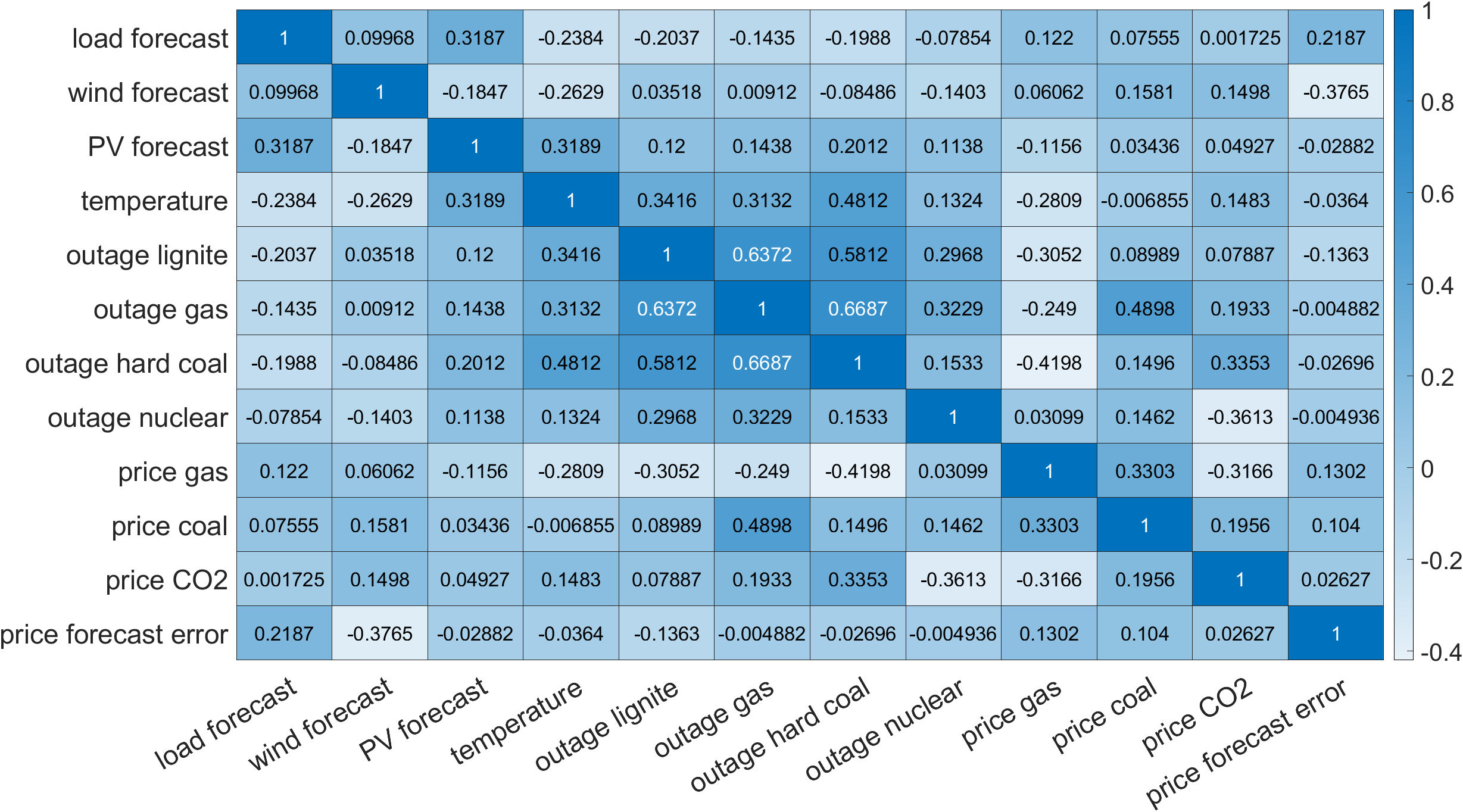}
	\caption{Correlation of possible exogenous variables and the price estimation error.}
	\label{H_fig_heatmap}
\end{figure}

Figure \ref{H_fig_heatmap} illustrates the correlation of the price forecast errors with several exogenous variables. Evidently, wind generation correlates significantly with the price forecast errors from the energy optimisation step and explains 14~\% of the overall forecast error’s volatility.

\subsection{Improvement of Price Estimators}
\label{H_sec.results.postprocessing}

The detected systematic deviations and seasonal patterns in the errors of the price estimators following the energy system optimisation step can be captured and predicted by the stochastic model in the post-processing step in a way that improves the price forecast's RMSE for 2016–2020 by 22~\% through the combination of both model classes. 
The statistical data of the hybrid model’s forecast error show that the error is distributed nearly symmetrically with a mean of -0.09~\euro{}/MWh and a median of -0.08~\euro{}/MWh. Post-processing centres the price forecast error to near zero. At the same time, both the minimum and maximum errors generated decrease each year and over the entire period -- as does overall volatility. Thus, the post-processing model achieves a reduction of 20~\% in the standard deviation of the day-ahead price forecast, compared to the one of the price estimators of the energy system optimisation step. 

\begin{table}[htb]
		%\bigskip
		
	\begin{center}
        \caption{Descriptive statistics of the error of the hybrid model in [€/MWh]}
	%\footnotesize
		\begin{tabular}{|l|r|r|r|r|r|r|}
			\hline
				    	&	all	years &	2016	&	2017	&	2018	&	2019	&	2020	\\ \hline
                mean	&	-0.09	&	0.30	&	-0.15	&	1.12	&	-0.73	&	-0.99	\\ \hline
                median	&	-0.08	&	0.08	&	-0.23	&	1.06	&	-0.68	&	-0.50	\\ \hline
                minimum	&	-133.44	&	-133.44	&	-81.29	&	-54.41	&	-79.27	&	-69.43	\\ \hline
                maximum	&	86.10	&	43.76	&	74.30	&	52.72	&	35.48	&	86.10	\\ \hline
                5\%-quantile	&	-9.58	&	-6.64	&	-10.94	&	-9.09	&	-9.36	&	-11.51	\\ \hline
                95\%-quantile	&	10.07	&	8.48	&	11.21	&	11.70	&	9.08	&	8.24	\\ \hline
                Std.	&	7.38	&	5.82	&	8.79	&	7.20	&	7.01	&	7.56	\\ \hline
		\end{tabular}
  	\label{H_table_descstat_hybrid}
	\end{center}

\end{table}

\begin{figure}[htbp]
	\centering
	\includegraphics[width=0.9\linewidth]{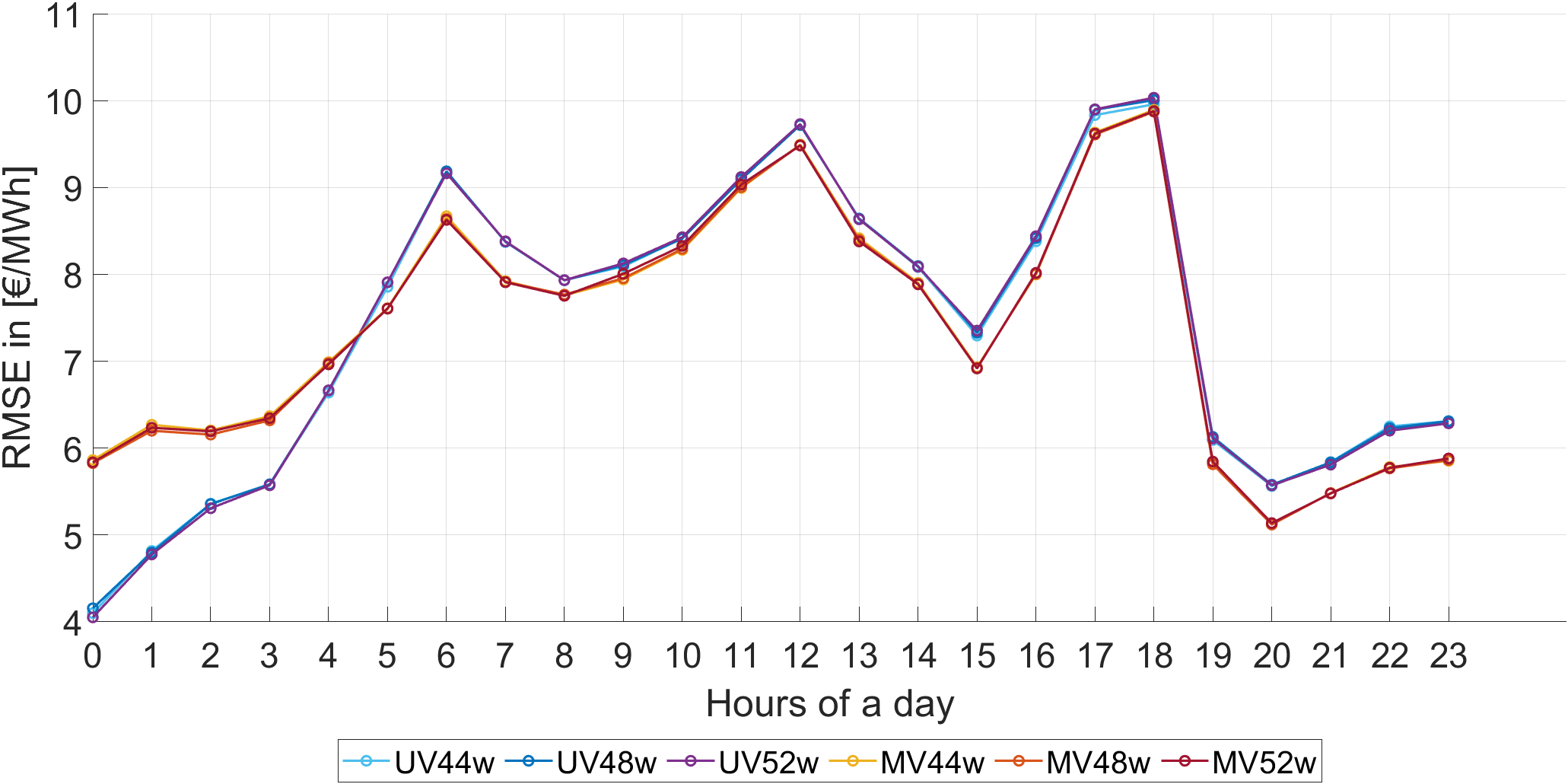}
	\caption{RMSE for univariate and multivariate post-processing sub-models for each hour of the day in [\euro{}/MWh].}
	\label{H_fig_RMSE_uvmv_hourly}
\end{figure}

The stochastic post-processing step consists of six individual sub-models, as described in Section \ref{H_sec.EIM}. Each of these sub-models improves the price estimators obtained after the energy system optimisation step, significantly increasing the forecast quality, as measured in Table \ref{H_table_result_EIM_all} by specifying the error measures RMSE and MAE. While the models with multivariate frameworks have lower error measures than those with univariate frameworks over the entire period, the price forecasts of the univariate models are qualitatively better than those of the multivariate models in 2019 and 2020. Figure \ref{H_fig_RMSE_uvmv_hourly} shows the hourly RMSE for each sub-model. Depending on the length of the calibration window, the models of the univariate framework dominate those of the multivariate framework in the first five hours of a day. From that point forward, the multivariate sub-models achieve lower error measures. The result aligns with the expectations of the model specifications. In the univariate framework, the time series is captured as a high-frequency data set, so the forecast includes data up to the last available hour. However, this also requires 24 future values to be forecasted, heightening inaccuracy. With the multivariate framework, a reverse effect is observable due to the split into 24 individual time series, as this split entails even just short steps into the future needing to be forecasted. Still, the most recent available information is only partially considered.

\begin{table}[htb]

    \caption{Error measurements RMSE and MAE of the error time series in [€/MWh].}
	%\begin{center}
	\scriptsize
		\begin{tabular}{|l|c||r|r|r|r|r|r|r|r c|}
			\hline
			\multicolumn{2}{|c||}{} &
			\multicolumn{1}{c|}{Initial} &
			\multicolumn{8}{c|}{Post-processing}\\
			\hline
			\multicolumn{2}{|c||}{} & \multicolumn{1}{c|}{ESM} & \multicolumn{1}{c|}{UV44w} & \multicolumn{1}{c|}{UV48w} & \multicolumn{1}{c|}{UV52w} & \multicolumn{1}{c|}{MV44w} & \multicolumn{1}{c|}{MV48w} & \multicolumn{1}{c|}{MV52w} & \multicolumn{2}{c|}{Combination} \\	\hline
		%&		&	ESM	&	UV44w	&	UV48w	&	UV52w	&	MV44w	&	MV48w	&	MV52w	&	Combination	&					\\	\hline
	&	All years &	9.50	&	7.57	&	7.59	&	7.59	&	7.51	&	7.50	&	7.51	&	7.38	&	\textcolor{green}{	22	\%	}	\\
	&	2016	&	7.60	&	6.07	&	6.13	&	6.09	&	5.95	&	5.94	&	5.92	&	5.82	&	\textcolor{green}{	23	\%	}	\\	
RMSE	&	2017	&	12.01	&	9.22	&	9.20	&	9.13	&	8.92	&	8.86	&	8.86	&	8.79	&	\textcolor{green}{	27	\%	}	\\	
	&	2018	&	8.61	&	7.49	&	7.47	&	7.54	&	7.40	&	7.38	&	7.40	&	7.28	&	\textcolor{green}{	15	\%	}	\\	
	&	2019	&	8.09	&	7.11	&	7.13	&	7.14	&	7.21	&	7.22	&	7.21	&	7.05	&	\textcolor{green}{	13	\%	}	\\	
	&	2020	&	10.49	&	7.65	&	7.69	&	7.71	&	7.77	&	7.79	&	7.85	&	7.63	&	\textcolor{green}{	27	\%	}	\\	\hline
	&	All	years &	6.00	&	4.73	&	4.75	&	4.75	&	4.75	&	4.74	&	4.74	&	4.60	&	\textcolor{green}{	23	\%	}	\\	
	&	2016	&	4.83	&	3.69	&	3.71	&	3.69	&	3.64	&	3.62	&	3.61	&	3.48	&	\textcolor{green}{	28	\%	}	\\	
MAE	&	2017	&	7.09	&	5.55	&	5.55	&	5.51	&	5.37	&	5.31	&	5.28	&	5.25	&	\textcolor{green}{	26	\%	}	\\	
	&	2018	&	5.91	&	5.20	&	5.18	&	5.21	&	5.24	&	5.23	&	5.25	&	5.07	&	\textcolor{green}{	14	\%	}	\\	
	&	2019	&	5.07	&	4.46	&	4.49	&	4.49	&	4.61	&	4.61	&	4.59	&	4.43	&	\textcolor{green}{	13	\%	}	\\	
	&	2020	&	7.13	&	4.77	&	4.80	&	4.83	&	4.91	&	4.94	&	4.99	&	4.77	&	\textcolor{green}{	33	\%	}	\\	\hline

		\end{tabular}
		%\bigskip

%	\end{center}
\label{H_table_result_EIM_all}
\end{table}

The individual characteristics of the univariate and multivariate frameworks also explain the average errors for each hour of the day (see Figure \ref{H_fig_error_uvmv_hourwise}) and each hour of the week (see Figure \ref{H_fig_error_uvmv_hourwise192}). While the forecasts of the multivariate sub-models do not indicate a clear daily pattern in the under- or overestimation of values, the univariate sub-models show a similar daily structure as the price estimators following the energy system optimisation step: underestimation during the hours of 4–8 a.m. and 2–4 p.m. and overestimation,e.g., during the hours of 17-18 p.m. and for the hours from 21 p.m. The average error for weekly hours displays a significantly higher overestimation and underestimation on public holidays. Notably, however, this trend is evident across all forecasts of the univariate and multivariate sub-models.

Table \ref{H_table_result_EIM_all} presents the error measures for the price forecast that stems from combining the forecasts of individual sub-models. Thus, it depicts the final price forecast of the hybrid model. The RMSE and MAE of this price forecast are lower than the error measures of the individual price forecasts over each individual year as well as the entire period. 

\begin{figure}[htbp]
	\centering
	\includegraphics[width=0.6\linewidth]{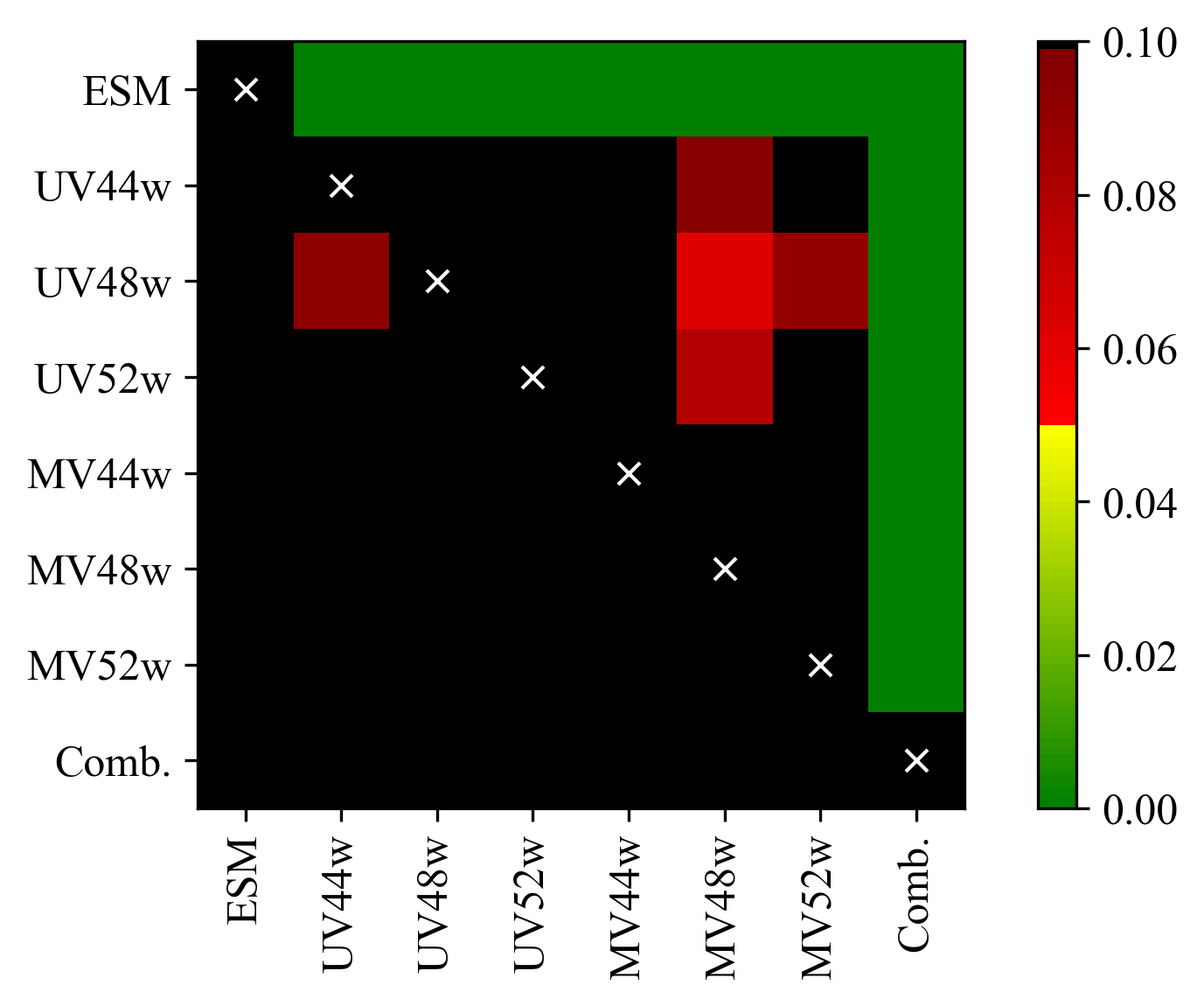}
	\caption{P-value of Diebold-Mariano test for all parts of the stochastic post-processing step (output derived from \textit{epftoolbox} by \cite{lago2021forecasting}).}
	\label{H_fig_result_DMtest}
\end{figure}

Using the multivariate Diebold-Mariano test from the \textit{epftoolbox} by \cite{lago2021forecasting} to compare different model forecasts via hypothesis testing and determine whether one forecast’s accuracy is significantly higher than that of the others, we show that the combination of the sub-models is significantly better than all six individual sub-models. The test results are shown in Figure \ref{H_fig_result_DMtest} as a heatmap illustrating the p-values of the hypothesis that the forecast of the model on the Y-axis is significantly more accurate than the forecast of the model on the X-axis. A p-value close to zero indicates that the model on the X-axis has a significantly higher forecast accuracy than the model on the Y-axis. The test indicates that, for forecasting day-ahead spot prices, the post-processing model and each stochastic sub-model are significantly better than the energy system model modified by stochastic pre-processing and interweaving, which means the hybrid model after the third step.

\begin{figure}[htbp]
	\centering
	\subfigure{\includegraphics[width=0.49\linewidth]{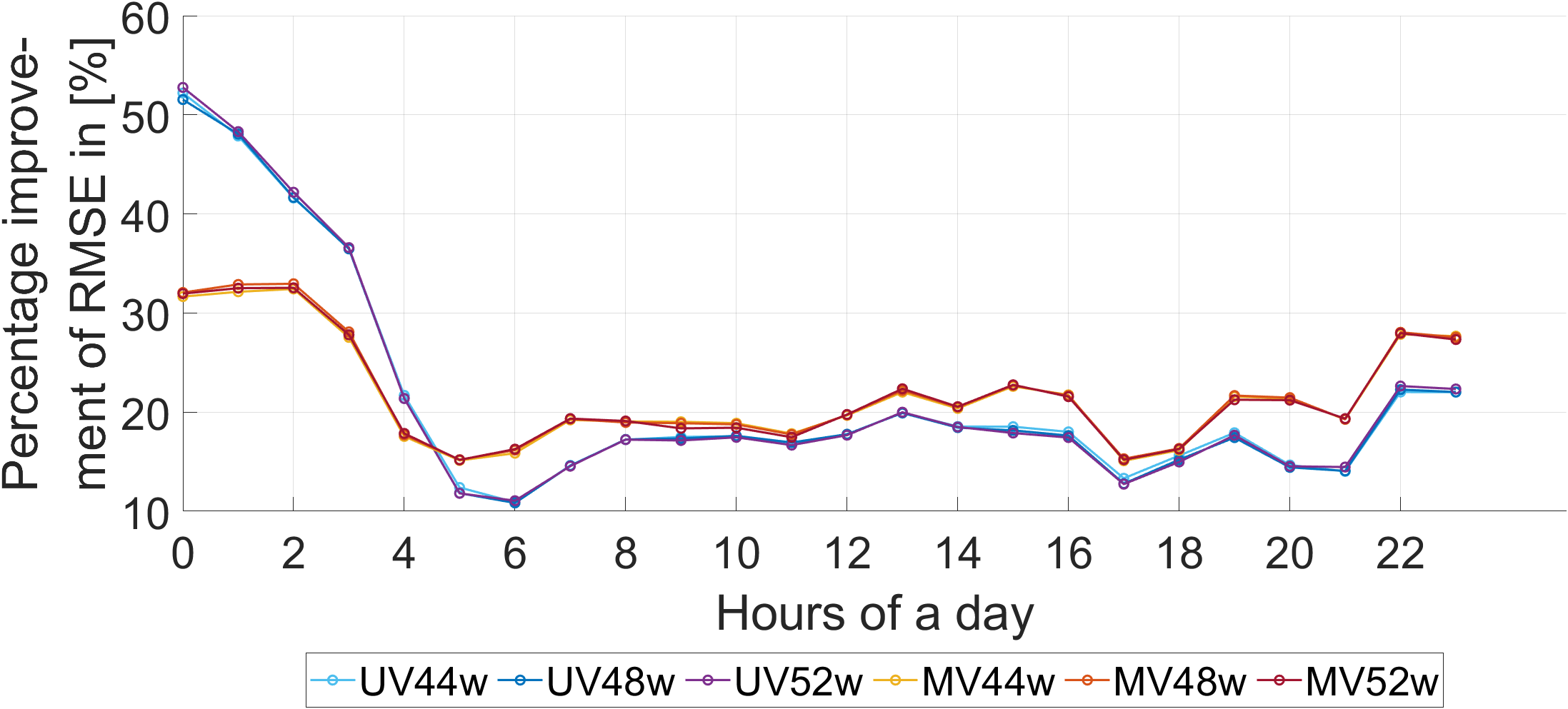}}
	\subfigure{\includegraphics[width=0.49\linewidth]{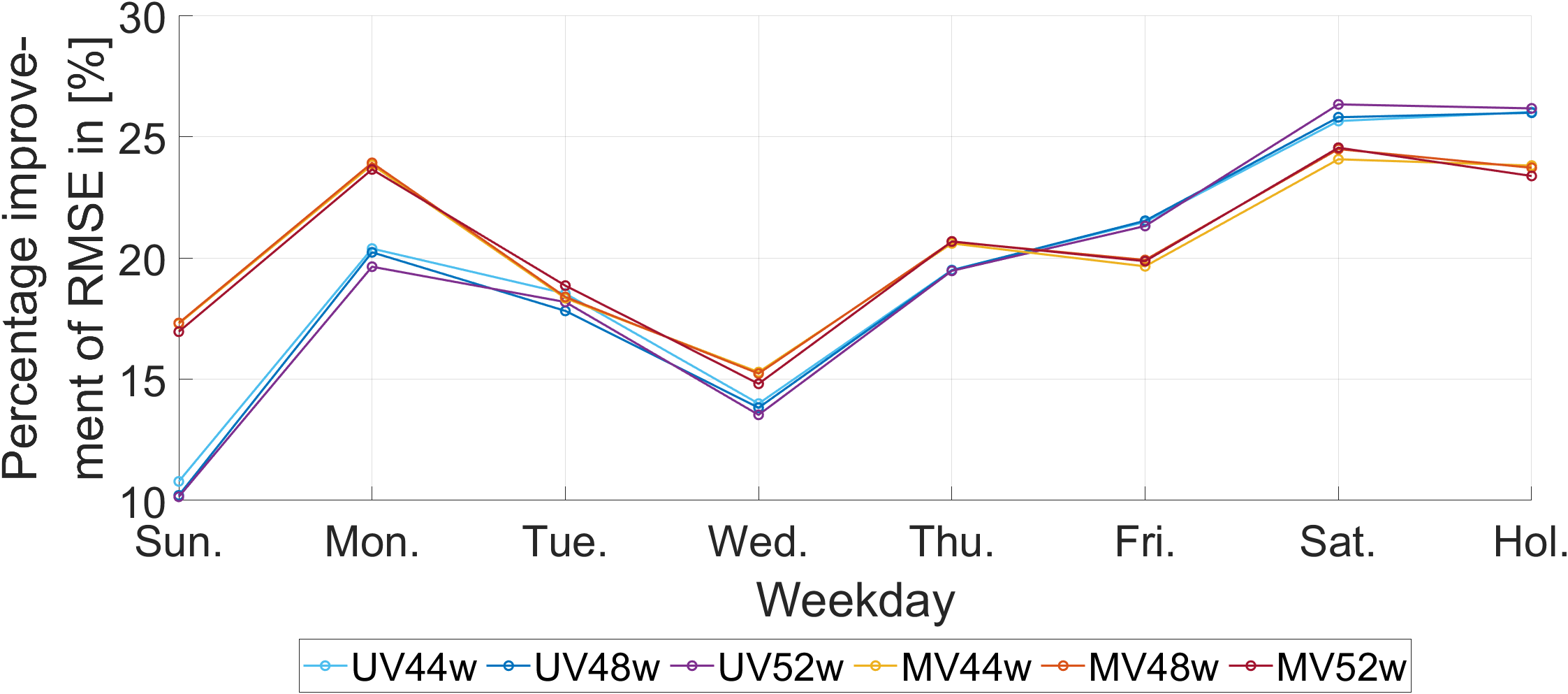}}
	\caption{Average RMSE improvement in day-ahead price forecasts for each hour of the day (left) and each day of the week (right).}
	\label{H_fig_impr_RMSE_hourly_weekday_EIM_all}
\end{figure}

To better attribute and understand the effect of the stochastic post-processing step, we also determine the improvement in RMSE for each hour of the day and day of the week, as shown in Figure \ref{H_fig_impr_RMSE_hourly_weekday_EIM_all}. Daytime and weekday structures, which can be observed in the price estimators’ errors after the energy system optimisation step, are evident in the improvement. This is due in large part to the seasonal components in the stochastic model and the affected  autoregressive structures. Notably, the most substantial improvements in terms of percentage are achieved at night. Additionally, weekends and especially holidays see relatively high levels of improvement. In the price estimators, these are the hours and days with the most significant mean error, meaning those with a sizeable mean error and, in turn, the most potential for improvement. Enhancing day-ahead price forecasts by reducing this error is a key goal of combining techno-economic energy system modelling and stochastic modelling.

\section{Conclusion}
\label{H_sec.concl}
This paper introduced a novel hybrid model for short-term electricity price forecasting, which combines stochastic modelling with (fundamental) techno-economic energy system modelling. The model consists of four steps. First, through stochastic data pre-processing, day-ahead load forecasts are significantly improved, and a load forecast for two days ahead is created. Second, this load forecast is extended by quantile forecasts in three different probable scenarios of hourly load consumption two days in advance using a parameter density forecast step. Third, the modelled quantities and input data are fed into the \emph{em.power dispatch}, a techno-economic market model adapted for predicting day-ahead price estimators. Fourth, the errors of the price estimators from the energy optimisation step are reduced by stochastic models through stochastic data post-processing and complemented by probabilistic forecasts.

The hybrid model presented in this paper combines the strengths of stochastic models (strength: trained with price data) and energy system models (strength: insights based on economic theory extending beyond prices). It can be parameterised with data from most energy markets worldwide, providing insights into a wide range of energy systems. We demonstrated its performance using German day-ahead electricity market data from January 2016 to December 2020. The price forecasts from the model had an annual average RMSE of 7.38~\euro{}/MWh and an annual average MAE of 4.60~\euro{}/MWh. The hybrid model's forecasting accuracy surpasses the majority of benchmarks in the literature and matches the best statistical benchmark identified in the literature. This confirms that techno-economic energy system models can provide valuable input for price forecasting, even for short-term forecasting.

We conducted an in-depth analysis of price forecast quality and identified a notable relationship between the error in price forecasts and the price level. Over the years, the highest RMSE has consistently been found in the hours with the lowest 20~\% and the highest 20~\% of prices. This result can be explained by start-up costs and their impact on hourly prices, as discussed in Section \ref{H_sec.results.Hybrid}, and naturally affects all forecast models.

Additionally, our model provides calibrated density forecasts alongside point forecasts. With these, it can, for example, quantify the probability of negative price events. Our analysis revealed that some hours on Sundays and public holidays have a probability of more than 10~\% of becoming negative. 

Future research should focus on determining the most suitable model type for cases with varying lead times in price forecasts. As discussed in the first part of this paper, the relative strength of techno-economic models in handling structural breaks suggests their greater suitability for long-term forecasts (e.g., years ahead), while stochastic models, whether time-series-based or artificial intelligence-based, are better suited for short-term forecasts (e.g., intra-day). Nevertheless, hybrid models like the one introduced in this paper for day-ahead forecasts or similar approaches that combine the advantages of both model classes mentioned above could, if properly adapted, be suitable for performing accurate forecasts with variable lead times. As a result, we think that it is crucial to further explore and develop hybrid models to harness their potential for delivering accurate forecasts across various lead times, making them a valuable addition to the existing forecasting methodologies.

\section*{Data Availability}
\label{H_sec.available}
The data and source code presented in this work are openly available through the following link: \hyperlink{https://github.com/ProKoMoProject/A-hybrid-model-for-day-ahead-electricity-price-forecasting-combining-fundamental-and-stochastic-mod}{https://github.com/ProKoMoProject/A-hybrid-model-for-day-ahead-electricity-price-forecasting-combining-fundamental-and-stochastic-mod}.

\section*{Acknowledgements}
The work was supported by the German Federal Ministry of Economic Affairs and Climate Action through the research project ProKoMo within the Systems Analysis Research Network of the 6th energy research program.

%------------------------------------------------------------------------------------------------------------------------
%  Bibliography
%************************************************************************************************************************
%------------------------------------------------------------------------------------------------------------------------
%\newpage

%\bibliographystyle{econometrica}
% \addbibresource{References.bib}
%\bibliographystyle{plainnat}
%\bibliography{bibliography}

\printcredits

% \bibliographystyle{cas-model2-names}
\bibliographystyle{unsrtnat}

\bibliography{bibliography}

\begin{thebibliography}{110}
\providecommand{\natexlab}[1]{#1}
\providecommand{\url}[1]{\texttt{#1}}
\expandafter\ifx\csname urlstyle\endcsname\relax
  \providecommand{\doi}[1]{doi: #1}\else
  \providecommand{\doi}{doi: \begingroup \urlstyle{rm}\Url}\fi

\bibitem[Weron(2014)]{Weronreview2014}
Rafał Weron.
\newblock {Electricity price forecasting: A review of the state-of-the-art with
  a look into the future}.
\newblock \emph{International Journal of Forecasting}, 30\penalty0
  (4):\penalty0 1030--1081, 2014.
\newblock ISSN 0169-2070.
\newblock \doi{10.1016/j.ijforecast.2014.08.008}.

\bibitem[Nowotarski and Weron(2018)]{Nowotarski2018}
Jakub Nowotarski and Rafał Weron.
\newblock Recent advances in electricity price forecasting: {A} review of
  probabilistic forecasting.
\newblock \emph{Renewable and Sustainable Energy Reviews}, 81:\penalty0
  1548--1568, 1 2018.
\newblock ISSN 18790690.
\newblock \doi{10.1016/j.rser.2017.05.234}.

\bibitem[Hong et~al.(2016)Hong, Pinson, Fan, Zareipour, Troccoli, and
  Hyndman]{Hong2016}
Tao Hong, Pierre Pinson, Shu Fan, Hamidreza Zareipour, Alberto Troccoli, and
  Rob~J. Hyndman.
\newblock {Probabilistic Global Energy Forecasting Competition 2014 and
  beyond}.
\newblock \emph{International Journal of Forecasting}, 32:\penalty0 896--913,
  2016.
\newblock ISSN 01692070.
\newblock \doi{10.1016/J.IJFORECAST.2016.02.001}.

\bibitem[Hong et~al.(2020)Hong, Pinson, Wang, Weron, Yang, and
  Zareipour]{Hong2020}
Tao Hong, Pierre Pinson, Yi~Wang, Rafał Weron, Dazhi Yang, and Hamidreza
  Zareipour.
\newblock {Energy Forecasting: A Review and Outlook}.
\newblock \emph{IEEE Open Access Journal of Power and Energy}, 7:\penalty0
  376--388, 2020.
\newblock \doi{10.1109/OAJPE.2020.3029979}.

\bibitem[Amjady and Hemmati(2006)]{Amjady2006EnergyPF}
N.~Amjady and M.~Hemmati.
\newblock Energy price forecasting - problems and proposals for such
  predictions.
\newblock \emph{IEEE Power and Energy Magazine}, 4\penalty0 (2):\penalty0
  20--29, 2006.
\newblock \doi{10.1109/MPAE.2006.1597990}.

\bibitem[Aggarwal et~al.(2009)Aggarwal, Saini, and Kumar]{AGGARWAL200913}
Sanjeev~Kumar Aggarwal, Lalit~Mohan Saini, and Ashwani Kumar.
\newblock {Electricity price forecasting in deregulated markets: A review and
  evaluation}.
\newblock \emph{International Journal of Electrical Power \& Energy Systems},
  31\penalty0 (1):\penalty0 13--22, 2009.
\newblock ISSN 0142-0615.
\newblock \doi{10.1016/j.ijepes.2008.09.003}.

\bibitem[Weron and Ziel(2019)]{Weron2019}
Rafa{\l} Weron and Florian Ziel.
\newblock Electricity price forecasting.
\newblock In \emph{Routledge Handbook of Energy Economics}, pages 506--521.
  Routledge, 9 2019.
\newblock \doi{10.4324/9781315459653-36}.

\bibitem[Petropoulos et~al.(2022)Petropoulos, Apiletti, Assimakopoulos, Babai,
  Barrow, {Ben Taieb}, Bergmeir, Bessa, Bijak, Boylan, Browell, Carnevale,
  Castle, Cirillo, Clements, Cordeiro, {Cyrino Oliveira}, {De Baets},
  Dokumentov, Ellison, Fiszeder, Franses, Frazier, Gilliland, Gönül, Goodwin,
  Grossi, Grushka-Cockayne, Guidolin, Guidolin, Gunter, Guo, Guseo, Harvey,
  Hendry, Hollyman, Januschowski, Jeon, Jose, Kang, Koehler, Kolassa,
  Kourentzes, Leva, Li, Litsiou, Makridakis, Martin, Martinez, Meeran, Modis,
  Nikolopoulos, Önkal, Paccagnini, Panagiotelis, Panapakidis, Pavía, Pedio,
  Pedregal, Pinson, Ramos, Rapach, Reade, Rostami-Tabar, Rubaszek, Sermpinis,
  Shang, Spiliotis, Syntetos, Talagala, Talagala, Tashman, Thomakos,
  Thorarinsdottir, Todini, {Trapero Arenas}, Wang, Winkler, Yusupova, and
  Ziel]{PETROPOULOS2022}
Fotios Petropoulos, Daniele Apiletti, Vassilios Assimakopoulos, Mohamed~Zied
  Babai, Devon~K. Barrow, Souhaib {Ben Taieb}, Christoph Bergmeir, Ricardo~J.
  Bessa, Jakub Bijak, John~E. Boylan, Jethro Browell, Claudio Carnevale,
  Jennifer~L. Castle, Pasquale Cirillo, Michael~P. Clements, Clara Cordeiro,
  Fernando~Luiz {Cyrino Oliveira}, Shari {De Baets}, Alexander Dokumentov,
  Joanne Ellison, Piotr Fiszeder, Philip~Hans Franses, David~T. Frazier,
  Michael Gilliland, M.~Sinan Gönül, Paul Goodwin, Luigi Grossi, Yael
  Grushka-Cockayne, Mariangela Guidolin, Massimo Guidolin, Ulrich Gunter,
  Xiaojia Guo, Renato Guseo, Nigel Harvey, David~F. Hendry, Ross Hollyman, Tim
  Januschowski, Jooyoung Jeon, Victor Richmond~R. Jose, Yanfei Kang, Anne~B.
  Koehler, Stephan Kolassa, Nikolaos Kourentzes, Sonia Leva, Feng Li,
  Konstantia Litsiou, Spyros Makridakis, Gael~M. Martin, Andrew~B. Martinez,
  Sheik Meeran, Theodore Modis, Konstantinos Nikolopoulos, Dilek Önkal,
  Alessia Paccagnini, Anastasios Panagiotelis, Ioannis Panapakidis, Jose~M.
  Pavía, Manuela Pedio, Diego~J. Pedregal, Pierre Pinson, Patrícia Ramos,
  David~E. Rapach, J.~James Reade, Bahman Rostami-Tabar, Michał Rubaszek,
  Georgios Sermpinis, Han~Lin Shang, Evangelos Spiliotis, Aris~A. Syntetos,
  Priyanga~Dilini Talagala, Thiyanga~S. Talagala, Len Tashman, Dimitrios
  Thomakos, Thordis Thorarinsdottir, Ezio Todini, Juan~Ramón {Trapero Arenas},
  Xiaoqian Wang, Robert~L. Winkler, Alisa Yusupova, and Florian Ziel.
\newblock Forecasting: theory and practice.
\newblock \emph{International Journal of Forecasting}, 38\penalty0
  (3):\penalty0 705--871, 2022.
\newblock ISSN 0169-2070.
\newblock \doi{https://doi.org/10.1016/j.ijforecast.2021.11.001}.
\newblock URL
  \url{https://www.sciencedirect.com/science/article/pii/S0169207021001758}.

\bibitem[Steinert and Ziel(2019)]{Steinert2019}
Rick Steinert and Florian Ziel.
\newblock Short- to mid-term day-ahead electricity price forecasting using
  futures.
\newblock \emph{Energy Journal}, 40:\penalty0 105--127, 2019.
\newblock ISSN 01956574.
\newblock \doi{10.5547/01956574.40.1.rste}.

\bibitem[Nowotarski and Weron(2016)]{Nowotarski2016}
Jakub Nowotarski and Rafał Weron.
\newblock On the importance of the long-term seasonal component in day-ahead
  electricity price forecasting.
\newblock \emph{Energy Economics}, 57:\penalty0 228--235, 6 2016.
\newblock ISSN 01409883.
\newblock \doi{10.1016/J.ENECO.2016.05.009}.

\bibitem[Ziel and Weron(2018)]{Ziel2018}
Florian Ziel and Rafał Weron.
\newblock {Day-ahead electricity price forecasting with high-dimensional
  structures: Univariate vs. multivariate modeling frameworks}.
\newblock \emph{Energy Economics}, 70:\penalty0 396--420, 2018.
\newblock ISSN 0140-9883.
\newblock \doi{10.1016/j.eneco.2017.12.016}.

\bibitem[Christensen et~al.(2012)Christensen, Hurn, and
  Lindsay]{Christensen2012}
T.~M. Christensen, A.~S. Hurn, and K.~A. Lindsay.
\newblock Forecasting spikes in electricity prices.
\newblock \emph{International Journal of Forecasting}, 28:\penalty0 400--411, 4
  2012.
\newblock ISSN 01692070.
\newblock \doi{10.1016/j.ijforecast.2011.02.019}.

\bibitem[Eichler et~al.(2014)Eichler, Grothe, Manner, and Dennis]{Manner2014}
Michael Eichler, Oliver Grothe, Hans Manner, and Tuerk Dennis.
\newblock Models for short-term forecasting of spike occurrences in
  {A}ustralian electricity markets: {A} comparative study.
\newblock \emph{The Journal of Energy Markets}, 7\penalty0 (1):\penalty0
  55--81, 3 2014.
\newblock \doi{10.21314/JEM.2014.104}.

\bibitem[Manner et~al.(2016)Manner, Türk, and Eichler]{MANNER2016255}
Hans Manner, Dennis Türk, and Michael Eichler.
\newblock Modeling and forecasting multivariate electricity price spikes.
\newblock \emph{Energy Economics}, 60:\penalty0 255--265, 2016.
\newblock ISSN 0140-9883.
\newblock \doi{10.1016/j.eneco.2016.10.006}.

\bibitem[Garcia et~al.(2005)Garcia, Contreras, van Akkeren, and
  Garcia]{Garcia2005}
R.C. Garcia, J.~Contreras, M.~van Akkeren, and J.B.C. Garcia.
\newblock A {GARCH} forecasting model to predict day-ahead electricity prices.
\newblock \emph{IEEE Transactions on Power Systems}, 20\penalty0 (2):\penalty0
  867--874, 2005.
\newblock \doi{10.1109/TPWRS.2005.846044}.

\bibitem[Hickey et~al.(2012)Hickey, Loomis, and Mohammadi]{HICKEY2012307}
Emily Hickey, David~G. Loomis, and Hassan Mohammadi.
\newblock {Forecasting hourly electricity prices using ARMAX–GARCH models: An
  application to MISO hubs}.
\newblock \emph{Energy Economics}, 34\penalty0 (1):\penalty0 307--315, 2012.
\newblock ISSN 0140-9883.
\newblock \doi{10.1016/j.eneco.2011.11.011}.

\bibitem[Bordignon et~al.(2013)Bordignon, Bunn, Lisi, and Nan]{Bordignon2013}
Silvano Bordignon, Derek~W. Bunn, Francesco Lisi, and Fany Nan.
\newblock Combining day-ahead forecasts for british electricity prices.
\newblock \emph{Energy Economics}, 35:\penalty0 88--103, 1 2013.
\newblock ISSN 01409883.
\newblock \doi{10.1016/j.eneco.2011.12.001}.

\bibitem[Kosater and Mosler(2006)]{KOSATER2006943}
Peter Kosater and Karl Mosler.
\newblock Can markov regime-switching models improve power-price forecasts?
  {E}vidence from {G}erman daily power prices.
\newblock \emph{Applied Energy}, 83\penalty0 (9):\penalty0 943--958, 2006.
\newblock ISSN 0306-2619.
\newblock \doi{10.1016/j.apenergy.2005.10.007}.

\bibitem[Bierbrauer et~al.(2004)Bierbrauer, Trück, and Weron]{Bierbrauer2004}
Michael Bierbrauer, Stefan Trück, and Rafał Weron.
\newblock Modeling electricity prices with regime switching models.
\newblock \emph{Lecture Notes in Computer Science}, 3039:\penalty0 859--867,
  2004.
\newblock ISSN 16113349.
\newblock \doi{10.1007/978-3-540-25944-2\_111}.

\bibitem[Mari and Mari(2022)]{Mari2022}
Carlo Mari and Emiliano Mari.
\newblock Deep learning based regime-switching models of energy commodity
  prices.
\newblock \emph{Energy Systems}, 2022.
\newblock \doi{10.1007/s12667-022-00515-6}.

\bibitem[Uniejewski et~al.(2016)Uniejewski, Nowotarski, and
  Weron]{Uniejewski2016}
Bartosz Uniejewski, Jakub Nowotarski, and Rafał Weron.
\newblock {Automated Variable Selection and Shrinkage for Day-Ahead Electricity
  Price Forecasting}.
\newblock \emph{Energies}, 9\penalty0 (8), 2016.
\newblock ISSN 1996-1073.
\newblock \doi{10.3390/en9080621}.

\bibitem[Lago et~al.(2021)Lago, Marcjasz, {De Schutter}, and
  Weron]{lago2021forecasting}
Jesus Lago, Grzegorz Marcjasz, Bart {De Schutter}, and Rafał Weron.
\newblock {Forecasting day-ahead electricity prices: A review of
  state-of-the-art algorithms, best practices and an open-access benchmark}.
\newblock \emph{Applied Energy}, 293:\penalty0 116983, 2021.
\newblock ISSN 0306-2619.
\newblock \doi{https://doi.org/10.1016/j.apenergy.2021.116983}.
\newblock URL
  \url{https://www.sciencedirect.com/science/article/pii/S0306261921004529}.

\bibitem[Panapakidis and Dagoumas(2016)]{Panapakidis2016}
Ioannis~P. Panapakidis and Athanasios~S. Dagoumas.
\newblock Day-ahead electricity price forecasting via the application of
  artificial neural network based models.
\newblock \emph{Applied Energy}, 172:\penalty0 132--151, 6 2016.
\newblock ISSN 03062619.
\newblock \doi{10.1016/j.apenergy.2016.03.089}.

\bibitem[Lago et~al.(2018)Lago, Ridder, Vrancx, and Schutter]{Lago2018}
Jesus Lago, Fjo~De Ridder, Peter Vrancx, and Bart~De Schutter.
\newblock Forecasting day-ahead electricity prices in {E}urope: {T}he
  importance of considering market integration.
\newblock \emph{Applied Energy}, 211:\penalty0 890--903, 2 2018.
\newblock ISSN 03062619.
\newblock \doi{10.1016/j.apenergy.2017.11.098}.

\bibitem[Amjady(2006)]{Amjady2006fuzzy}
Nima Amjady.
\newblock Day-ahead price forecasting of electricity markets by a new fuzzy
  neural network.
\newblock \emph{IEEE Transactions on Power Systems}, 21\penalty0 (2):\penalty0
  887--896, 2006.
\newblock \doi{10.1109/TPWRS.2006.873409}.

\bibitem[Marcjasz et~al.(2019)Marcjasz, Uniejewski, and
  Weron]{MARCJASZ20191520}
Grzegorz Marcjasz, Bartosz Uniejewski, and Rafał Weron.
\newblock On the importance of the long-term seasonal component in day-ahead
  electricity price forecasting with {NARX} neural networks.
\newblock \emph{International Journal of Forecasting}, 35\penalty0
  (4):\penalty0 1520--1532, 2019.
\newblock ISSN 0169-2070.
\newblock \doi{10.1016/j.ijforecast.2017.11.009}.

\bibitem[Lehna et~al.(2022)Lehna, Scheller, and Herwartz]{LEHNA2022105742}
Malte Lehna, Fabian Scheller, and Helmut Herwartz.
\newblock Forecasting day-ahead electricity prices: {A} comparison of time
  series and neural network models taking external regressors into account.
\newblock \emph{Energy Economics}, 106:\penalty0 105742, 2022.
\newblock ISSN 0140-9883.
\newblock \doi{10.1016/j.eneco.2021.105742}.

\bibitem[Qussous et~al.(2022)Qussous, Harder, and Weidlich]{Qussous2022}
Ramiz Qussous, Nick Harder, and Anke Weidlich.
\newblock {Understanding Power Market Dynamics by Reflecting Market
  Interrelations and Flexibility-Oriented Bidding Strategies}.
\newblock \emph{Energies}, 15\penalty0 (2), 2022.
\newblock ISSN 1996-1073.
\newblock \doi{10.3390/en15020494}.

\bibitem[Muesgens(2006{\natexlab{a}})]{Muesgens2006}
Felix Muesgens.
\newblock {Quantifying Market Power in the German Wholesale Electricity Market
  Using a Dynamic Multi-Regional Dispatch Model}.
\newblock \emph{The Journal of Industrial Economics}, 54\penalty0 (4):\penalty0
  471--498, 2006{\natexlab{a}}.
\newblock \doi{10.1111/j.1467-6451.2006.00297.x}.

\bibitem[Borenstein et~al.(2002)Borenstein, Bushnell, and
  Wolak]{Borenstein2002}
Severin Borenstein, James~B. Bushnell, and Frank~A. Wolak.
\newblock {Measuring Market Inefficiencies in California's Restructured
  Wholesale Electricity Market}.
\newblock \emph{American Economic Review}, 92\penalty0 (5):\penalty0
  1376--1405, December 2002.
\newblock \doi{10.1257/000282802762024557}.

\bibitem[Keles et~al.(2013)Keles, Genoese, Möst, Ortlieb, and
  Fichtner]{KELES2013213}
Dogan Keles, Massimo Genoese, Dominik Möst, Sebastian Ortlieb, and Wolf
  Fichtner.
\newblock A combined modeling approach for wind power feed-in and electricity
  spot prices.
\newblock \emph{Energy Policy}, 59:\penalty0 213--225, 2013.
\newblock ISSN 0301-4215.
\newblock \doi{https://doi.org/10.1016/j.enpol.2013.03.028}.
\newblock URL
  \url{https://www.sciencedirect.com/science/article/pii/S0301421513001821}.

\bibitem[Hirth(2013)]{HIRTH2013218}
Lion Hirth.
\newblock The market value of variable renewables: {T}he effect of solar wind
  power variability on their relative price.
\newblock \emph{Energy Economics}, 38:\penalty0 218--236, 2013.
\newblock ISSN 0140-9883.
\newblock \doi{10.1016/j.eneco.2013.02.004}.

\bibitem[Sensfuß et~al.(2008)Sensfuß, Ragwitz, and Genoese]{SENSFU20083086}
Frank Sensfuß, Mario Ragwitz, and Massimo Genoese.
\newblock The merit-order effect: {A} detailed analysis of the price effect of
  renewable electricity generation on spot market prices in {G}ermany.
\newblock \emph{Energy Policy}, 36\penalty0 (8):\penalty0 3086--3094, 2008.
\newblock ISSN 0301-4215.
\newblock \doi{10.1016/j.enpol.2008.03.035}.

\bibitem[Pape et~al.(2016)Pape, Hagemann, and Weber]{Pape2016}
Christian Pape, Simon Hagemann, and Christoph Weber.
\newblock Are fundamentals enough? {E}xplaining price variations in the
  {G}erman day-ahead and intraday power market.
\newblock \emph{Energy Economics}, 54:\penalty0 376--387, 2 2016.
\newblock ISSN 0140-9883.
\newblock \doi{10.1016/J.ENECO.2015.12.013}.

\bibitem[Muesgens(2020)]{MUSGENS2020104107}
Felix Muesgens.
\newblock Equilibrium prices and investment in electricity systems with
  co2-emission trading and high shares of renewable energies.
\newblock \emph{Energy Economics}, 86:\penalty0 104107, 2020.
\newblock ISSN 0140-9883.
\newblock \doi{10.1016/j.eneco.2018.07.028}.

\bibitem[Green and Vasilakos(2010)]{GREEN20103211}
Richard Green and Nicholas Vasilakos.
\newblock Market behaviour with large amounts of intermittent generation.
\newblock \emph{Energy Policy}, 38\penalty0 (7):\penalty0 3211--3220, 2010.
\newblock ISSN 0301-4215.
\newblock \doi{10.1016/j.enpol.2009.07.038}.
\newblock Large-scale wind power in electricity markets with Regular Papers.

\bibitem[Lamont(2008)]{LAMONT20081208}
Alan~D. Lamont.
\newblock Assessing the long-term system value of intermittent electric
  generation technologies.
\newblock \emph{Energy Economics}, 30\penalty0 (3):\penalty0 1208--1231, 2008.
\newblock ISSN 0140-9883.
\newblock \doi{10.1016/j.eneco.2007.02.007}.

\bibitem[Misconel et~al.(2021)Misconel, Zöphel, and Möst]{MISCONEL2021117326}
Steffi Misconel, Christoph Zöphel, and Dominik Möst.
\newblock Assessing the value of demand response in a decarbonized energy
  system – {A} large-scale model application.
\newblock \emph{Applied Energy}, 299:\penalty0 117326, 2021.
\newblock ISSN 0306-2619.
\newblock \doi{10.1016/j.apenergy.2021.117326}.

\bibitem[Kirchem et~al.(2020)Kirchem, Lynch, Bertsch, and
  Casey]{KIRCHEM2020114321}
Dana Kirchem, Muireann~Á. Lynch, Valentin Bertsch, and Eoin Casey.
\newblock Modelling demand response with process models and energy systems
  models: Potential applications for wastewater treatment within the
  energy-water nexus.
\newblock \emph{Applied Energy}, 260:\penalty0 114321, 2020.
\newblock ISSN 0306-2619.
\newblock \doi{10.1016/j.apenergy.2019.114321}.

\bibitem[{van der Weijde} and Hobbs(2012)]{VANDERWEIJDE20122089}
Adriaan~Hendrik {van der Weijde} and Benjamin~F. Hobbs.
\newblock The economics of planning electricity transmission to accommodate
  renewables: Using two-stage optimisation to evaluate flexibility and the cost
  of disregarding uncertainty.
\newblock \emph{Energy Economics}, 34\penalty0 (6):\penalty0 2089--2101, 2012.
\newblock ISSN 0140-9883.
\newblock \doi{10.1016/j.eneco.2012.02.015}.

\bibitem[Scheller and Bruckner(2019)]{SCHELLER2019444}
Fabian Scheller and Thomas Bruckner.
\newblock Energy system optimization at the municipal level: An analysis of
  modeling approaches and challenges.
\newblock \emph{Renewable and Sustainable Energy Reviews}, 105:\penalty0
  444--461, 2019.
\newblock ISSN 1364-0321.
\newblock \doi{10.1016/j.rser.2019.02.005}.

\bibitem[Lynch et~al.(2019)Lynch, Devine, and Bertsch]{LYNCH20191197}
Muireann Lynch, Mel~T. Devine, and Valentin Bertsch.
\newblock The role of power-to-gas in the future energy system: Market and
  portfolio effects.
\newblock \emph{Energy}, 185:\penalty0 1197--1209, 2019.
\newblock ISSN 0360-5442.
\newblock \doi{10.1016/j.energy.2019.07.089}.

\bibitem[Sgarciu et~al.(2023)Sgarciu, Scholz, and Muesgens]{SGARCIU2023113375}
Smaranda Sgarciu, Daniel Scholz, and Felix Muesgens.
\newblock How co2 prices accelerate decarbonisation – {T}he case of
  coal-fired generation in {G}ermany.
\newblock \emph{Energy Policy}, 173:\penalty0 113375, 2023.
\newblock ISSN 0301-4215.
\newblock \doi{10.1016/j.enpol.2022.113375}.

\bibitem[Plaga and Bertsch(2023)]{PLAGA2023120384}
Leonie~Sara Plaga and Valentin Bertsch.
\newblock Methods for assessing climate uncertainty in energy system models —
  a systematic literature review.
\newblock \emph{Applied Energy}, 331:\penalty0 120384, 2023.
\newblock ISSN 0306-2619.
\newblock \doi{10.1016/j.apenergy.2022.120384}.

\bibitem[Ventosa et~al.(2005)Ventosa, Álvaro Ba\'{i}llo, Ramos, and
  Rivier]{VENTOSA2005897}
Mariano Ventosa, Álvaro Ba\'{i}llo, Andr\'{e}s Ramos, and Michel Rivier.
\newblock Electricity market modeling trends.
\newblock \emph{Energy Policy}, 33\penalty0 (7):\penalty0 897--913, 2005.
\newblock ISSN 0301-4215.
\newblock \doi{10.1016/j.enpol.2003.10.013}.

\bibitem[Aggarwal and Tripathi(2017)]{Aggarwal2017}
Abhinav Aggarwal and M~M Tripathi.
\newblock A novel hybrid approach using wavelet transform, time series time
  delay neural network, and error predicting algorithm for day-ahead
  electricity price forecasting.
\newblock In \emph{2017 6th International Conference on Computer Applications
  In Electrical Engineering-Recent Advances (CERA)}, pages 199--204, 2017.
\newblock \doi{10.1109/CERA.2017.8343326}.

\bibitem[Chang et~al.(2019)Chang, Zhang, and Chen]{CHANG2019115804}
Zihan Chang, Yang Zhang, and Wenbo Chen.
\newblock {Electricity price prediction based on hybrid model of adam optimized
  LSTM neural network and wavelet transform}.
\newblock \emph{Energy}, 187:\penalty0 115804, 2019.
\newblock ISSN 0360-5442.
\newblock \doi{10.1016/j.energy.2019.07.134}.

\bibitem[Cheng et~al.(2019)Cheng, Ding, Zhou, and Ding]{CHENG2019653}
Hangyang Cheng, Xiangwu Ding, Wuneng Zhou, and Renqiang Ding.
\newblock {A hybrid electricity price forecasting model with Bayesian
  optimization for German energy exchange}.
\newblock \emph{International Journal of Electrical Power \& Energy Systems},
  110:\penalty0 653--666, 2019.
\newblock ISSN 0142-0615.
\newblock \doi{10.1016/j.ijepes.2019.03.056}.

\bibitem[Nazar et~al.(2018)Nazar, Fard, Heidari, Shafie-khah, and
  Catalão]{NAZAR2018214}
Mehrdad~Setayesh Nazar, Ashkan~Eslami Fard, Alireza Heidari, Miadreza
  Shafie-khah, and João~P.S. Catalão.
\newblock Hybrid model using three-stage algorithm for simultaneous load and
  price forecasting.
\newblock \emph{Electric Power Systems Research}, 165:\penalty0 214--228, 2018.
\newblock ISSN 0378-7796.
\newblock \doi{10.1016/j.epsr.2018.09.004}.

\bibitem[Yang et~al.(2017)Yang, Ce, and Lian]{YANG2017291}
Zhang Yang, Li~Ce, and Li~Lian.
\newblock {Electricity price forecasting by a hybrid model, combining wavelet
  transform, ARMA and kernel-based extreme learning machine methods}.
\newblock \emph{Applied Energy}, 190:\penalty0 291--305, 2017.
\newblock ISSN 0306-2619.
\newblock \doi{10.1016/j.apenergy.2016.12.130}.

\bibitem[Olamaee et~al.(2016)Olamaee, Mohammadi, Noruzi, and
  Hosseini]{Olamaee2016}
Javad Olamaee, Mohsen Mohammadi, Alireza Noruzi, and Seyed Mohammad~Hassan
  Hosseini.
\newblock Day-ahead price forecasting based on hybrid prediction model.
\newblock \emph{Complexity}, 21\penalty0 (S2):\penalty0 156--164, 2016.
\newblock \doi{10.1002/cplx.21792}.

\bibitem[Zhang et~al.(2020)Zhang, Tan, and Wei]{ZHANG2020114087}
Jinliang Zhang, Zhongfu Tan, and Yiming Wei.
\newblock An adaptive hybrid model for short term electricity price
  forecasting.
\newblock \emph{Applied Energy}, 258:\penalty0 114087, 2020.
\newblock ISSN 0306-2619.
\newblock \doi{10.1016/j.apenergy.2019.114087}.

\bibitem[{de Marcos} et~al.(2019){de Marcos}, Bello, and
  Reneses]{DEMARCOS2019240}
Rodrigo~A. {de Marcos}, Antonio Bello, and Javier Reneses.
\newblock Electricity price forecasting in the short term hybridising
  fundamental and econometric modelling.
\newblock \emph{Electric Power Systems Research}, 167:\penalty0 240--251, 2019.
\newblock ISSN 0378-7796.
\newblock \doi{10.1016/j.epsr.2018.10.034}.

\bibitem[Gonzalez et~al.(2012)Gonzalez, Contreras, and Bunn]{Gonzalez2012}
Virginia Gonzalez, Javier Contreras, and Derek~W. Bunn.
\newblock {Forecasting Power Prices Using a Hybrid Fundamental-Econometric
  Model}.
\newblock \emph{IEEE Transactions on Power Systems}, 27\penalty0 (1):\penalty0
  363--372, 2012.
\newblock \doi{10.1109/TPWRS.2011.2167689}.

\bibitem[Möbius et~al.(2023)Möbius, Watermeyer, Grothe, and
  Muesgens]{Preprocessing}
Thomas Möbius, Mira Watermeyer, Oliver Grothe, and Felix Muesgens.
\newblock {Enhancing Energy System Models Using Better Load Forecasts}.
\newblock \emph{arXiv}, 2023.
\newblock \doi{10.48550/ARXIV.2302.11017}.

\bibitem[Maciejowska and Nowotarski(2016)]{MACIEJOWSKA20161051}
Katarzyna Maciejowska and Jakub Nowotarski.
\newblock A hybrid model for {GEFCom2014} probabilistic electricity price
  forecasting.
\newblock \emph{International Journal of Forecasting}, 32\penalty0
  (3):\penalty0 1051--1056, 2016.
\newblock ISSN 0169-2070.
\newblock \doi{10.1016/j.ijforecast.2015.11.008}.

\bibitem[Weron and Misiorek(2008)]{WERON2008744}
Rafał Weron and Adam Misiorek.
\newblock Forecasting spot electricity prices: {A} comparison of parametric and
  semiparametric time series models.
\newblock \emph{International Journal of Forecasting}, 24\penalty0
  (4):\penalty0 744--763, 2008.
\newblock ISSN 0169-2070.
\newblock \doi{10.1016/j.ijforecast.2008.08.004}.
\newblock Energy Forecasting.

\bibitem[Chen et~al.(2012)Chen, Dong, Meng, Xu, Wong, and Ngan]{Chen2012}
Xia Chen, Zhao~Yang Dong, Ke~Meng, Yan Xu, Kit~Po Wong, and H.~W. Ngan.
\newblock {Electricity Price Forecasting With Extreme Learning Machine and
  Bootstrapping}.
\newblock \emph{IEEE Transactions on Power Systems}, 27\penalty0 (4):\penalty0
  2055--2062, 2012.
\newblock \doi{10.1109/TPWRS.2012.2190627}.

\bibitem[Wan et~al.(2014)Wan, Xu, Wang, Dong, and Wong]{Wan2014}
Can Wan, Zhao Xu, Yelei Wang, Zhao~Yang Dong, and Kit~Po Wong.
\newblock {A Hybrid Approach for Probabilistic Forecasting of Electricity
  Price}.
\newblock \emph{IEEE Transactions on Smart Grid}, 5\penalty0 (1):\penalty0
  463--470, 2014.
\newblock \doi{10.1109/TSG.2013.2274465}.

\bibitem[Rafiei et~al.(2017)Rafiei, Niknam, and Khooban]{Rafiei2017}
Mehdi Rafiei, Taher Niknam, and Mohammad Khooban.
\newblock Probabilistic electricity price forecasting by improved clonal
  selection algorithm and wavelet preprocessing.
\newblock \emph{Neural Computing and Applications}, 28, 12 2017.
\newblock \doi{10.1007/s00521-016-2279-7}.

\bibitem[Khosravi et~al.(2013)Khosravi, Nahavandi, and
  Creighton]{KHOSRAVI2013120}
Abbas Khosravi, Saeid Nahavandi, and Doug Creighton.
\newblock Quantifying uncertainties of neural network-based electricity price
  forecasts.
\newblock \emph{Applied Energy}, 112:\penalty0 120--129, 2013.
\newblock ISSN 0306-2619.
\newblock \doi{10.1016/j.apenergy.2013.05.075}.

\bibitem[Zhao et~al.(2008)Zhao, Dong, Xu, and Wong]{Zhao2008}
Jun~Hua Zhao, Zhao~Yang Dong, Zhao Xu, and Kit~Po Wong.
\newblock {A Statistical Approach for Interval Forecasting of the Electricity
  Price}.
\newblock \emph{IEEE Transactions on Power Systems}, 23\penalty0 (2):\penalty0
  267--276, 2008.
\newblock \doi{10.1109/TPWRS.2008.919309}.

\bibitem[Zhou et~al.(2006)Zhou, Yan, Ni, Li, and Nie]{Zou2006}
Ming Zhou, Z.~Yan, Y.X. Ni, Gengyin Li, and Y.~Nie.
\newblock Electricity price forecasting with confidence-interval estimation
  through an extended arima approach.
\newblock \emph{Generation, Transmission and Distribution, IEE Proceedings-},
  pages 187 -- 195, 04 2006.
\newblock \doi{10.1049/ip-gtd:20045131}.

\bibitem[Panagiotelis and Smith(2008)]{Panagiotelis2008}
Anastasios Panagiotelis and Michael Smith.
\newblock Bayesian density forecasting of intraday electricity prices using
  multivariate skew t distributions.
\newblock \emph{International Journal of Forecasting}, 24:\penalty0 710--727,
  10 2008.
\newblock \doi{10.1016/j.ijforecast.2008.08.009}.

\bibitem[Manner et~al.(2019)Manner, {Alavi Fard}, Pourkhanali, and
  Tafakori]{MANNER2019143}
Hans Manner, Farzad {Alavi Fard}, Armin Pourkhanali, and Laleh Tafakori.
\newblock Forecasting the joint distribution of {A}ustralian electricity prices
  using dynamic vine copulae.
\newblock \emph{Energy Economics}, 78:\penalty0 143--164, 2019.
\newblock ISSN 0140-9883.
\newblock \doi{https://doi.org/10.1016/j.eneco.2018.10.034}.

\bibitem[Grothe et~al.(2023)Grothe, Kächele, and Krüger]{Kaechele2022}
Oliver Grothe, Fabian Kächele, and Fabian Krüger.
\newblock From point forecasts to multivariate probabilistic forecasts: {T}he
  {S}chaake shuffle for day-ahead electricity price forecasting.
\newblock \emph{Energy Economics}, page 106602, 2023.
\newblock ISSN 0140-9883.
\newblock \doi{10.1016/j.eneco.2023.106602}.

\bibitem[Nowotarski and Weron(2015)]{Nowotarski2015}
Jakub Nowotarski and Rafał Weron.
\newblock Computing electricity spot price prediction intervals using quantile
  regression and forecast averaging.
\newblock \emph{Computational Statistics}, 30:\penalty0 791--803, 9 2015.
\newblock ISSN 16139658.
\newblock \doi{10.1007/S00180-014-0523-0/FIGURES/3}.

\bibitem[Misiorek et~al.(2006)Misiorek, Trueck, and
  Weron]{MisiorekTrueckWeron2006}
Adam Misiorek, Stefan Trueck, and Rafal Weron.
\newblock {Point and Interval Forecasting of Spot Electricity Prices: Linear
  vs. Non-Linear Time Series Models}.
\newblock \emph{Studies in Nonlinear Dynamics \& Econometrics}, 10\penalty0
  (3), 2006.
\newblock \doi{10.2202/1558-3708.1362}.

\bibitem[Dudek(2016)]{DUDEK20161057}
Grzegorz Dudek.
\newblock Multilayer perceptron for {GEFCom2014} probabilistic electricity
  price forecasting.
\newblock \emph{International Journal of Forecasting}, 32\penalty0
  (3):\penalty0 1057--1060, 2016.
\newblock ISSN 0169-2070.
\newblock \doi{10.1016/j.ijforecast.2015.11.009}.

\bibitem[Maciejowska et~al.(2016)Maciejowska, Nowotarski, and
  Weron]{MACIEJOWSKA2016957}
Katarzyna Maciejowska, Jakub Nowotarski, and Rafał Weron.
\newblock Probabilistic forecasting of electricity spot prices using factor
  quantile regression averaging.
\newblock \emph{International Journal of Forecasting}, 32\penalty0
  (3):\penalty0 957--965, 2016.
\newblock ISSN 0169-2070.
\newblock \doi{10.1016/j.ijforecast.2014.12.004}.

\bibitem[Weron(2006)]{WeronLoadbook2006}
Rafał Weron.
\newblock \emph{Modeling and forecasting electricity loads and prices: a
  statistical approach}.
\newblock Wiley finance series. Wiley \& Sons, Chichester [u.a.], 2006.
\newblock ISBN 047005753X; 9780470057537.

\bibitem[Nowotarski and Weron(2014)]{Nowotarski2014}
Jakub Nowotarski and Rafał Weron.
\newblock Merging quantile regression with forecast averaging to obtain more
  accurate interval forecasts of {N}ord {P}ool spot prices.
\newblock In \emph{11th International Conference on the European Energy Market
  (EEM14)}, pages 1--5, 2014.
\newblock \doi{10.1109/EEM.2014.6861285}.

\bibitem[Marcjasz et~al.(2020)Marcjasz, Uniejewski, and Weron]{MARCJASZ2020466}
Grzegorz Marcjasz, Bartosz Uniejewski, and Rafał Weron.
\newblock Probabilistic electricity price forecasting with {NARX} networks:
  {C}ombine point or probabilistic forecasts?
\newblock \emph{International Journal of Forecasting}, 36\penalty0
  (2):\penalty0 466--479, 2020.
\newblock ISSN 0169-2070.
\newblock \doi{10.1016/j.ijforecast.2019.07.002}.

\bibitem[Uniejewski et~al.(2019)Uniejewski, Marcjasz, and
  Weron]{UNIEJEWSKI2019171}
Bartosz Uniejewski, Grzegorz Marcjasz, and Rafał Weron.
\newblock On the importance of the long-term seasonal component in
  day-ahead electricity price forecasting: {P}art {II} — {P}robabilistic
  forecasting.
\newblock \emph{Energy Economics}, 79:\penalty0 171--182, 2019.
\newblock ISSN 0140-9883.
\newblock \doi{10.1016/j.eneco.2018.02.007}.
\newblock Energy Markets Dynamics in a Changing Environment.

\bibitem[Uniejewski and Weron(2021)]{Uniejewski2021}
Bartosz Uniejewski and Rafał Weron.
\newblock Regularized quantile regression averaging for probabilistic
  electricity price forecasting.
\newblock \emph{Energy Economics}, 95:\penalty0 105121, 2021.
\newblock \doi{10.1016/j.eneco.2021.105121}.

\bibitem[{Sandbag}(2020)]{Sandbag2020}
{Sandbag}.
\newblock Co2 emission allowance, 2020.
\newblock URL \url{https://sandbag.be/index.php/carbon-price-viewer/}.
\newblock Accessed on 03-01-2022.

\bibitem[{Regelleistung.net}(2018)]{Regelleistung}
{Regelleistung.net}.
\newblock List of tenders capacity, 2018.
\newblock URL \url{https://www.regelleistung.net/ext/}.
\newblock Accessed on 12-12-2021.

\bibitem[Schröder et~al.(2013)Schröder, Kunz, Meiss, Mendelevitch, and von
  Hirschhausen]{Schroeder2013}
Andreas Schröder, Friedrich Kunz, Jan Meiss, Roman Mendelevitch, and Christian
  von Hirschhausen.
\newblock {Current and Prospective Costs of Electricity Generation until 2050}.
\newblock \emph{DIW Data Documentation}, 68, 2013.
\newblock ISSN 1861-1532.

\bibitem[{Open Power System Data}(2020{\natexlab{a}})]{OPSDb}
{Open Power System Data}.
\newblock {Data Package Weather Data. Version 2020-09-16.}, 2020{\natexlab{a}}.
\newblock URL \url{https://doi.org/10.25832/weather\_data/2020-09-16}.
\newblock Accessed on 15-05-2020.

\bibitem[{ENTSO-E Transparency Platform}(2021{\natexlab{a}})]{EntsoeTPe}
{ENTSO-E Transparency Platform}.
\newblock {Total Load - Day Ahead / Actual}, 2021{\natexlab{a}}.
\newblock URL \url{https://transparency.entsoe.eu/}.
\newblock Accessed on 20-12-2021.

\bibitem[{Destatis Statistisches Bundesamt}(2021)]{Destatis2020}
{Destatis Statistisches Bundesamt}.
\newblock {Erzeugerpreise gewerblicher Produkte (Inlandsabsatz). Preise für
  leichtes Heizöl, Motorenbenzin und Diesel}, 2021.
\newblock URL
  \url{https://www.destatis.de/DE/Themen/Wirtschaft/Preise/Erzeugerpreisindex-gewerbliche-Produkte/_inhalt.html}.
\newblock {Accessed on 25-01-2021}.

\bibitem[{ENTSO-E}(2018)]{EntsoS}
{ENTSO-E}.
\newblock {TYNDP 2018 Scenario Report}, 2018.
\newblock URL \url{https://tyndp.entsoe.eu/tyndp2018/scenario-report}.
\newblock Accessed on 23-02-2023.

\bibitem[{EEX}(2021)]{EEX2021}
{EEX}.
\newblock {European Energy Exchange: Historic gas price data}, 2021.
\newblock Accessed on 03-01-2022.

\bibitem[{BNetzA}(2021)]{BNetzA2021}
{BNetzA}.
\newblock {Kraftwerksliste der Bundesnetzagentur}, 2021.
\newblock URL \url{https://www.bundesnetzagentur.de/}.
\newblock Accessed on 03-01-2022.

\bibitem[{UBA}(2020)]{UBA2020}
{UBA}.
\newblock {Umweltbundesamt: Datenbank “Kraftwerke in Deutschland"}, 2020.
\newblock URL
  \url{https://www.umweltbundesamt.de/dokument/datenbank-kraftwerke-in-deutschland}.
\newblock Accessed on 03-01-2022.

\bibitem[{EBC}(2021)]{EBC2021}
{EBC}.
\newblock {Europe Beyond Coal: European Coal Plant Database, 25 Jan 2021},
  2021.
\newblock URL \url{https://beyond-coal.eu/database/}.
\newblock Accessed on 25-01-2021.

\bibitem[{ENTSO-E Transparency Platform}(2021{\natexlab{b}})]{EntsoeTPa}
{ENTSO-E Transparency Platform}.
\newblock {Installed Capacities per Production Type}, 2021{\natexlab{b}}.
\newblock URL \url{https://transparency.entsoe.eu/}.
\newblock Accessed on 20-12-2021.

\bibitem[{Open Power System Data}(2020{\natexlab{b}})]{OPSDa}
{Open Power System Data}.
\newblock {Data Package National Generation Capacity. Version 2019-12-02.},
  2020{\natexlab{b}}.
\newblock URL
  \url{https://doi.org/10.25832/national\_generation\_capacity/2019-12-02}.
\newblock Accessed on 15-05-2020.

\bibitem[{European Commission}(2021)]{EC2021}
{European Commission}.
\newblock {Eurostat Statistics Database}, 2021.
\newblock URL \url{https://ec.europa.eu/eurostat/data/database}.
\newblock Accessed on 12-12-2021.

\bibitem[{ENTSO-E Transparency Platform}(2021{\natexlab{c}})]{EntsoeTPd}
{ENTSO-E Transparency Platform}.
\newblock {Actual Generation per Production Type}, 2021{\natexlab{c}}.
\newblock URL \url{https://transparency.entsoe.eu/}.
\newblock Accessed on 20-12-2021.

\bibitem[{ENTSO-E Transparency Platform}(2021{\natexlab{d}})]{EntsoeTPf}
{ENTSO-E Transparency Platform}.
\newblock {Forecasted Transfer Capacities - Day Ahead}, 2021{\natexlab{d}}.
\newblock URL \url{https://transparency.entsoe.eu/}.
\newblock Accessed on 20-12-2021.

\bibitem[{JAO Joint Allocation Office}(2021)]{JAO2021}
{JAO Joint Allocation Office}.
\newblock {ATC for Shadow Auction}, 2021.
\newblock URL \url{https://www.jao.eu/implict-allocation}.
\newblock Accessed on 20-12-2021.

\bibitem[{ENTSO-E Transparency Platform}(2021{\natexlab{e}})]{EntsoeTPb}
{ENTSO-E Transparency Platform}.
\newblock {Unavailability of Production and Generation Units},
  2021{\natexlab{e}}.
\newblock URL \url{https://transparency.entsoe.eu/}.
\newblock Accessed on 20-12-2021.

\bibitem[{ENTSO-E Transparency Platform}(2021{\natexlab{f}})]{EntsoeTPc}
{ENTSO-E Transparency Platform}.
\newblock {Generation Forecast - Day ahead}, 2021{\natexlab{f}}.
\newblock URL \url{https://transparency.entsoe.eu/}.
\newblock Accessed on 20-12-2021.

\bibitem[{ENTSO-E Transparency Platform}(2021{\natexlab{g}})]{EntsoeTPg}
{ENTSO-E Transparency Platform}.
\newblock Day-ahead prices, 2021{\natexlab{g}}.
\newblock URL \url{https://transparency.entsoe.eu/}.
\newblock Accessed on 20-12-2021.

\bibitem[Hellwig(2003)]{Hellwig2013}
Mark Hellwig.
\newblock \emph{{Entwicklung und Anwendung parametrisierter
  Standard-Lastprofile}}.
\newblock {Dissertation, Technische Universität München}, 2003.
\newblock {Dissertation, Technische Universität München}.

\bibitem[DENA(2010)]{Dena2010}
DENA.
\newblock {dena-Netzstudie II. Integration erneuerbarer Energien in die
  deutsche Stromversorgung im Zeitraum 2015 – 2020 mit Ausblick 2025}, 2010.
\newblock URL
  \url{https://www.dena.de/fileadmin/user\_upload/Download/Dokumente/Studien\_\_\_Umfragen/Endbericht\_dena-Netzstudie\_II.PDF}.

\bibitem[Kunz et~al.(2017)Kunz, Kendziorski, Schill, Weibezahn, Zepter, von
  Hirschhausen, Hauser, Zech, M\"{o}st, Heidari, Felten, and
  Weber]{Kunz2017Electricity}
Friedrich Kunz, Mario Kendziorski, Wolf-Peter Schill, Jens Weibezahn, Jan
  Zepter, Christian~R. von Hirschhausen, Philipp Hauser, Matthias Zech, Dominik
  M\"{o}st, Sina Heidari, Bj\"{o}rn Felten, and Christoph Weber.
\newblock Electricity, heat, and gas sector data for modeling the german
  system.
\newblock DIW Data Documentation~92, 2017.
\newblock URL \url{http://hdl.handle.net/10419/173388}.

\bibitem[Maciejowska et~al.(2021)Maciejowska, Nitka, and
  Weron]{Maciejowska2021}
Katarzyna Maciejowska, Weronika Nitka, and Tomasz Weron.
\newblock Enhancing load, wind and solar generation for day-ahead forecasting
  of electricity prices.
\newblock \emph{Energy Economics}, 99:\penalty0 105273, 7 2021.
\newblock ISSN 01409883.
\newblock \doi{10.1016/j.eneco.2021.105273}.

\bibitem[Lütkepohl(2005)]{Luetkepohl2005}
Helmut Lütkepohl.
\newblock \emph{New introduction to multiple time series analysis}.
\newblock Springer, Berlin, 2005.
\newblock ISBN 3540401725; 3540262393; 9783540262398; 9783540401728.

\bibitem[Hyndman and Athanasopoulos(2021)]{Hyndman21}
Rob~J. Hyndman and George Athanasopoulos.
\newblock \emph{Forecasting: principles and practice}.
\newblock Otexts: Melbourne, Australia, Lexington, Ky., 2021.
\newblock ISBN 0987507133.
\newblock Accessed on 04-02-2022.

\bibitem[Box et~al.(2015)Box, Jenkins, Reinsel, and Ljung]{BoxJenkins2016}
George E.~P. Box, Gwilym~M. Jenkins, Gregory~C. Reinsel, and Greta~M. Ljung.
\newblock \emph{Time series analysis: forecasting and control}.
\newblock Wiley series in probability and statistics. John Wiley and Sons Inc.,
  Hoboken, New Jersey, fifth edition edition, 2015.
\newblock ISBN 9781118675021.
\newblock \doi{10.1111/jtsa.12194}.

\bibitem[Koenker and Bassett(1978)]{Koenker1978}
Roger Koenker and Gilbert Bassett.
\newblock Regression {Q}uantiles.
\newblock \emph{Econometrica}, 46\penalty0 (1):\penalty0 33, jan 1978.
\newblock \doi{10.2307/1913643}.

\bibitem[Grothe(2013)]{Grothe2013}
Oliver Grothe.
\newblock A higher order correlation unscented {K}alman filter.
\newblock \emph{Applied mathematics and computation}, 219\penalty0
  (17):\penalty0 9033--9042, 2013.
\newblock ISSN 0096-3003, 1873-5649.
\newblock \doi{10.1016/j.amc.2013.03.019}.

\bibitem[Muesgens(2006{\natexlab{b}})]{Muesgens:2006}
Felix Muesgens.
\newblock {Quantifying Market Power in the German Wholesale Electricity Market
  Using a Dynamic Multi-Regional Dispatch Model}.
\newblock \emph{The Journal of Industrial Economics}, 54\penalty0 (4):\penalty0
  471--498, 2006{\natexlab{b}}.
\newblock \doi{10.1111/j.1467-6451.2006.00297.x}.

\bibitem[Muesgens and Neuhoff(2006)]{MuesgensandNeuhoff:2006}
Felix Muesgens and Karsten Neuhoff.
\newblock {Modelling Dynamic Constraints in Electricity Markets and the Costs
  of Uncertain Wind Output}, 2006.

\bibitem[Marcjasz et~al.(2018)Marcjasz, Serafin, and Weron]{Marcjasz2018}
Grzegorz Marcjasz, Tomasz Serafin, and Rafał Weron.
\newblock {Selection of Calibration Windows for Day-Ahead Electricity Price
  Forecasting}.
\newblock \emph{Energies}, 11:\penalty0 2364, 9 2018.
\newblock ISSN 1996-1073.
\newblock \doi{10.3390/EN11092364}.

\bibitem[DeMiguel et~al.(2009)DeMiguel, Garlappi, and
  Uppal]{demiguel2009optimal}
Victor DeMiguel, Lorenzo Garlappi, and Raman Uppal.
\newblock Optimal versus naive diversification: How inefficient is the 1/n
  portfolio strategy.
\newblock \emph{Review of Financial Studies}, 22\penalty0 (5):\penalty0
  1915--1953, 2009.
\newblock \doi{10.1093/rfs/hhm075}.

\bibitem[Kuntz and Muesgens(2007)]{Kuntz2007}
Ludwig Kuntz and Felix Muesgens.
\newblock Modelling start-up costs of multiple technologies in electricity
  markets.
\newblock \emph{Mathematical Methods of Operations Research}, 66\penalty0
  (1):\penalty0 21--32, 2007.
\newblock ISSN 14322994.
\newblock \doi{10.1007/s00186-007-0148-y}.

\bibitem[Gneiting et~al.(2007)Gneiting, Balabdaoui, and
  Raftery]{gneiting2007probabilistic}
Tilmann Gneiting, Fadoua Balabdaoui, and Adrian~E Raftery.
\newblock Probabilistic forecasts, calibration and sharpness.
\newblock \emph{Journal of the Royal Statistical Society: Series B (Statistical
  Methodology)}, 69\penalty0 (2):\penalty0 243--268, 2007.
\newblock \doi{10.1111/j.1467-9868.2007.00587.x}.

\end{thebibliography}

% % Biography
% % Here goes the biography details.

% % Here goes the biography details.

%------------------------------------------------------------------------------------------------------------------------
%  Appendix
%************************************************************************************************************************
%------------------------------------------------------------------------------------------------------------------------
\appendix
\renewcommand{\thetable}{\Alph{section}.\arabic{table}}
\setcounter{table}{0}
\renewcommand{\thefigure}{\Alph{section}.\arabic{figure}}
\setcounter{figure}{0}
\setcounter{section}{1}

%\newpage
\begin{appendix}
	
	\section{Additional Analysis Material}
	This appendix provides additional figures for Chapters \ref{H_sec.results.Hybrid} and \ref{H_sec.analysis model steps}. 
	\begin{figure}[h!]
		\centering
		\includegraphics[width=0.9\linewidth]{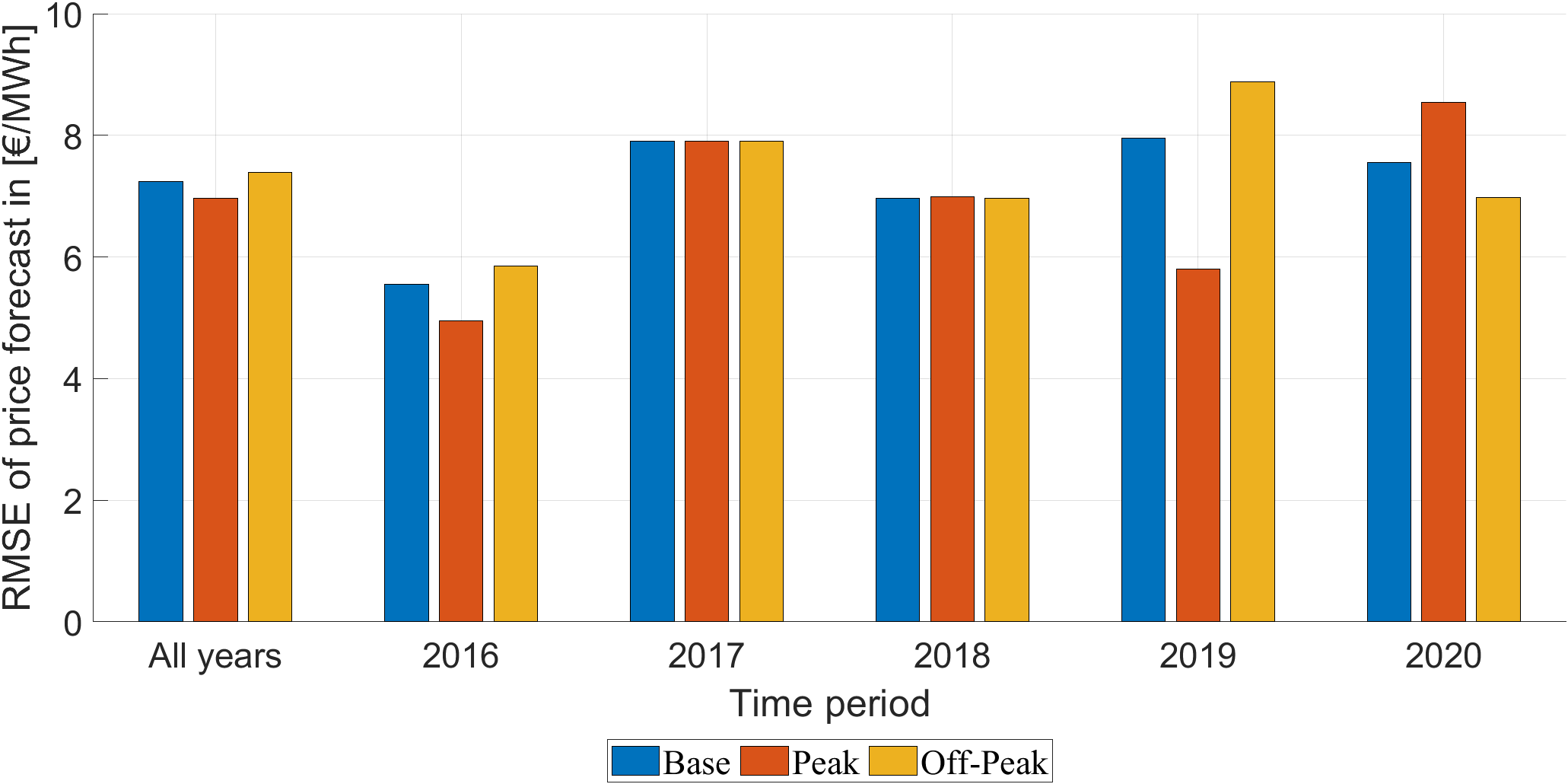}
		\caption{RMSE of the day-ahead price forecast generated with the LEAR benchmark model for base, peak and off-peak hours.}
		\label{H_fig_RMSE_BPOP_LEAR}
	\end{figure}
	
	\begin{figure}[h!]
		\centering
		\includegraphics[width=0.9\linewidth]{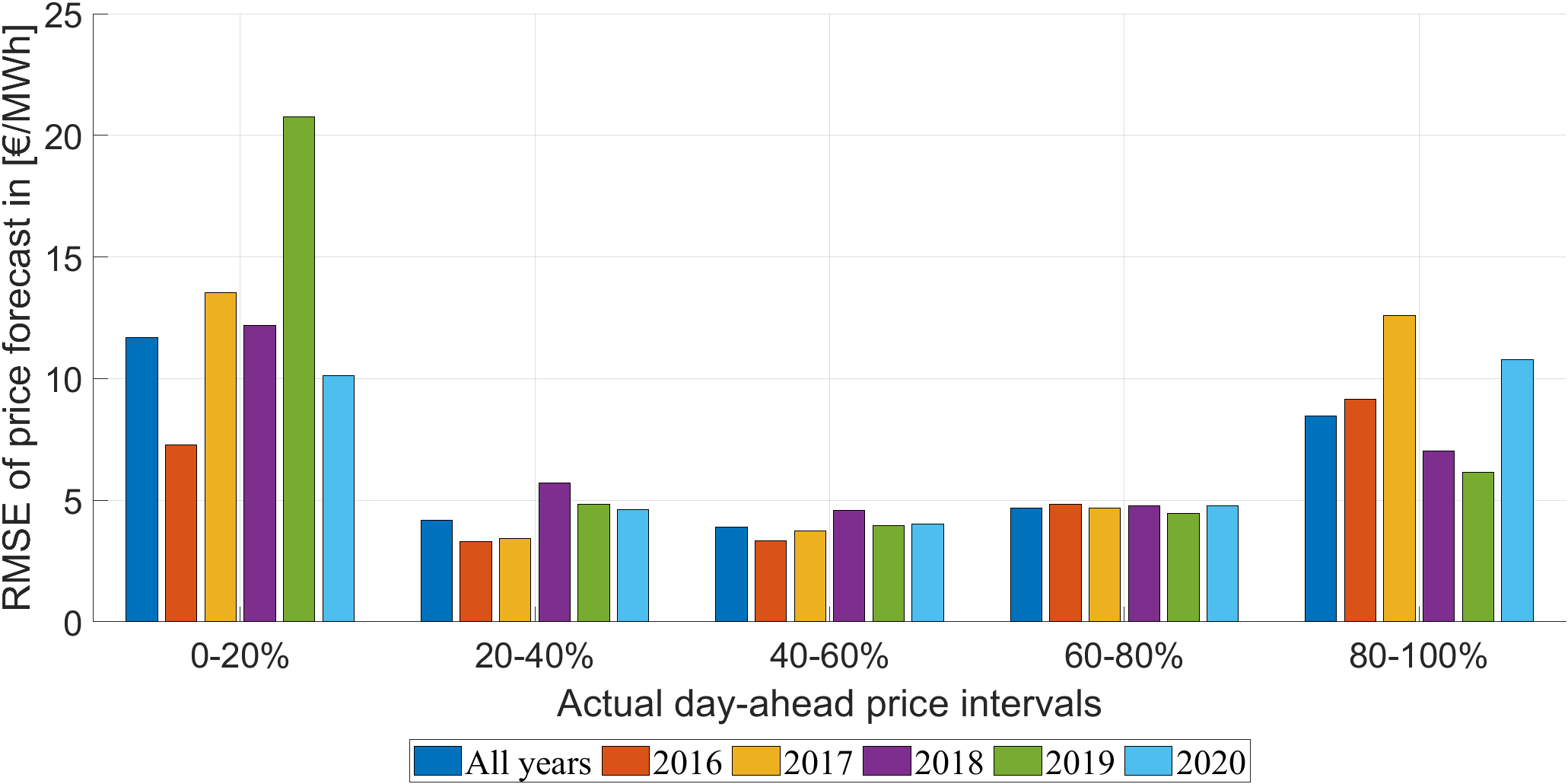}
		\caption{RMSE of the day-ahead price forecast generated with the LEAR benchmark model for hours at different day-ahead price quantiles.}
		\label{H_fig_RMSE_pricequantiles_LEAR}
	\end{figure}
	
	\begin{figure}[h!]
		\centering
		\includegraphics[width=0.9\linewidth]{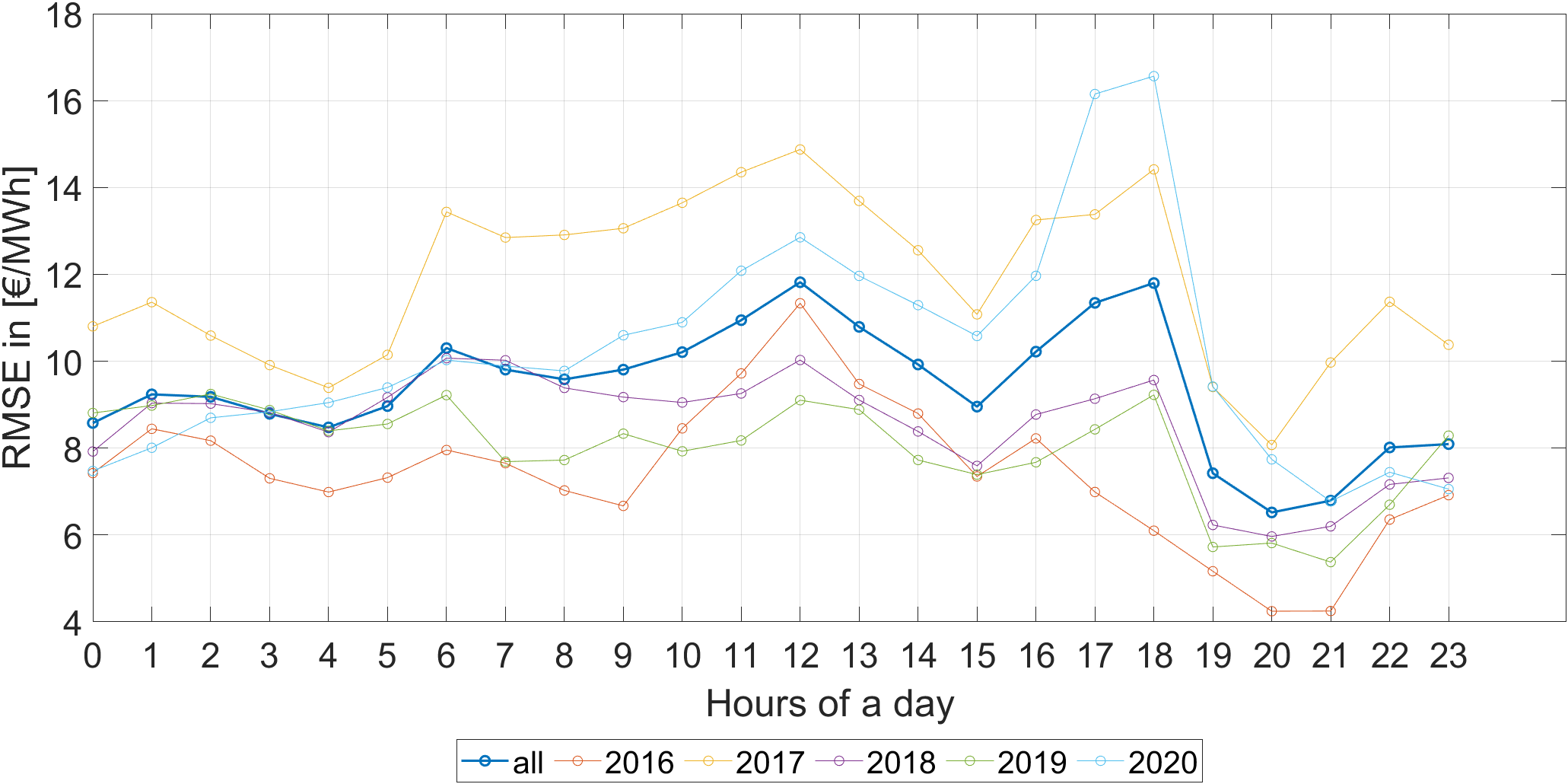}
		\caption{RMSE for price estimators for each hour of the day and each year after the energy system optimisation step in [\euro{}/MWh].}
		\label{H_fig_RMSE_ESM_hourly}
	\end{figure}
	
	\begin{figure}[h!]
		\centering
		\includegraphics[width=0.9\linewidth]{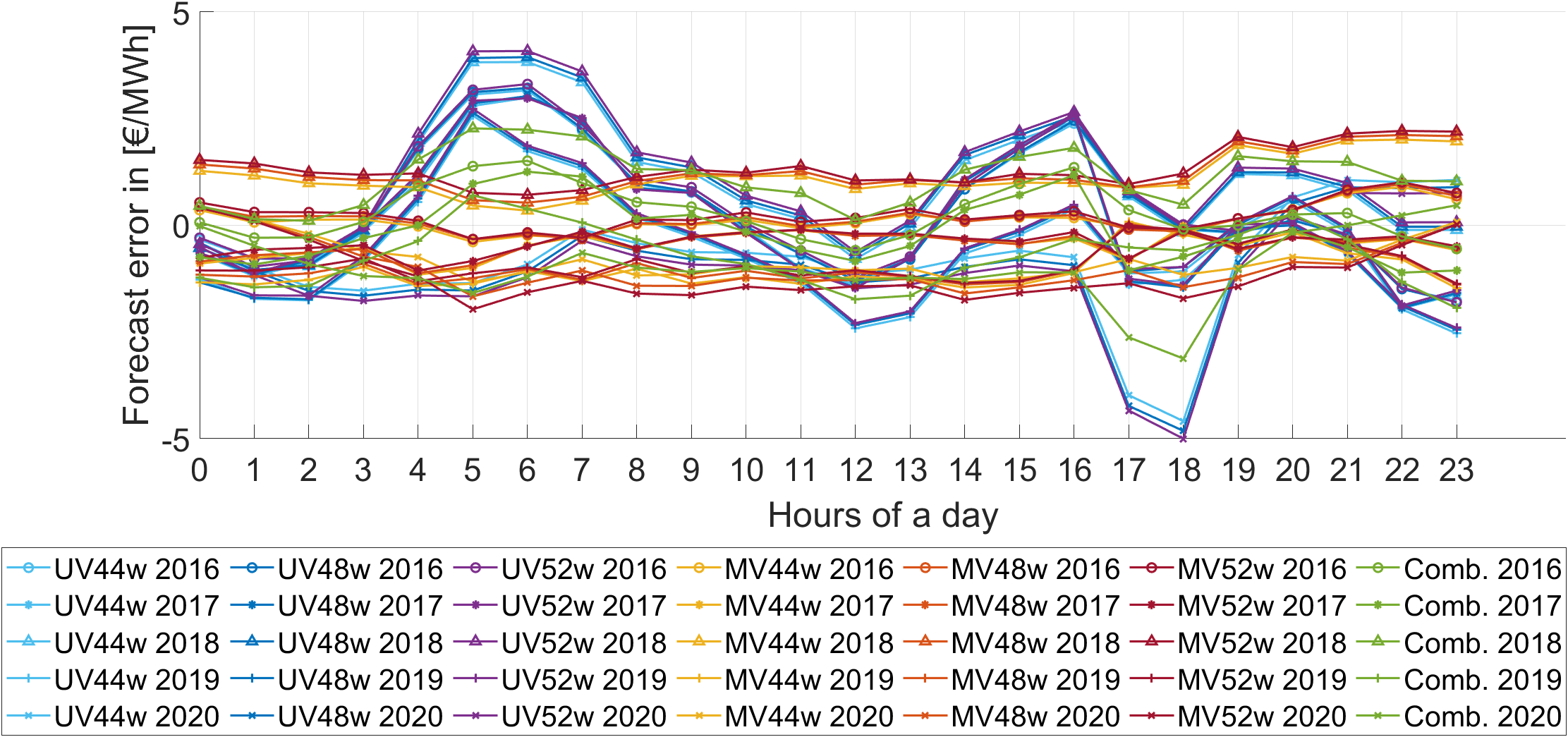}
		\caption{Mean price forecast errors of the individual sub-models and the combined sub-models for the hours of a day in each year in [\euro{}/MWh].}
		\label{H_fig_error_uvmv_hourwise}
	\end{figure}
	
\begin{figure}[h!]
	\centering
	\includegraphics[width=0.9\linewidth]{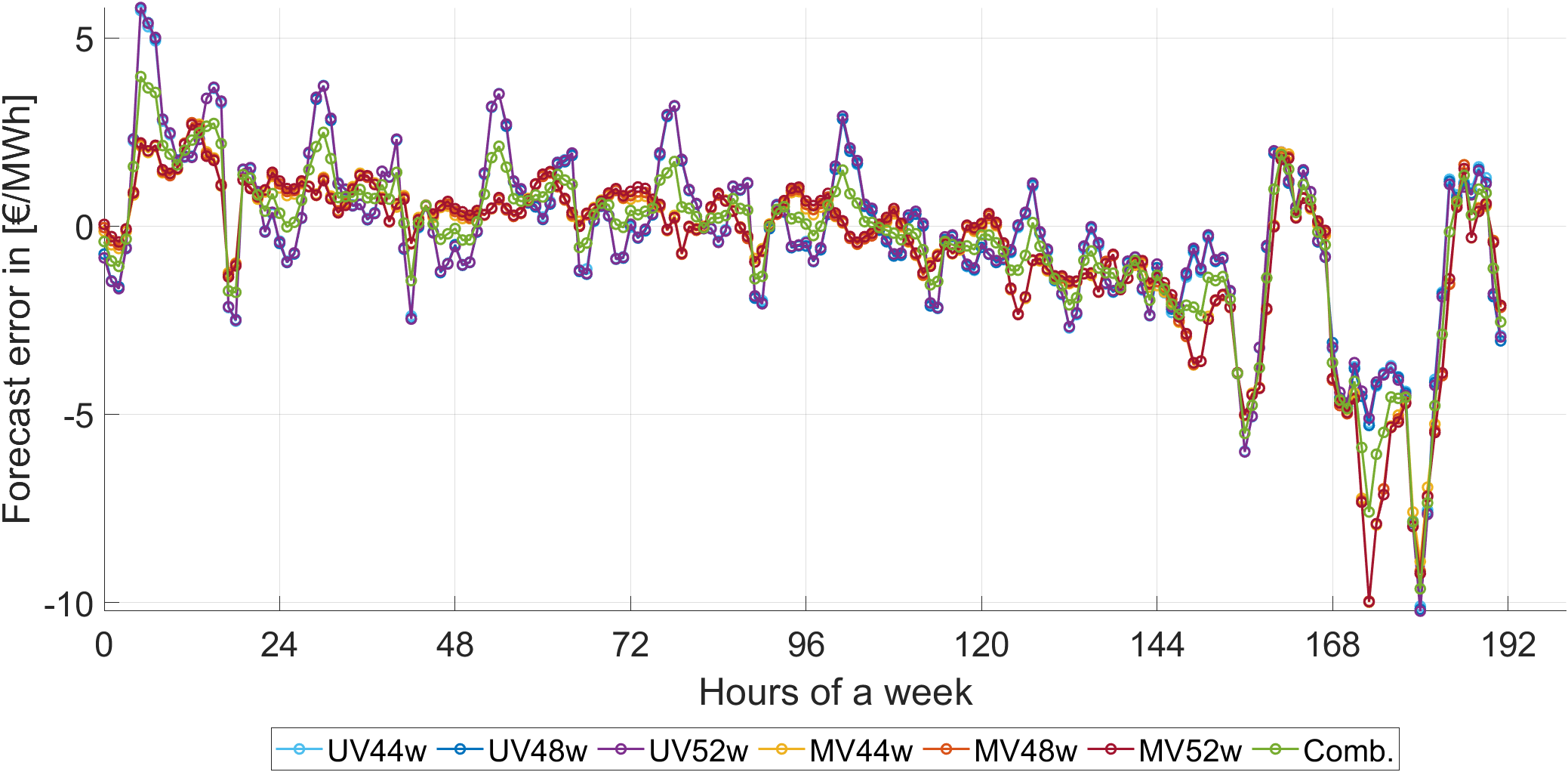}
	\caption{Mean price forecast errors of the individual sub-models and the combined sub-models for the hours of a week (including holidays, hour 168 to 192) in [\euro{}/MWh].}
	\label{H_fig_error_uvmv_hourwise192}
\end{figure}

\cleardoublepage
\section{Nomenclature Energy System Model}
\label{H_Nomenclature}
\begin{figure}[h!]
	\centering
	\includegraphics[width=1\linewidth]{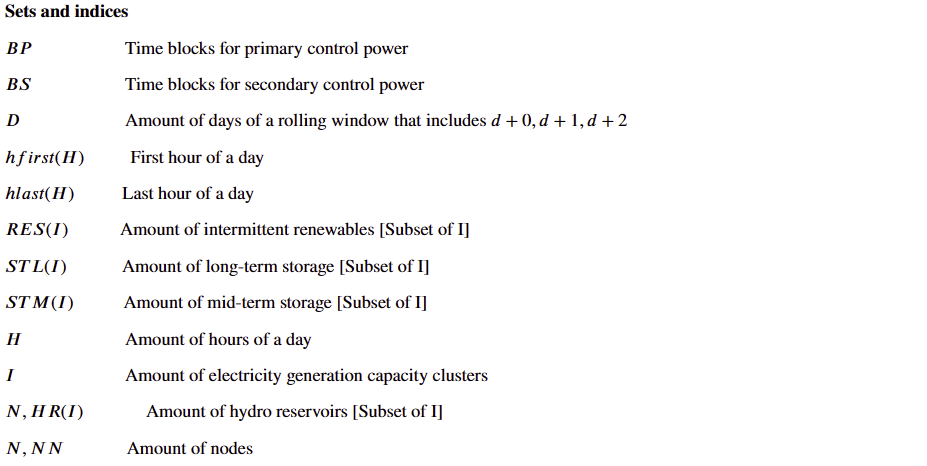}
\end{figure}
\begin{figure}[h!]
	\centering
	\includegraphics[width=1\linewidth]{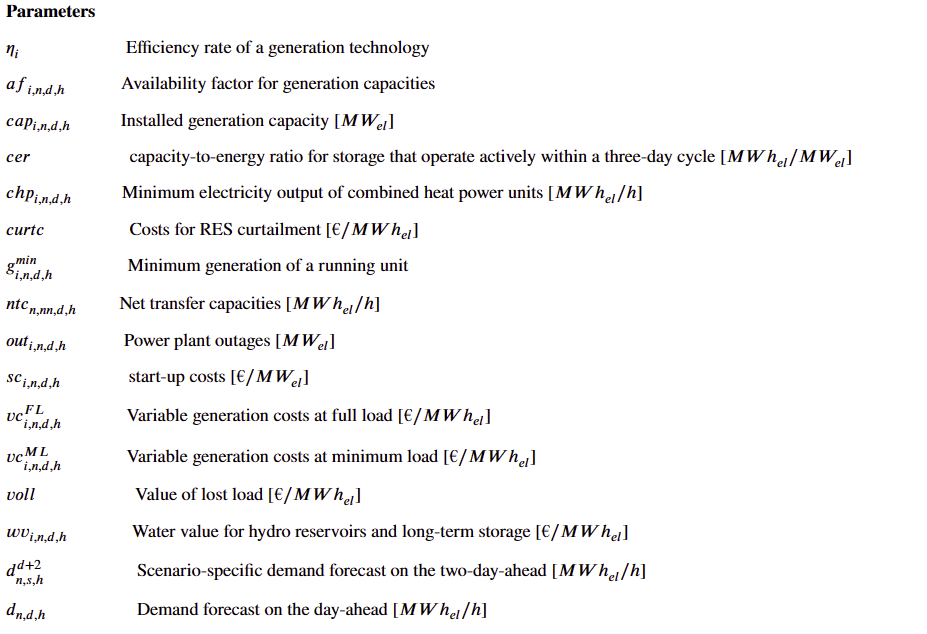}
\end{figure}
\begin{figure}[h!]
	
	\includegraphics[width=1\linewidth]{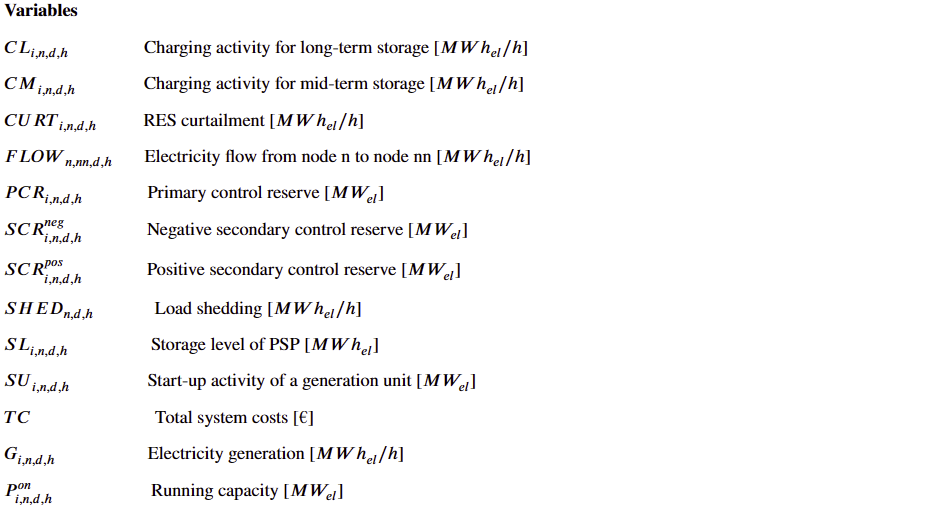}
\end{figure}
	
\end{appendix}

\end{document}